\pdfoutput=1
\documentclass[a4paper]{article}
\usepackage[compatibility=false]{caption}
\usepackage{jheppub}
\usepackage{dsfont}
\usepackage{amsmath,amssymb,amscd,amsfonts,mathtools,graphicx}
\usepackage{comment}
\usepackage{tabu}

\usepackage[toc,page]{appendix}
\usepackage{epsfig}
\usepackage{epstopdf}
\usepackage{latexsym}
\usepackage{enumerate} 
\usepackage{placeins}
\usepackage{floatrow}
\usepackage{caption}
\usepackage{booktabs}
\usepackage{bbm}
\usepackage{color}
\usepackage{physics}
\usepackage{tensor}
\usepackage{tikz}
\usepackage{cancel}
\usetikzlibrary{matrix}
\usetikzlibrary{decorations.markings,calc,shapes,decorations.pathmorphing,patterns,decorations.pathreplacing,arrows.meta}
\usetikzlibrary{positioning}
\usepackage{hyperref}
\usepackage{tabularx}
\newcolumntype{Y}{>{\centering\arraybackslash}X}
\renewcommand{\arraystretch}{1.5}

\newcommand{\be}{\begin{equation}}
\newcommand{\ee}{\end{equation}}
\newcommand{\bea}{\begin{eqnarray}}
\newcommand{\eea}{\end{eqnarray}}

\usepackage{soul}

\definecolor{mycolor}{RGB}{255,238,140}

\DeclareMathOperator\vol{Vol}
\newcommand{\overbar}[1]{\mkern 1.5mu\overline{\mkern-1.5mu#1\mkern-1.5mu}\mkern 1.5mu}

\begin{document}

\begin{flushright}HIP-2026-8/TH\\
CCTP-2026-12\\
ITCP-2026-12\end{flushright}

\title{Covariant unification of holographic c-functions}

\author[a,b]{Niko Jokela,}
\affiliation[a]{Department of Physics, University of Helsinki, P.O. Box 64, FI-00014, University of Helsinki, Finland}
\affiliation[b]{Helsinki Institute of Physics, P.O. Box 64, FIN-00014 University of Helsinki, Finland}
\emailAdd{niko.jokela@helsinki.fi}
\author[c]{Jani Kastikainen,}
\affiliation[c]{Institute for Theoretical Physics and Astrophysics and W\"urzburg-Dresden Cluster of Excellence
ctd.qmat, Julius-Maximilians-Universit\"at W\"urzburg, Am Hubland, 97074 W\"urzburg, Germany}
\emailAdd{jani.kastikainen@uni-wuerzburg.de}
\author[d]{Carlos Nunez,}
\affiliation[d]{Centre for Quantum Fields and Gravity, Department of Physics, Swansea University, Swansea SA2 8PP, United Kingdom}
\emailAdd{c.nunez@swansea.ac.uk}
\author[e]{Jos\'e Manuel Pen\'in,}
\affiliation[e]{Crete Center for Theoretical Physics, Institute for Theoretical and Computational Physics, Department of Physics, University of Crete 71003 Heraklion, Greece}
\emailAdd{jmanpen@gmail.com} 
\author[a,b]{\\ Helime Ruotsalainen,}
\emailAdd{helime@firstqfm.com}

\abstract{We propose a covariant holographic c-function, defined directly in a top-down background and constructed from the extrinsic curvature of codimension-two slices of the bulk geometry. The definition does not rely on a special choice of coordinates or on the existence of a consistent dimensional reduction. We show that it unifies previous foliation-based holographic c-functions into a single covariant formula, reducing to them in the appropriate limits. We evaluate the covariant expression in a range of top-down string backgrounds, including conformal models, confining geometries, flows across dimensions, and the Klebanov--Murugan geometry, in which the holographic radial direction mixes with internal coordinates and which is not the uplift of a lower-dimensional solution. In all cases, the c-function behaves as expected: it interpolates monotonically between AdS fixed points when they are present and decreases towards zero in gapped infrared regions, while in the Klebanov--Murugan case we recover the correct fixed-point values and find evidence for monotonicity. We highlight open conceptual issues, including: the lack of a universal covariant definition of the holographic radial direction in the presence of a nontrivial internal manifold; the derivation of the flow from a bulk action; and the relation to the entanglement c-function.}

\maketitle
\flushbottom
\setcounter{page}{2}

\newpage

\section{Introduction}\label{sec:introduction}

Quantifying the number of degrees of freedom in a quantum field theory (QFT) is central to understanding its long-distance physics. Along renormalization group (RG) flow, the effective number of degrees of freedom should decrease, and this intuition is sharpened in two dimensions by the c-theorem of Zamolodchikov~\cite{Zamolodchikov:1986gt}. Extending this idea to higher dimensions has proven considerably more subtle~\cite{Jafferis:2010un,Jafferis:2011zi,Klebanov:2011gs,Komargodski:2011vj}. Holography has played a significant role in advancing the subject: in the AdS/CFT correspondence an RG flow is represented as a higher-dimensional spacetime whose radial direction encodes the energy scale, and c-theorems become geometric statements about the bulk~\cite{deBoer:1999tgo}.

Early holographic proposals for candidate measures of degrees of freedom, so-called c-functions, focused largely on energy conditions, especially the null energy condition (NEC) in gravity~\cite{Hawking:1973uf}. The NEC often ensures the monotonicity of various geometric quantities that underlie many constructions of classical holographic c-functions in both bottom-up and top-down settings. On the bottom-up side, the first example of this kind applicable to domain-wall geometries in Einstein gravity was introduced in~\cite{Freedman:1999gp} and later generalized to include higher-derivative corrections in~\cite{Myers:2010tj,Myers:2010xs}. A covariant construction based on the expansion of null congruences was proposed in~\cite{Sahakian:1999bd}; see also~\cite{Alvarez:1998wr,Bousso:1999xy}. More recently, the study of anisotropic and boomerang RG flows~\cite{Donos:2017ljs,Donos:2017sba,Hoyos:2021vhl} has led to modified versions of these proposals tailored to backgrounds with reduced spatial symmetry or nontrivial IR structure~\cite{Giataganas:2017koz,Hoyos:2020zeg,Cremonini:2020rdx,Caceres:2022hei,Caceres:2023mqz}. In addition, non-covariant top-down extensions were proposed in~\cite{Macpherson:2014eza,Bea:2015fja,Merrikin:2022yho}.

However, it is now understood that energy conditions are not a robust guide. At the quantum level, the NEC is generically violated even in flat space~\cite{Kontou:2020bta}. In holography, $1/N$ corrections correspond to quantum corrections to the bulk stress tensor, where one should not expect a simple classical energy condition to persist. The situation is further complicated in top-down models. Ten-dimensional and lower-dimensional NECs are independent requirements. Indeed, perfectly healthy 10-dimensional supergravity backgrounds can exhibit lower-dimensional NEC violations~\cite{Hoyos:2021vhl}. This lower-dimensional violation implies that foliation-based holographic c-functions are inevitably non-monotonic, highlighting a significant limitation of relying solely on the reduced effective theory. Taken together, these observations suggest that classical energy conditions provide too restrictive a foundation for constructing general holographic c-functions.

In this work, we instead take geometry and covariance as guiding principles and construct a covariant holographic c-function directly in the top-down spacetime; see equations \eqref{eq:covariant_c_function_V2} and \eqref{eq:covariant_c_function} below. The construction relies only on the staticity and spatial translation invariance of the Einstein-frame bulk metric in the field theory directions, on the existence of a radial foliation dual to the energy scale of the boundary theory, and on the identification of a compact internal manifold. It is defined in terms of the extrinsic curvature of codimension-two intersections of the constant-time and constant-radius hypersurfaces that foliate the geometry. The resulting c-function has the expected properties: it is constant and stationary at end-points of the flow described by asymptotically locally AdS spacetimes, and, when dimensional reduction is possible, its value is invariant under this reduction.

Our proposal unifies the c-function constructions discussed above within a single covariant formula, uncovering their common geometric origin. When a dimensional reduction exists, we show that our c-function is proportional both to the covariant c-function of~\cite{Sahakian:1999bd} and to the non-covariant proposal of~\cite{Caceres:2023mqz}. In top-down geometries where the non-covariant constructions of~\cite{Macpherson:2014eza,Bea:2015fja,Merrikin:2022yho} apply, our formula reproduces them. At the same time, our proposal remains well-defined in genuinely top-down geometries that are not obtained as uplifts of known lower-dimensional solutions, where the assumptions underlying~\cite{Macpherson:2014eza,Bea:2015fja,Merrikin:2022yho} may fail. Finally, our construction also applies to holographic RG flows with anisotropic scaling, such as hyperscaling-violating flows, although such examples will not be the focus of this work.

We test the proposal on a sequence of top-down backgrounds of increasing complexity. We first perform simple checks in geometries dual to conformal theories, realized as warped products of an AdS factor and a compact internal space, including $\text{AdS}_5\times \mathcal{X}_5$ solutions, M5-branes, M2-branes, and the D1-D5 system. We also study D-brane backgrounds. We then study several top-down backgrounds dual to confining theories, where the covariant c-function recovers the correct UV fixed-point value and decreases monotonically towards zero in the deep IR, reflecting the disappearance of deconfined degrees of freedom. We then extend this to the baryonic branch of the Klebanov--Strassler theory. In all these cases, we recover the previous results obtained using the non-covariant proposal of~\cite{Macpherson:2014eza,Bea:2015fja,Merrikin:2022yho}.

Finally, we apply our covariant c-function to a top-down geometry where previously known proposals cannot be used: the Klebanov--Murugan solution~\cite{Klebanov:2007us}. This solution describes an RG flow from the Klebanov--Witten theory~\cite{Klebanov:1998hh} in the UV to $\mathcal{N}=4$ SYM theory in the IR, but it has unusual features from the bulk perspective. In particular, the radial component of the bulk metric depends on the internal coordinates in a way that obstructs any dimensional reduction to lower-dimensional supergravity. The absence of such a reduction is further reflected in the fact that the IR region of the geometry is located at a different value of an internal coordinate than the UV region~\cite{Klebanov:2007us}. This raises a previously unnoticed puzzle in the covariant definition of the holographic radial direction, which in top-down geometries can mix with internal directions along the flow. We show how this feature can be incorporated into our covariant proposal, which then naturally reproduces the correct fixed-point values in both the UV and the IR. Furthermore, we establish monotonicity at leading order in the UV expansion of the Klebanov--Murugan solution.

The paper is organized as follows. Section~\ref{sec:proposal} introduces the covariant c-function, establishes its main properties, and relates it to previous constructions. Section~\ref{sec:tests} tests the proposal on a sequence of holographic backgrounds. Section~\ref{sec:KlebanovMurugan} focuses on the Klebanov--Murugan solution, illustrating the usefulness of the covariant approach in settings where dimensional reduction is absent. Section~\ref{sec:discussion} presents a broader conceptual discussion, including field-theoretic aspects, the relation to holographic entanglement entropy, and the possibility of a bulk-action interpretation. Appendices~\ref{app:Schur_complement}, \ref{app:newton_constant}, \ref{app:null_expansion}, and \ref{sahakian-examples} collect technical details, while Appendices~\ref{appendix-details-confining} and \ref{app:C5} provide additional examples of holographic RG flows to which our c-function applies.

\section{A proposal for a covariant holographic c-function}\label{sec:proposal}

We begin by defining a covariant holographic c-function associated with a choice of radial foliation of the bulk spacetime. The novelty of the construction is that it is formulated directly in the top-down spacetime and therefore naturally incorporates the geometry of the compact internal manifold.

\subsection{Definition of a covariant c-function}\label{section-definitions}

Consider a $(d+1+p)$-dimensional spacetime $(\mathcal{M},\mathcal{G})$ with a Lorentzian metric $\mathcal{G}_{\mu\nu}$ where $\mu,\nu = 0,\ldots, d+p$. We assume that the spacetime is asymptotically locally AdS$_{d+1}\times \mathcal{X}_p$, where $\mathcal{X}_p$ is a $p$-dimensional compact internal manifold with finite volume. Thus there is a flat Lorentzian $\mathbb{R}^{d-1,1}$ conformal boundary where a dual $d$-dimensional UV CFT is supported. The spacetime $(\mathcal{M},\mathcal{G})$ is taken to represent a renormalization group flow of the dual theory that preserves staticity and $(d-1)$-dimensional spatial translation invariance. Thus we assume that the metric $\mathcal{G}$ is static and that the $(d-1)$-dimensional spatial translation group acts on $\mathcal{M}$ as diffeomorphisms with $\mathcal{G}$ invariant under this action. As a result, there are $d$ mutually commuting Killing vectors $(\xi^\mu_t,\xi^\mu_a)$, with $a = 1, \ldots, d-1$, where $\xi^\mu_t$ is timelike and $\xi^\mu_a$ are spacelike. Together, they determine the field theory directions on $\mathcal{M}$ which we parametrize by coordinates $(t,x^a)$.

\paragraph{The bulk metric.} We denote the coordinates spanning directions orthogonal to the field theory directions by $(r,y^i)$ with $i = 1,\ldots, p$. We take $r$ to be the Fefferman--Graham (FG) radial coordinate while $y^i$ parametrizes the internal manifold. We assume that the metric asymptotes at $r\rightarrow \infty$ to the warped product $\text{AdS}_{d+1}\times \mathcal{X}_p$ form
\begin{equation}
    ds^2_{d+1+p} \sim \Delta(y)^2\,\biggl[ r^2\,(-dt^2 + \delta_{ab}\,dx^adx^b) + \frac{dr^2}{r^2}\biggr] + h_{(0)ij}(y)\,(dy^i + A^i_{(0)a}(y)\,dx^a)(dy^j + A^j_{(0)b}(y)\,dx^b)\,,
    \label{eq:UV_asymptotics_full_metric}
\end{equation}
where $h_{(0)ij}(y)$ is the metric of $\mathcal{X}_p$, $\Delta(y)$ is an internal coordinate dependent warp factor and $\delta_{ab}$ is the Kronecker delta. This is the most general bulk metric in FG gauge that has an asymptotic $\text{AdS}_{d+1}$ factor, and therefore describes a UV fixed-point CFT on the conformal boundary $r = \infty$, equipped with a flat Minkowski background metric. In the usual case, $\Delta(y) = L_{\text{UV}}$ is constant and determines the curvature radius of the asymptotic AdS factor.

The unit normalized Killing vector $t^\mu\propto \xi^\mu_t$ provides a foliation of $\mathcal{M}$ with spacelike slices $Q_t$ which we parametrize by coordinates $X^\alpha = (x^a,r,y^i) $ and denote its induced Euclidean metric by $G_{\alpha\beta}$. Then the full metric $\mathcal{G}_{\mu\nu}$ takes the form
\begin{equation}
    ds^2_{d+1+p} = -\mathcal{N}(r,y)^2\, dt^2 + G_{\alpha\beta}(r,y)\,dX^\alpha dX^\beta\,,
    \label{eq:full_metric_0}
\end{equation}
where staticity and spatial translation invariance in field theory directions imply that the metric components are independent of the field theory coordinates $(t,x^a)$ and that the shift vector vanishes. 

All of the top-down geometries considered in this paper take the following simple form in $(r,y^i)$ coordinates. Globally for all $r$, the spatial metric is assumed to be of the form
\begin{equation}
    G_{\alpha\beta}(r,y)\,dX^\alpha dX^\beta = B(r,y)\,dr^2 + H_{AB}(r,y)\,dY^A dY^B\,,
    \label{eq:G_metric}
\end{equation}
where the induced metric $H_{AB}$ on the constant-$r$ slices inside $Q_0$ in $Y^A = (x^a,y^i)$ coordinates takes the form
\begin{equation}
    H_{AB}(r,y)\,dY^A dY^B= g_{ab}(r,y)\,dx^adx^b + h_{ij}(r,y)\,(dy^i + A^i_a(r,y)\,dx^a)(dy^j + A^j_b(r,y)\,dx^b)\,,
    \label{eq:H_metric}
\end{equation}
which is the most general warped product metric on $\mathbb{R}^{d-1}\times \mathcal{X}_p$ whose components do not depend on the field theory directions $(t,x^a)$. Thus the $(d+1+p)$-dimensional Lorentzian metric in $(t,x^a,r,y^i)$ coordinates takes the form
\begin{equation}
     ds^2_{d+1+p} = -\mathcal{N}^2 dt^2  +g_{ab}\,dx^adx^b+ B\,dr^2 + h_{ij}\,(dy^i + A^i_a\,dx^a)(dy^j + A^j_b\,dx^b)\,,
     \label{eq:full_metric}
\end{equation}
where all components $\mathcal{N}$, $g_{ab}$, $B$, $h_{ij}$, and $A^i_a$ are generic functions of $(r,y^i)$. This class of metrics is extremely general and encompasses practically all top-down solutions in supergravity.

\paragraph{The holographic radial direction.} The dual field theory is equipped with an energy scale $E>0$ which parametrizes the RG flow. In holography, the RG flow is realized geometrically in the bulk as a foliation of $\mathcal{M}$ by codimension-one timelike slices $M_E$ parametrized by the energy scale $E$. The slices correspond to the equipotential surfaces $E = \rho(r,y)$ of a scalar field $\rho $ which is the covariant description of the energy scale in the bulk. In the UV $r\rightarrow \infty$, the energy scale is parametrized by the $r$-coordinate defined by \eqref{eq:UV_asymptotics_full_metric} so that we require $\lim_{r\rightarrow \infty}\rho(r,y) = r$, but this is not true all along the flow in general and there can be mixing with the internal directions $y$ in the interior. To our knowledge, there has not been a general proposal for the energy scale in terms of a scalar field $\rho = \rho(r,y)$ in top-down geometries, and therefore, in what follows, we keep the discussion generic and assume it can be found. Furthermore, we simplify the notation and denote the slices as $M_\rho$ with $\rho$ as a parameter instead of $E$.

Given a bulk dual $\rho = \rho(r,y)$ of the field theory energy scale $E$, we define the holographic radial direction by the vector field orthogonal to constant-$\rho$ surfaces. The spacelike unit vector field pointing in the holographic radial direction is therefore given by
\begin{equation}
    n^\mu = \frac{\mathcal{G}^{\mu\nu}\partial_\nu \rho}{\sqrt{\mathcal{G}^{\mu\nu}\,\partial_\mu \rho\,\partial_\nu \rho}}\,,
    \label{eq:n_vector_field}
\end{equation}
where $\mathcal{G}_{\mu\nu}$ is the full Lorentzian metric and we assume $\mathcal{G}^{\mu\nu}\,\partial_\mu \rho\,\partial_\nu \rho>0$. Note that $n^\mu$ is not proportional to $\delta^\mu_r$ in general and can have components in the internal directions. It follows that we can define coordinates $\hat{y}^i$ such that $(\rho,\hat{y}^i)$ parametrize the directions orthogonal to the field theory directions $(t,x^a)$. In other words, the coordinates $\hat{y}^i(r,y)$ are constant in the holographic radial direction, \emph{i.e.}, solve the equations
\begin{equation}
    n^\mu\,\partial_\mu \hat{y}^i = 0\,.
    \label{eq:yhat}
\end{equation}
The coordinates $(\rho,\hat{y}^i)$ are related to $(r,y^i)$, by a coordinate transformation
\begin{equation}
    \rho = \rho(r,y)\ ,\quad \hat{y}^i = \hat{y}^i(r,y)\,,\label{coord-transf}
\end{equation}
which satisfy $\rho(r,y) \rightarrow r$ and $\hat{y}^i(r,y)\rightarrow y^i$ as $r\rightarrow \infty$. When $\rho = \rho(r)$ such that $\rho'(r)>0$ is an invertible function, the complete flow can be equivalently parametrized by $r$ and we can set $\hat{y}^i = y^i$. Below, the Klebanov--Murugan background of Section \ref{sec:KlebanovMurugan} provides an explicit top-down example in which the direction of the holographic radial coordinate is not simply $r$ but instead mixes with internal coordinates along the flow.

\begin{figure}[t]
    \centering
    \begin{tikzpicture}[
        x=0.014cm,
        y=-0.014cm, 
        line cap=round,
        line join=round,
        >={Stealth[length=8pt,width=7pt]}
    ]

        \draw[line width=1.6pt] (68,268) -- (511,268);

        \draw[line width=1.6pt] (389,97) -- (389,444);
        \draw[line width=1.6pt] (511,97) -- (511,444);

        \fill[black] (389,268) circle (2.5pt);

        \draw[->, line width=0.9pt] (347,22) -- (427,22);

        \draw[->, line width=0.9pt] (389,153) -- (447,153) ;

        \draw[->, line width=0.9pt] (209,268) -- (209,210);

        \draw[->, line width=0.9pt] (536,307) -- (536,225);

        \draw[->, line width=0.9pt]
            (300,329) .. controls (327,325) and (357,306) .. (378,284);

        \node[anchor=east] at (360,350) {$\Sigma_\rho = Q_0 \cap M_\rho$};

        \node[anchor=south] at (209,205) {$t^\mu$};
        \node[anchor=east]  at (160,240) {$Q_0$};
        \node[anchor=east]  at (132,290) {$t=0$};

        \node[anchor=south] at (360,121) {$M_\rho$};
        \node[anchor=west]  at (414,125) {$n^\mu$};

        \node[anchor=south] at (387,16) {$\rho$};
        \node[anchor=west]  at (520,132) {$\rho = r=\infty$};
        \node[anchor=west]  at (543,270) {$t$};

    \end{tikzpicture}

    \caption{Visualization of the geometric setup for the definition of the covariant c-function. We consider a $(d+1+p)$-dimensional spacetime with a unit normalized timelike Killing vector $t^\mu$ which provides a foliation with spacelike slices $Q_t$. The holographic radial direction is a spacelike vector field $n^\mu$ which provides a slicing with timelike surfaces $M_\rho$ where $\rho$ is a scalar field dual to field theory energy scale via $E = \rho(r,y)$. The c-function is defined as a local integral on the codimension-two intersection $\Sigma_\rho$.}
    \label{fig:sect2}
\end{figure}
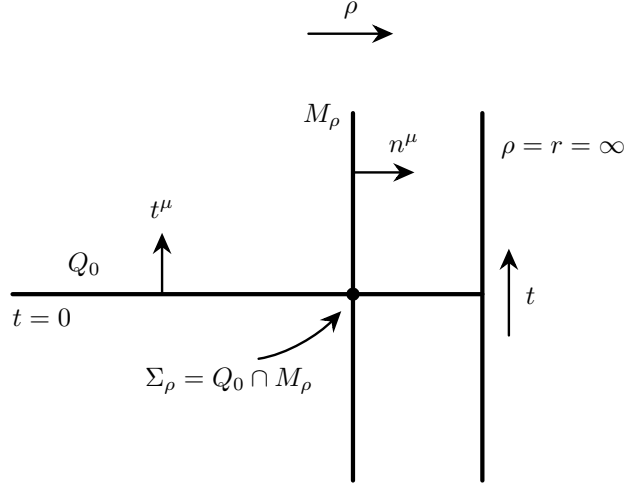

\paragraph{The c-function.} Now, together with the foliation induced by the holographic radial direction $n^\mu$, the initial $t = 0$ slice $Q_0$ is foliated by codimension-two intersections $\Sigma_\rho\equiv Q_0\cap M_\rho$. The covariant c-function is constructed from geometric quantities associated to $\Sigma_\rho $. To this end, we parametrize $\Sigma_\rho$ with coordinates $\hat{Y}^A= (x^a,\hat{y}^i) $, with $\hat{y}^i$ defined in \eqref{eq:yhat}, and denote the components of the induced metric of $\Sigma_\rho$ in these coordinates by $\hat{H}_{AB}$. We assume that this induced metric takes the form
\begin{equation}
    \hat{H}_{AB}(\rho,\hat{y})\,d\hat{Y}^A d\hat{Y}^B= \hat{g}_{ab}(\rho,\hat{y})\,dx^adx^b + \hat{h}_{ij}(\rho,\hat{y})\,(d\hat{y}^i + \hat{A}^i_a(\rho,\hat{y})\,dx^a)(d\hat{y}^j + \hat{A}^j_b(\rho,\hat{y})\,dx^b)\,,
    \label{eq:H_metric_rho}
\end{equation}
where $\hat{g}_{ab}(\rho,\hat{y}) = g_{ab}(r,y)$ via \eqref{coord-transf}, but $\hat{h}_{ij}\neq h_{ij}$ and $\hat{A}^i_a \neq A^i_a$ are different from the components $H_{AB}$ of the induced metric \eqref{eq:H_metric} of the constant-$r$ slices: they coincide only when $\rho(r,y) = r$.

In our conventions, $n^\mu $ is outward-pointing inside $Q_0$ so that the trace of the extrinsic curvature of $\Sigma_\rho $ understood as a submanifold of $Q_0$ is defined by
\begin{equation}
	K = D_\alpha n^{\alpha} = \frac{1}{\sqrt{G}}\,\partial_\alpha(\sqrt{G}\,n^\alpha)\,,\label{extrinsiccurvature}
\end{equation}
where $D_{\alpha}$ is the covariant derivative compatible with the spatial induced metric $G_{\alpha\beta}$ of the $Q_0$-slice, $G \equiv \det G_{\alpha\beta}$ is its determinant and $n^{\alpha} \propto G^{\alpha\beta}\,\partial_\beta \rho$ are the spatial components of the normal vector field.  Then we define our covariant c-function as the local integral
\begin{equation}
	\mathcal{C}(\rho) \equiv \frac{1}{G_{\text{N}}^{(d+1+p)}}\int_{\mathcal{X}_p}d^p\hat{y}\sqrt{\hat{h}(\rho,\hat{y})}\,\biggl(\frac{d-1}{2K(\rho,\hat{y})}\biggr)^{d-1}\,,
    \label{eq:covariant_c_function_V2}
\end{equation}
where $\hat{h}\equiv \det \hat{h}_{ij}$ and $G_{\text{N}}^{(D)}$ is the $D$-dimensional Newton's constant. Here we understand that $\mathcal{X}_p$ denotes the internal manifold at any value of $\rho$, not only in the asymptotic limit $\rho\rightarrow \infty$. The expression \eqref{eq:covariant_c_function_V2} is covariant in the sense that it is independent of the coordinates $\hat{y}$ chosen on the internal directions, however, it is dependent on a choice of foliation coming from the definition of the holographic energy scale $E = \rho(r,y)$. Note, to simplify notation, we parametrize the c-function directly with the scalar field $\rho$ instead of $E$.

The formula \eqref{eq:covariant_c_function_V2} only relies on staticity and $(d-1)$-dimensional translation invariance of the full metric, and on the warped product form \eqref{eq:H_metric_rho} of the codimension-two induced metric $\hat{H}_{AB}$ of the $\Sigma_\rho$-slices. In principle, we do not have to assume that the metric has the global form \eqref{eq:full_metric} in $(t,x,r,y)$-coordinates, but this is the case in practice in all our examples. Furthermore, we do not have to assume that the metric is asymptotically AdS \eqref{eq:UV_asymptotics_full_metric} in general either, because the expression \eqref{eq:covariant_c_function_V2} remains valid for any $r$-dependence in $g_{ab}(r,y)$, including asymptotics that give anisotropic scaling in the dual field theory.

Before concluding this part of the section, we would like to point out that $\mathcal{C}(\rho)$ as defined by \eqref{eq:covariant_c_function_V2} is only independent of the field theory spatial directions $x$ due to the assumed $(d-1)$-dimensional spatial translation invariance of the full metric. In the more general case, where spatial translation invariance is broken completely or partially, either by inhomogeneities from the field theory $\partial_cg_{ab}\neq 0$ or the internal $\partial_a\hat{h}_{ij}\neq 0$ directions, the c-function \eqref{eq:covariant_c_function_V2} is a function $\mathcal{C} = \mathcal{C}(\rho,x)$ also of $x$. Stationary and time-dependent geometries are discussed in Section \ref{sec:discussion}. Thus our definition \eqref{eq:covariant_c_function_V2} is more precisely a measure of the \textit{density} of degrees of freedom of the dual field theory: under diffeomorphisms in the spatial $x$-directions, it transforms as a scalar. Hence, to count the total number of degrees of freedom, denoted below as $\text{DOFs}$, it is natural to integrate over the spatial directions as
\begin{equation}
    \text{total DOFs} =  \int d^{d-1}x\,\mathcal{C}(\rho,x)\,.
    \label{eq:total_dofs}
\end{equation}
However, in writing down this integral, a question arises: what is the spatial volume element to be included to make it covariant? Here, in \eqref{eq:total_dofs}, we have chosen it to be the volume element of the flat spatial background metric of the dual QFT. Under spatial translation invariance as we have assumed above, the integral thus produces only a constant infinite volume factor $\vol{(\mathbb{R}^{d-1})}$ and does not have practical significance, which is why we have not included it in the definition \eqref{eq:covariant_c_function_V2} of the c-function.\footnote{In confining flows across dimensions studied in Section \ref{sec:confiningmodels}, where one of the field theory directions is compactified on a circle, the factor is $\vol{(\mathbb{R}^{d-2})}\vol{(S^1)}$.} In Appendix \ref{app:C5}, we consider a flow from $\mathrm{AdS}_5$ to $\mathrm{AdS}_3\times \mathbb{H}_2\slash \Gamma$ which is inhomogeneous in the field theory directions due to the presence of a hyperbolic metric. However, we find that this type of inhomogeneity is harmless and does not lead to any $x$-dependence in the c-function itself. In more complicated inhomogeneous cases, it is possible that a different volume element must be chosen and we do not have a universal proposal on how to choose it. A more detailed study is left for future work.

\paragraph{The special case \texorpdfstring{$\rho=r$}{rho = r}.} In general, the holographic radial coordinate $\rho(r,y)\neq r$, but let us now consider the special case in which the holographic radial direction is simply $\rho(r,y) = r$ and thus does not mix with the internal directions. In this case, we simply have $\hat{y} = y$. The c-function \eqref{eq:covariant_c_function_V2}, now parametrized by $r$, reduces to
\begin{equation}
	\mathcal{C}(r) = \frac{1}{G_{\text{N}}^{(d+1+p)}}\int_{\mathcal{X}_p}d^py\sqrt{h(r,y)}\,\biggl(\frac{d-1}{2K(r,y)}\biggr)^{d-1}\,.
    \label{eq:covariant_c_function}
\end{equation}
For the metric \eqref{eq:full_metric}, the vector field \eqref{eq:n_vector_field} takes the form $n^\mu = \frac{1}{\sqrt{B}}\,\delta^\mu_r $. Similarly, the determinant of \eqref{eq:G_metric} is given by $\sqrt{G} = \sqrt{B}\,\sqrt{H}$, where $H \equiv \det H_{AB}$ is the determinant of \eqref{eq:H_metric}, so that the extrinsic curvature \eqref{extrinsiccurvature} becomes
\begin{equation}
    K = \frac{1}{\sqrt{B}}\frac{\partial_r\sqrt{H}}{\sqrt{H}} = \frac{1}{2}\frac{1}{\sqrt{B}}\frac{\partial_r H}{H}\,.
    \label{eq:top_down_K}
\end{equation}
Substituting into the c-function \eqref{eq:covariant_c_function}, we obtain
\begin{equation}
	\mathcal{C}(r) = \frac{(d-1)^{d-1}}{G_{\text{N}}^{(d+1+p)}}\int_{\mathcal{X}_p}d^{p}y\sqrt{h}\,\biggl(\sqrt{B}\,\frac{H}{\partial_r H}\biggr)^{d-1}\,.\label{eq:covariant_c_function-2}
\end{equation}
The expression can be further simplified using the Schur complement formula which implies that  the determinant factorizes $H = gh$ where $g = \det g_{ab}$ and $h = \det h_{ij}$ (see Appendix \ref{app:Schur_complement} for the proof). We obtain from \eqref{eq:covariant_c_function-2} explicitly that
\begin{equation}
	\mathcal{C}(r) = \frac{(d-1)^{d-1}}{G_{\text{N}}^{(d+1+p)}}\int_{\mathcal{X}_p}d^{p}y\sqrt{h}\,\biggl(\frac{\sqrt{B}}{\partial_r\log(gh)}\,\biggr)^{d-1}\,.
    \label{eq:covariant_c_function_simplified}
\end{equation}
The calculation of the c-function in this case is summarized in table \ref{fig:c_function_summary}.

The c-function transforms as a scalar also under reparametrizations $r\rightarrow D(r)$ of the radial coordinate. Under this reparametrization, we have $B(r,y)\rightarrow D'(r)^2\,B(D(r),y)$, $H(r,y)\rightarrow H(D(r),y)$ so that $\mathcal{C}(r)\rightarrow \mathcal{C}(D(r))$ after using the chain rule for the derivative $\partial_r$. Therefore, as long as the diffeomorphism is orientation-preserving $D'(r)>0$, the monotonicity properties of the c-function are independent of the parametrization.

\paragraph{Dimensional reduction.} Our definition of the covariant c-function above is given directly in the $(d+1+p)$-dimensional top-down geometry. However, the same formula is valid in any dimension smaller than $d+1+p$, for example, in the $(d+1)$-dimensional bottom-up geometry without an internal space, and the resulting value of the c-function is the same as long as the metrics in different dimensions are related by a dimensional reduction. In this sense, the c-function is invariant under dimensional reduction. We will show this for a complete reduction over the full internal space $\mathcal{X}_p$ to $d+1$ dimensions, so that the reduced geometry has no remaining internal directions. The proof can be adapted to a reduction to any intermediate dimension $d+1+q$ with $0<q<p$.

Consider the $(d+1+p)$-dimensional metric \eqref{eq:full_metric} and assume it can be written as
\begin{equation}
    ds^{2}_{d+1+p} = W\,ds^{2}_{d+1} + h_{ij}\,(dy^i + A^i_a\, dx^a)(dy^j + A^j_b\, dx^b)\,,
    \label{eq:top_down_reduction}
\end{equation}
where $ds^{2}_{d+1}$ is the $(d+1)$-dimensional reduced metric that we assume to take the form
\begin{equation}
    ds^2_{d+1} = -\overbar{\mathcal{N}}(r)^2\, dt^2+\overbar{G}_{\overbar{\alpha}\overbar{\beta}}\,d\overbar{X}^{\overbar{\alpha}}d\overbar{X}^{\overbar{\beta}}  = -\overbar{\mathcal{N}}(r)^2\, dt^2+\overbar{B}(r)\,dr^2 + \overbar{g}_{ab}(r)\,dx^a dx^b\,,
    \label{eq:bottom_up_metric}
\end{equation}
where $\overbar{X}^{\overbar{\alpha}} = (r,x^a)$ are the reduced spatial coordinates, and the warp factor $W = W(r,y)$ is such that
\begin{equation}
 W(r,y)\,\overbar{\mathcal{N}}(r)^2 = \mathcal{N}(r,y)^2\,,\quad W(r,y)\,\overbar{B}(r) = B(r,y)\,,\quad W(r,y)\,\overbar{g}_{ab}(r) = g_{ab}(r,y)\,.
 \label{eq:metric_uplift}
\end{equation}
We assume here that the components of the dimensionally reduced metric \eqref{eq:bottom_up_metric} are independent of the internal coordinates $y$ which is necessary for a dimensional reduction to exist. This implies that all $y$-dependence in the components of the top-down metric along the reduced spacetime directions arises through the warp factor $W(r,y)$.

We focus on the case in which the holographic radial direction $\rho(r,y) = r$. The c-function $\mathcal{C}(r)$ in the top-down geometry is \eqref{eq:covariant_c_function} while in the reduced geometry it is
\begin{equation}
    \overbar{\mathcal{C}}(r) = \frac{1}{G_{\text{N}}^{(d+1)}}\biggl(\frac{d-1}{2\overbar{K}(r)}\biggr)^{d-1}\,,
    \label{eq:bottom_up_c}
\end{equation}
where the trace of the extrinsic curvature of the constant-$r$ slices in the metric $ds^{2}_{d+1}$ is
\begin{equation}
	\overbar{K} = \overbar{D}_{\overbar{\alpha}}\overbar{n}^{\overbar{\alpha}} = \frac{1}{\sqrt{\overbar{G}}}\,\partial_{\overbar{\alpha}}(\sqrt{\overbar{G}}\,\overbar{n}^{\overbar{\alpha}})\,.\label{extrinsiccurvature_bottom_up}
\end{equation}
Here $\overbar{D}_{\overbar{\alpha}}$ is the covariant derivative compatible with $\overbar{G}_{\overbar{\alpha}\overbar{\beta}}$ and $\overbar{n}^{\overbar{\alpha}}$ is the $(d+1)$-dimensional vector field pointing in the $r$-direction unit normalized using $\overbar{G}_{\overbar{\alpha}\overbar{\beta}}$. For the metric \eqref{eq:bottom_up_metric}, we obtain explicitly
\begin{equation}
    \overbar{n}^{\overbar{\alpha}} = \frac{1}{\sqrt{\overbar{B}}}\,\delta^{\overbar{\alpha}}_r\,,\quad \overbar{K} = \frac{1}{2\sqrt{\overbar{B}}}\,\partial_r\log{\overbar{g}}
\end{equation}
so that
\begin{equation}
	\overbar{\mathcal{C}}(r) = \frac{(d-1)^{d-1}}{G_{\text{N}}^{(d+1)}}\biggl(\frac{\sqrt{\overbar{B}}}{\partial_r\log{\overbar{g}}}\biggr)^{d-1}\,.
    \label{eq:bottom_up_covariant_c_function}
\end{equation}
Let us show how this expression is obtained as the dimensional reduction of \eqref{eq:covariant_c_function} in the  top-down metric \eqref{eq:top_down_reduction}. Starting from \eqref{eq:covariant_c_function_simplified} and using \eqref{eq:metric_uplift}, we obtain
\begin{equation}
    \mathcal{C}(r) = \frac{(d-1)^{d-1}}{G_{\text{N}}^{(d+1+p)}}
    \int_{\mathcal{X}_{p}} d^{p} y\,\sqrt{h}\;W^{\frac{d-1}{2}}\biggl(\frac{\sqrt{\overbar{B}}}{\partial_r\log{(\overbar{g}hW^{d-1})}}\biggr)^{d-1}\,.
    \label{eq:C_reduced}
\end{equation}
In Appendix \ref{app:newton_constant}, we show that the Newton's constants in different dimensions are related according to,
\begin{equation}
    \frac{1}{G_{\text{N}}^{(d+1)}}
    =
    \frac{1}{G_{\text{N}}^{(d+1+p)}}
    \int_{\mathcal{X}_{p}} d^{p} y\,\sqrt{h}\;W^{\frac{d-1}{2}}\,,
    \label{eq:Newton_constant_relation}
\end{equation}
which naturally requires that the product $ T \equiv h W^{d-1}$ does not depend on the radial coordinate $\partial_r T = 0$, because otherwise the lower-dimensional Newton's constant would not be independent of the radial coordinate. Strictly speaking, only $r$-independence of the integral \eqref{eq:Newton_constant_relation} is required, but this holds pointwise in $y$ in all examples below. Thus, we obtain $\partial_r\log{(g h)} = \partial_r\log{(\overbar{g} hW^{d-1})} = \partial_r\log{\overbar{g}} $. Substituting into \eqref{eq:C_reduced} and comparing with \eqref{eq:bottom_up_covariant_c_function}, we obtain
\begin{equation}
    \mathcal{C}(r) = \overbar{\mathcal{C}}(r) \,.
    \label{eq:c_bar_equals_c}
\end{equation}
Thus, when the covariant formula \eqref{eq:covariant_c_function} is applied to the dimensionally reduced metric \eqref{eq:bottom_up_metric} without an internal space $p = 0$, it yields the same result as when it is applied directly to the top-down metric \eqref{eq:top_down_reduction}. In this sense, the value of the c-function is invariant under dimensional reduction.

\paragraph{The UV limit.} Let us then consider the UV limit $r\rightarrow \infty$ of the c-function \eqref{eq:covariant_c_function_V2}. In this limit, the holographic radial direction approaches $r$ as $\rho\sim r$ and the transverse directions $\hat{y}^i\sim y^i$. Thus in the UV, the full metric takes the form \eqref{eq:full_metric} and the c-function \eqref{eq:covariant_c_function_V2} is simply given by \eqref{eq:covariant_c_function_simplified} at leading order in $r\rightarrow \infty$. The metric is locally AdS$_{d+1}\times \mathcal{X}_p$ \eqref{eq:UV_asymptotics_full_metric} so that the components of \eqref{eq:full_metric} have the asymptotics
\begin{equation}
    B(r,y) \sim \frac{\Delta(y)^2}{r^2} \,,\quad h_{ij}(r,y) \sim h_{(0)ij}(y) \,,\quad A_a^i(r,y) \sim A_{(0)a}^i(y) \,,\quad r \rightarrow \infty\,,
    \label{eq:UV_limit_1}
\end{equation}
together with
\begin{equation}
    \mathcal{N}(r,y)^2 \sim \Delta(y)^2\,r^2 \,,\quad g_{ab}(r,y) \sim \Delta(y)^2\,r^2 \,\delta_{ab} \,,\quad r\rightarrow \infty\,.
    \label{eq:UV_limit_2}
\end{equation}
Here, often in explicit examples, the UV value $A_{(0)}$ of the gauge field is a pure gauge holonomy. By substituting \eqref{eq:UV_limit_1} and \eqref{eq:UV_limit_2} to \eqref{eq:covariant_c_function_simplified}, we obtain the finite result
\begin{equation}
    \lim_{r\rightarrow \infty}\mathcal{C}(r) = \frac{1}{2^{d-1}\,G_{\text{N}}^{(d+1+p)} }\int_{\mathcal{X}_p}d^py\,\sqrt{h_{\smash{(0)}}}\,\Delta^{d-1}\,.
    \label{eq:covariant_c_function_UV_0}
\end{equation}
In the usual simple case of $\Delta(y) = L_{\text{UV}}$, this reduces to
\begin{equation}
    \lim_{r\rightarrow \infty}\mathcal{C}(r) = \frac{\vol{(\mathcal{X}_p)}\,L_{\text{UV}}^{d+p-1}}{2^{d-1}\,G_{\text{N}}^{(d+1+p)} }\ ,\quad \vol{(\mathcal{X}_p)} \equiv \frac{1}{L_{\text{UV}}^p}\int_{\mathcal{X}_p}d^py\,\sqrt{h_{\smash{(0)}}}\ ,
    \label{eq:covariant_c_function_UV}
\end{equation}
where $\vol{(\mathcal{X}_p)}$ is the (asymptotic) volume of the internal space.

The UV limit \eqref{eq:covariant_c_function_UV_0} can also be written in terms of a $(d+1)$-dimensional Newton's constant of a dimensionally reduced theory. Assuming the lower-dimensional metric \eqref{eq:bottom_up_metric} becomes asymptotically AdS$_{d+1}$ with radius $L_{\text{UV}}$, we see that $W(r,y)\sim \Delta(y)^2\slash L_{\text{UV}}^2$ as $r\rightarrow \infty$ in order for the higher-dimensional metric \eqref{eq:top_down_reduction} to obey the asymptotics \eqref{eq:UV_asymptotics_full_metric}. Therefore, we obtain $\sqrt{h}\,W^{\frac{d-1}{2}} \sim \sqrt{h_{\smash{(0)}}}\,(\Delta\slash L_{\text{UV}})^{d-1} $ when $r\rightarrow \infty$, however, since $\partial_rT=\partial_r(hW^{d-1}) = 0$, this is true for all $r$. Substituting into \eqref{eq:Newton_constant_relation}, we find that the Newton's constants are related as
\begin{equation}
    \frac{L_{\text{UV}}^{d-1}}{G_{\text{N}}^{(d+1)}}
    =
    \frac{1}{G_{\text{N}}^{(d+1+p)}}\int_{\mathcal{X}_p}d^py\,\sqrt{h_{\smash{(0)}}}\,\Delta^{d-1}
    \,.
    \label{eq:Newton_constant_relation_constant}
\end{equation}
Substituting further into \eqref{eq:covariant_c_function_UV_0}, we obtain
\begin{equation}
    \lim_{r\rightarrow \infty}\mathcal{C}(r) = \frac{L_{\text{UV}}^{d-1}}{2^{d-1}\,G_{\text{N}}^{(d+1)} }\,,
\end{equation}
which for even $d$ is proportional to the type A Weyl anomaly coefficient~\cite{Henningson:1998gx,Henningson:1998ey,Myers:2010tj}.

\paragraph{Stationarity at fixed points.} Consider a flow as a function of $\rho>0$ which starts from the UV fixed-point at $\rho\rightarrow \infty$ and ends in an IR fixed-point at $\rho = 0$. The UV fixed-point $\rho\rightarrow \infty$ is described by a radial coordinate $\rho \sim r\rightarrow \infty$ where the metric takes the form \eqref{eq:UV_asymptotics_full_metric}. On the other hand, the IR fixed-point is described by a second radial coordinate $\hat{r} $ such that $\hat{r}\rightarrow \infty$ when $\rho\rightarrow 0$ with constant-$\rho$ slices coinciding with constant-$\hat{r}$ slices. In this limit, the IR metric takes the form
\begin{equation}
    ds^2_{d+1+p} \sim \Delta_{*}(y)^2\,\biggl[ \hat{r}^2\,(-dt^2 + \delta_{ab}\,dx^adx^b) + \frac{d\hat{r}^2}{\hat{r}^2}\biggr] + h_{*ij}(y)\,(dy^i + A^i_{*a}(y)\,dx^a)(dy^j + A^j_{*b}(y)\,dx^b)
    \label{eq:IR_asymptotics_full_metric}
\end{equation}
with a different set of component functions. For these AdS metrics, the trace of the extrinsic curvature of the constant-$\rho$ slices in the UV and IR coincides with that of the constant-$r$ or constant-$\hat{r}$ slices. At leading order,
\begin{equation}
    K \sim (d-1)\times\begin{dcases}
    \Delta^{-1}\,, & \rho\rightarrow \infty\\
    \Delta_{*}^{-1}\,, & \rho\rightarrow 0
    \end{dcases} \ .
\end{equation}
This leading fixed-point contribution is independent of the radial coordinate, although it may depend on the internal coordinates through $\Delta$ or $\Delta_*$. Thus, for flows that approach the fixed-point geometries asymptotically, the leading AdS contribution gives a constant value for the c-function. The c-function is stationary, $\mathcal{C}'(\rho) = 0$, at endpoints, provided the subleading corrections vanish sufficiently fast, as happens in the standard asymptotic expansions considered below.

\paragraph{String frame.} Let us consider 10-dimensional supergravity with $d+1+p = 10$ and $p = 9-d$. In this case, there are two metrics of interest which are the Einstein and string frame metrics. We emphasize that the metric $G$ used to define the c-function \eqref{eq:covariant_c_function_V2} is assumed to be in the Einstein frame, but we can also translate the formula to the string frame as follows.

The string frame metric $G_{\text{s}}$ on the time-slice $Q_0$ is given by $(G_{\text{s}})_{\alpha\beta} = e^{\Phi\slash 2}\,G_{\alpha\beta} $ where $\Phi$ is the dilaton. Thus in the string frame $\hat{h}_{ij} = e^{-\Phi\slash 2}\,(\hat{h}_{\text{s}})_{ij}$ so that we obtain
\begin{equation}
	\sqrt{\hat{h}}= \sqrt{\hat{h}_{\text{s}}}\,e^{-(9-d)\,\Phi\slash 4}\,,\quad K= e^{\Phi\slash 4}\,(K_{\text{s}} - 2\,n^{\alpha}_{\text{s}}\partial_\alpha\Phi)\,,
\end{equation}
where $K_{\text{s}}$ is the extrinsic curvature and $n^{\mu}_{\text{s}}$ is the outward-pointing unit normal vector of $\Sigma_\rho\subset Q_0$ in the string frame metric $(G_{\text{s}})_{\alpha\beta}$. Substituting into \eqref{eq:covariant_c_function_V2} gives the c-function in terms of string frame quantities
\begin{equation}
	\mathcal{C}(\rho) = \frac{(d-1)^{d-1}}{G_{\text{N}}^{(10)}}\int_{\mathcal{X}_{9-d}} d^{9-d}\hat{y}\sqrt{\hat{h}_{\text{s}}}\,\frac{e^{-2\Phi}}{(2K_{\text{s}} -4\, n^{\alpha}_{\text{s}}\,\partial_\alpha\Phi)^{d-1}}\,.
\end{equation}
In the special case $\rho(r,y) = r$ above, this reduces to
\begin{equation}
    \mathcal{C}(r)=
\frac{(d-1)^{d-1}}{G_{\mathrm N}^{(10)}}
\int_{\mathcal{X}_{9-d}} d^{9-d}y\,\sqrt{h_{\mathrm s}}\,
e^{-2\Phi}
\biggl(
\frac{\sqrt{B_{\mathrm s}}}
{\partial_r\log(g_{\mathrm s}h_{\mathrm s})-4\,\partial_r\Phi}
\biggr)^{d-1}\,.
\end{equation}
We see that the derivative of the dilaton appears.

\begin{table}[t]
\centering
{\tabulinesep=4mm
\renewcommand{\arraystretch}{1.2}
\begin{tabu} {|c|c|}
    \hline
    metric &$\displaystyle
\begin{gathered}
ds^2_{d+1+p} = -\mathcal{N}^2 dt^2  +g_{ab}\,dx^adx^b+ B\,dr^2 + h_{ij}\,(dy^i + A^i_a\,dx^a)(dy^j + A^j_b\,dx^b)
\\[1mm]
\mathcal{N} = \mathcal{N}(r,y)\,,\quad  g_{ab} = g_{ab}(r,y)\,,\quad B = B(r,y)\,,\quad h_{ij} = h_{ij}(r,y)\,,\quad A^i_a = A^i_a(r,y)
\end{gathered}$\\
    \hline
    c-function & 
    $\displaystyle
\begin{gathered}
\mathcal{C}(r)=\frac{(d-1)^{d-1}}{G_{\mathrm N}^{(d+1+p)}}
\int_{\mathcal{X}_p} d^p y\,\sqrt{h}\,
\biggl(\frac{\sqrt{B}}{\partial_r \log(gh)}\biggr)^{d-1}
\\[1mm]
h = \det h_{ij}\,,\quad g = \det g_{ab}
\end{gathered}$ \\
\hline
    $\displaystyle
    \begin{gathered}
    \text{string-frame}\\
    \text{c-function}
    \end{gathered}$ &
    $\displaystyle
\begin{gathered}
\mathcal{C}(r)=
\frac{(d-1)^{d-1}}{G_{\mathrm N}^{(10)}}
\int_{\mathcal{X}_{9-d}} d^{9-d}y\,\sqrt{h_{\mathrm s}}\,
e^{-2\Phi}
\biggl(
\frac{\sqrt{B_{\mathrm s}}}
{\partial_r\log(g_{\mathrm s}h_{\mathrm s})-4\,\partial_r\Phi}
\biggr)^{d-1}
\\[1mm]
h_{\mathrm s}=\det\, (h_{\mathrm s})_{ij}\,,\quad
g_{\mathrm s}=\det\, (g_{\mathrm s})_{ab}\,,\quad
d+1+p=10\,,\quad p=9-d
\end{gathered}$\\
    \hline
\end{tabu}
}
\caption{Summary of the calculation of the covariant c-function when the holographic radial direction satisfies $\rho(r,y)=r$. The first expression is written in Einstein frame. The final row gives the equivalent 10-dimensional string-frame formula, with $(G_{\mathrm s})_{\alpha\beta}=e^{\Phi/2}G_{\alpha\beta}$. When $\rho(r,y)\neq r$, the c-function must be computed using the general formula \eqref{eq:covariant_c_function_V2}.}
\label{fig:c_function_summary}
\end{table}

\subsection{Relations to previous c-function proposals}\label{subsec:relation_to_old}

In the literature, there are various proposals for c-functions of the type described in the previous section. This includes the covariant formula proposed in~\cite{Sahakian:1999bd} and non-covariant formulas in increasing levels of generality~\cite{Freedman:1999gp,Myers:2010xs,Myers:2010tj,Caceres:2022hei,Caceres:2023mqz}. We will now discuss how our c-function \eqref{eq:covariant_c_function_V2} relates to these formulas.

\paragraph{Top-down non-covariant c-function.} A non-covariant ``flow'' c-function defined using a radial slicing of the type discussed above was first introduced in~\cite{Freedman:1999gp} and generalized to more general situations in~\cite{Macpherson:2014eza}. It was recently studied further in~\cite{Bea:2015fja,Merrikin:2022yho}. In its most general form~\cite{Merrikin:2022yho}, it is defined for Einstein frame metrics of the form \eqref{eq:full_metric} for which $g_{ab}$ is diagonal
\begin{equation}
    g_{ab}(r,y)\,dx^adx^b = \sum_{a=1}^{d-1} P_a(r,y)\,(dx^a)^2\,,
    \label{eq:restriction}
\end{equation}
where the components $P_a$ are such that the ratio
\begin{equation}
    \widetilde{B}(r) = \frac{B(r,y)}{\prod_{a = 1}^{d-1} P_{a}(r,y)^{1\slash (d-1)}}
    \label{eq:B_tilde}
\end{equation}
is independent of the internal coordinates $y^i$. Under these assumptions, the flow c-function is defined as~\cite{Merrikin:2022yho}
\begin{equation}
	\mathcal{C}_{\text{flow}}(r) = \frac{(d-1)^{d-1}}{G_{\text{N}}^{(d+1+p)}}\,\widetilde{B}(r)^{\frac{d-1}{2}}\frac{\mathcal{H}(r)^{\frac{2d-1}{2}}}{\mathcal{H}'(r)^{d-1}}\,,
    \label{eq:C_flow_first}
\end{equation}
where the function\footnote{In~\cite{Merrikin:2022yho}, the formula is defined in the string frame of supergravity with $d+1+p = 10$. Our Einstein frame expression \eqref{eq:Btilde_curly_H} for $\mathcal{H}$ matches their string frame expression since $\sqrt{H} = e^{-(d+p-1)\,\Phi\slash 4}\,\sqrt{H_{\text{s}}} = e^{-2\Phi}\sqrt{H_{\text{s}}}$. In addition, $\widetilde{B} = \widetilde{B}_{\text{s}}$ is the same in both frames so that \eqref{eq:C_flow_first} is equal to the expression of~\cite{Merrikin:2022yho}.}
\begin{equation}
	 \sqrt{\mathcal{H}(r)} \equiv \int_{\mathbb{R}^{d-1}} d^{d-1}x\int_{\mathcal{X}_p} d^{p}y\,\sqrt{H(r,y)}\,.
    \label{eq:Btilde_curly_H}
\end{equation}
This formula is non-covariant as it is built from components of the metric in a specific coordinate system. Furthermore, it assumes that the energy scale is parametrized by the $r$-coordinate along the whole flow.

We can rewrite the c-function \eqref{eq:C_flow_first} using the identity
\begin{equation}
	\mathcal{H}(r)^{\frac{2d-1}{2}} = \mathcal{H}(r)^{d-1}\,\sqrt{\mathcal{H}(r)}
\end{equation}
to obtain
\begin{equation}
	\mathcal{C}_{\text{flow}}(r) = \frac{(d-1)^{d-1}}{G_{\text{N}}^{(d+1+p)}}\int_{\mathbb{R}^{d-1}} d^{d-1}x\int_{\mathcal{X}_p} d^{p}y\,\sqrt{H(r,y)}\,\biggl(\sqrt{\widetilde{B}(r)}\,\frac{\mathcal{H}(r)}{\mathcal{H}'(r)}\biggr)^{d-1}\,,
\end{equation}
where we also used the second equation in \eqref{eq:Btilde_curly_H}. Further, using $\sqrt{H} = \sqrt{g}\sqrt{h}$ (see Appendix \ref{app:Schur_complement}) and that $\sqrt{g} = \prod_{a=1}^{d-1}\sqrt{P_a}$ by \eqref{eq:restriction}, we obtain
\begin{equation}
	\mathcal{C}_{\text{flow}}(r) = \frac{(d-1)^{d-1}}{G_{\text{N}}^{(d+1+p)}}\int_{\mathbb{R}^{d-1}} d^{d-1}x\int_{\mathcal{X}_p} d^{p}y\,\sqrt{h}\,\biggl(\sqrt{B}\,\frac{\mathcal{H}}{\partial_r\mathcal{H}}\biggr)^{d-1}\,.
    \label{eq:flow_c_function_final}
\end{equation}
This is very similar to the formula \eqref{eq:covariant_c_function-2} for our covariant c-function in the $\rho = r$ case except that the ratio inside the brackets involves the integral $\mathcal{H} $ instead of the determinant $H = \det H_{AB}$. Therefore in general our c-function \eqref{eq:covariant_c_function-2} is different from \eqref{eq:flow_c_function_final}, \emph{i.e.}, $\mathcal{C}(r)\neq \mathcal{C}_{\text{flow}}(r)$.

However, there is a situation in which they agree given the diagonal metric \eqref{eq:restriction}. This occurs when the determinant $H = \det H_{AB}$ of the codimension-two induced metric factorizes as $H(r,y) = H_y(y)\,H_r(r)$. In this case, we have
\begin{equation}
    \sqrt{\mathcal{H}(r)} = \sqrt{H_r(r)}\int_{\mathbb{R}^{d-1}} d^{d-1}x\int_{\mathcal{X}_p} d^{p}y\,\sqrt{H_{\smash{y}}(y)} \equiv \sqrt{H_r(r)}\;\mathcal{I}\,,
\end{equation}
where $\mathcal{I}$ is the $r$-independent value of the integral. Thus it follows that the ratio in \eqref{eq:flow_c_function_final} is given by
\begin{equation}
    \frac{\mathcal{H}}{\partial_r\mathcal{H}} = \frac{1}{2}\frac{\sqrt{\mathcal{H}}}{\partial_r \sqrt{\mathcal{H}}} = \frac{1}{2}\frac{\sqrt{H_r}}{\partial_r \sqrt{H_r}} = \frac{H_r}{\partial_rH_r} = \frac{H}{\partial_rH}\,,
\end{equation}
where in the second equality the $r$-independent $\mathcal{I}$ integral has canceled in the ratio. In this situation, comparing \eqref{eq:flow_c_function_final} with \eqref{eq:covariant_c_function-2}, we obtain
\begin{equation}
    \mathcal{C}_{\text{flow}}(r) = \vol{(\mathbb{R}^{d-1})}\,\mathcal{C}(r)\,,\quad \text{when}\quad H(r,y) = H_r(r)\,H_y(y)\,.
\end{equation}
We believe that our covariant c-function \eqref{eq:covariant_c_function_V2} provides the correct generalization of $\mathcal{C}_{\text{flow}}$ for metrics without determinant factorization $H(r,y) \neq H_r(r)\,H_y(y)$. Furthermore, our c-function also works for geometries where $\widetilde{B} = \widetilde{B}(r,y)$ depends on the internal coordinates, for which the non-covariant flow c-function of~\cite{Bea:2015fja, Merrikin:2022yho} is ill-defined.

All explicit top-down backgrounds considered below, except for the Klebanov--Murugan solution of Section \ref{sec:KlebanovMurugan}, belong to the $\widetilde{B}= \widetilde{B}(r)$ class of metrics described by \eqref{eq:restriction} which also have a factorizing determinant $H(r,y) = H_r(r)\,H_y(y)$. Thus our formula reproduces previous results in these cases as we show in detail in Section \ref{sec:tests}. Previous c-function proposals cannot be used in the case of the Klebanov--Murugan background because in that case $\widetilde{B}=\widetilde{B}(r,y)$ depends on the internal coordinates. Our covariant c-function is nevertheless applicable and is studied in Section~\ref{sec:KlebanovMurugan}.

\paragraph{Bottom-up covariant c-function.} Let us consider the covariant c-function of~\cite{Sahakian:1999bd} which is defined for $(d+1)$-dimensional spacetimes $(\overbar{\mathcal{M}},\overbar{\mathcal{G}})$ without an internal space. Let $\overbar{X}^{\overbar{\mu}}$, with $\overbar{\mu} = 0,1,\ldots,d$, denote coordinates on $\overbar{\mathcal{M}}$ in which the components of the $(d+1)$-dimensional Lorentzian metric are $\overbar{\mathcal{G}}_{\overbar{\mu}\overbar{\nu}}$. We assume that $\overbar{\mathcal{M}}$ is foliated by spacelike $\overbar{Q}_t$ and timelike surfaces $\overbar{M}_r$ with unit normal vectors $\overbar{t}^{\overbar{\mu}}$ (which we take to be future-pointing) and $\overbar{n}^{\overbar{\mu}}$ (which is outward-pointing as above) respectively. Let $\overbar{k}^{\overbar{\mu}} $ be the future-directed $(d+1)$-dimensional null vector field $\overbar{\mathcal{G}}_{\overbar{\mu}\overbar{\nu}}\,\overbar{k}^{\overbar{\mu}}\overbar{k}^{\overbar{\nu}} = 0$ orthogonal to the intersection $\overbar{\Sigma}_r = \overbar{Q}_0\cap\overbar{M}_r$ and satisfying the affine condition 
\begin{equation}
    \overbar{k}^{\overbar{\nu}}\,\overbar{\nabla}_{\overbar{\nu}}\overbar{k}^{\overbar{\mu}} = 0\,,
    \label{eq:affine_condition}
\end{equation}
where $\overbar{\nabla}_{\overbar{\mu}}$ is the covariant derivative compatible with $\overbar{\mathcal{G}}$. Since $\overbar{t}^{\overbar{\mu}}\,\overbar{n}_{\overbar{\mu}} = 0$ and $\overbar{t}^{\overbar{\mu}}\,\overbar{t}_{\overbar{\mu}} = -1 = -\overbar{n}^{\overbar{\mu}}\,\overbar{n}_{\overbar{\mu}}$, we can write $\overbar{k}^{\overbar{\mu}} = \Omega\,( \overbar{n}^{\overbar{\mu}} + \overbar{t}^{\overbar{\mu}})$ where the proportionality function $\Omega$ is fixed by the affine condition up to a factor which only changes the overall normalization of the c-function defined below. In this way, we can define the expansion of the null congruence in the metric $\overbar{\mathcal{G}}$ as
\begin{equation}
    \overbar{\Theta} \equiv \overbar{\nabla}_{\overbar{\mu}}\overbar{k}^{\overbar{\mu}}\,. 
    \label{eq:expansion_definition}
\end{equation}
The Sahakian c-function~\cite{Sahakian:1999bd} is then defined as
\begin{equation}
	\mathcal{C}_{\text{S}}(r) \equiv \frac{1}{G_{\text{N}}^{(d+1)}}\frac{1}{\sqrt{\overbar{g}(r)}}\biggl(\frac{d-1}{2\overbar{\Theta}(r)}\biggr)^{d-1}\,,
    \label{eq:sahakian_covariant_c_function}
\end{equation}
where $\overbar{g} = \det \overbar{g}_{ab}$ is the determinant of the induced metric $\overbar{g}_{ab}$ of the intersection $ \overbar{\Sigma}_r \equiv \overbar{Q}_0\cap \overbar{M}_r$. Note that in contrast to~\cite{Sahakian:1999bd}, we have fixed the normalization of the c-function and not integrated over the $x$-directions (see also the discussion around equation \eqref{eq:total_dofs}).

In the conceptual picture of~\cite{Sahakian:1999bd}, the RG flow is taken to be parametrized by an affine parameter $\tau$ of the null-congruence and \eqref{eq:sahakian_covariant_c_function} should be understood to be a function of $\tau$. By the Raychaudhuri equation, the function \eqref{eq:sahakian_covariant_c_function} is a monotonic function of $\tau$ when the bulk metric obeys Einstein's equations and the stress tensor of matter obeys the null energy condition~\cite{Sahakian:1999bd}. However, with the modern understanding of holographic RG flows, which is also the view adopted above, the RG flow is instead parametrized by timelike hypersurfaces labeled here by $r$, as first made concrete in~\cite{deBoer:1999tgo}. Thus, we will treat the formula \eqref{eq:sahakian_covariant_c_function} as a function of $r$ instead of $\tau$.

The expansion decomposes as (see Appendix \ref{app:null_expansion})
\begin{equation}
    \overbar{\Theta}  = \Omega\,(\overbar{K} + \overbar{L} - \overbar{n}^{\overbar{\mu}}\overbar{n}^{\overbar{\nu}} \overbar{L}_{\overbar{\mu}\overbar{\nu}})\,,
    \label{eq:Theta_decomposition}
\end{equation}
where we have defined the extrinsic curvatures
\begin{equation}
    \overbar{K}_{\overbar{\mu}\overbar{\nu}} = \overbar{H}_{\overbar{\mu}}^{
	\overbar{\rho}}\overbar{H}_{\overbar{\nu}}^{\overbar{\sigma}}\,\overbar{\nabla}_{\overbar{\rho}}\overbar{n}_{\overbar{\sigma}} \,,\quad \overbar{L}_{\overbar{\mu}\overbar{\nu}} = \overbar{G}_{\overbar{\mu}}^{
	\overbar{\rho}}\overbar{G}_{\overbar{\nu}}^{\overbar{\sigma}}\,\overbar{\nabla}_{\overbar{\rho}}\overbar{t}_{\overbar{\sigma}}\,,
\end{equation}
and their traces $\overbar{K} = \overbar{\mathcal{G}}^{\overbar{\mu}\overbar{\nu}}\,\overbar{K}_{\overbar{\mu}\overbar{\nu}} $, $\overbar{L} = \overbar{\mathcal{G}}^{\overbar{\mu}\overbar{\nu}}\,\overbar{L}_{\overbar{\mu}\overbar{\nu}}$, where $\overbar{H}^{\overbar{\mu}}_{\overbar{\nu}} = \delta_{\overbar{\nu}}^{\overbar{\mu}}- \overbar{n}^{\overbar{\mu}} \overbar{n}_{\overbar{\nu}}+\overbar{t}^{\overbar{\mu}}\overbar{t}_{\overbar{\nu}} $ and $\overbar{G}_{\overbar{\nu}}^{\overbar{\mu}} = \delta_{\overbar{\nu}}^{\overbar{\mu}} +\overbar{t}^{\overbar{\mu}}\overbar{t}_{\overbar{\nu}}$ are the projectors onto $\overbar{\Sigma}_r$ and $\overbar{Q}_t$ respectively. When the metric $\overbar{\mathcal{G}}_{\overbar{\mu}\overbar{\nu}}$ is static $\overbar{L}_{\overbar{\mu}\overbar{\nu}} = \frac{1}{2}\pounds_{\overbar{t}} \overbar{G}_{\overbar{\mu}\overbar{\nu}}  =  0$, where $\pounds_{\overbar{t}}$ is the Lie derivative in $\overbar{t}^{\overbar{\mu}}$ direction. Thus \eqref{eq:sahakian_covariant_c_function} becomes
\begin{equation}
	\mathcal{C}_{\text{S}}(r) = \frac{1}{G_{\text{N}}^{(d+1)}}\frac{1}{\sqrt{\overbar{g}(r)}}\biggl(\frac{d-1}{2\,\Omega(r)\,\overbar{K}(r)}\biggr)^{d-1}\,.
    \label{eq:sahakian_covariant_c_function_2}
\end{equation}
Let us compute this explicitly for a $(d+1)$-dimensional static metric of the form \eqref{eq:bottom_up_metric}. In this case, we have
\begin{equation}
    \overbar{t}^{\overbar{\mu}} = \frac{1}{\overbar{\mathcal{N}}}\,\delta^{\overbar{\mu}}_t\,,\quad \overbar{n}^{\overbar{\mu}} = \frac{1}{\sqrt{\overbar{B}}}\,\delta^{\overbar{\mu}}_r\,,\quad \overbar{K} = \frac{1}{2\sqrt{\overbar{B}}}\,\partial_r\log{\overbar{g}}\,,\quad \overbar{L}_{\overbar{\mu}\overbar{\nu}} = 0\,,
\end{equation}
where $\overbar{g} = \det \overbar{g}_{ab}$. It follows that the affine condition \eqref{eq:affine_condition} is equivalent to the equation
\begin{equation}
    \Omega'(r)  + \Omega(r)\;\frac{\overbar{\mathcal{N}}'(r)}{\overbar{\mathcal{N}}(r)} = 0\,,
\end{equation}
which has the solution $\Omega = \frac{1}{\overbar{\mathcal{N}}}$. The c-function \eqref{eq:sahakian_covariant_c_function_2} becomes
\begin{equation}
	\mathcal{C}_{\text{S}}(r) = \frac{(d-1)^{d-1}}{G_{\text{N}}^{(d+1)}}\frac{1}{\sqrt{\overbar{g}}}\biggl(\frac{\overbar{\mathcal{N}}\sqrt{\overbar{B}}}{\partial_r\log{\overbar{g}}}\biggr)^{d-1}\,.
    \label{eq:Sahakian_c_function_simplified}
\end{equation}
The c-function \eqref{eq:sahakian_covariant_c_function_2} is directly related to our covariant c-function via dimensional reduction when the $(d+1)$-dimensional metric $\overbar{\mathcal{G}}$ is part of the $(d+1+p)$-dimensional top-down metric as in \eqref{eq:top_down_reduction}. First, we see that \eqref{eq:sahakian_covariant_c_function_2} is proportional to our covariant c-function $\overbar{\mathcal{C}}(r)$ \eqref{eq:bottom_up_c} computed using the bottom-up metric. Second, we proved in \eqref{eq:c_bar_equals_c} that $\overbar{\mathcal{C}}(r) = \mathcal{C}(r)$ is equal to the covariant c-function \eqref{eq:covariant_c_function} computed using the top-down metric. Thus we obtain
\begin{equation}
    \mathcal{C}_{\text{S}}(r) = \biggl(\frac{\overbar{\mathcal{N}}^{d-1}}{\sqrt{\overbar{g}}}\biggr)\;\overbar{\mathcal{C}}(r)=\biggl(\frac{\mathcal{N}^{d-1}}{\sqrt{g}}\biggr)\;\mathcal{C}(r)\,,
    \label{eq:Sahakian_equals_covariant}
\end{equation}
where we used the equality \eqref{eq:c_bar_equals_c} and that the ratio
\begin{equation}
    \frac{\overbar{\mathcal{N}}^{d-1}}{\sqrt{\overbar{g}}} = \frac{\mathcal{N}^{d-1}}{\sqrt{g}}
    \label{eq:factor}
\end{equation}
is a function of the components of the lower-dimensional metric alone.

It is remarkable that the Sahakian c-function is related to the dimensional reduction of our covariant formula by the simple prefactor \eqref{eq:factor}. This relation shows that the monotonicity of $\mathcal{C}_{\text{S}}(r)$ does not automatically imply the monotonicity of $\mathcal{C}(r)$ unless the prefactor in \eqref{eq:factor} is constant. In most examples considered in this paper, the prefactor is unity, and the two notions of monotonicity coincide. In Appendix \ref{sahakian-examples}, we discuss the expressions above in various examples.

\paragraph{Bottom-up non-covariant c-function.} Let us discuss the connection with other non-covariant c-functions which are only defined in a bottom-up context. The non-covariant holographic c-function was proposed in~\cite{Myers:2010xs,Myers:2010tj} which has been generalized to more general bottom-up geometries in~\cite{Caceres:2022hei,Caceres:2023mqz}. The c-function of~\cite{Caceres:2023mqz}, which we will denote by $\mathcal{C}_{\text{CCLS}}$, applies to $(d+1)$-dimensional metrics without an internal space of the form 
\begin{equation}
    ds^2_{d+1} = -e^{2\mathcal{A}(r)}f(r)^2\,dt^2 + dr^2 + e^{2\mathcal{A}(r)}\biggl((dx^1)^2 + (dx^2)^2 + 2\mathcal{W}(r)\,dx^1dx^2 +\sum_{a=3}^{d-1}\mathcal{Z}(r)\,(dx^a)^2\biggr)
    \label{eq:bottom_up_metricbis}
\end{equation}
and it is defined by~\cite{Caceres:2023mqz}
\begin{equation}
    \mathcal{C}_{\text{CCLS}}(r) \equiv \frac{1}{G_{\text{N}}^{(d+1)}}\frac{1}{[(1-\mathcal{W}(r)^2)\,\mathcal{Z}(r)^{d-3}]^{1\slash 2}}\biggl(\frac{(d-1)\,f(r)}{(d-1)\,\mathcal{A}'(r)+\frac{d-3}{2}\frac{\mathcal{Z}'(r)}{\mathcal{Z}(r)}+\frac{\mathcal{W}(r)\,\mathcal{W}'(r)}{\mathcal{W}(r)^2-1}}\biggr)^{d-1}\, ,
     \label{eq:caceres_c_function}
\end{equation}
where we have refined the definition by including the Newton's constant as an overall multiplicative factor. In particular, when $\mathcal{W} = 0$ and $f = \mathcal{Z} = 1$, {\emph{i.e.}}, in the homogeneous and isotropic geometry, \eqref{eq:caceres_c_function} reduces to the c-function of~\cite{Freedman:1999gp,Myers:2010xs,Myers:2010tj},
\begin{equation}
    \mathcal{C}_{\text{CCLS}}(r)  = \frac{1}{G_{\text{N}}^{(d+1)}}\frac{1}{\mathcal{A}'(r)^{d-1}}\,.
\end{equation}
The c-function \eqref{eq:caceres_c_function} is a monotonically increasing function of $r$ when the $(d+1)$-dimensional NEC (NEC$_{d+1}$) is satisfied~\cite{Caceres:2023mqz}. This is not surprising, because it turns out that \eqref{eq:caceres_c_function} is a special case of the Sahakian c-function \eqref{eq:sahakian_covariant_c_function}, a detail not noted in~\cite{Caceres:2023mqz}. To see this, we notice that the metric \eqref{eq:bottom_up_metricbis} is a special case of \eqref{eq:bottom_up_metric} with
\begin{equation}
    \overbar{\mathcal{N}} = e^{\mathcal{A}} f(r)\,,\quad \overbar{B} = 1\,,\quad \sqrt{\overbar{g}} = e^{(d-1)\mathcal{A}(r)}\,[(1-\mathcal{W}(r)^2)\,\mathcal{Z}(r)^{d-3}]^{1\slash 2}\,.
\end{equation}
It follows that
\begin{equation}
    \frac{1}{2}\,\partial_r\log{\overbar{g}} = (d-1)\,\mathcal{A}'(r)+\frac{d-3}{2}\frac{\mathcal{Z}'(r)}{\mathcal{Z}(r)}+\frac{\mathcal{W}(r)\,\mathcal{W}'(r)}{\mathcal{W}(r)^2-1}
\end{equation}
so that the Sahakian c-function \eqref{eq:Sahakian_c_function_simplified} for the metric \eqref{eq:bottom_up_metric} matches \eqref{eq:caceres_c_function} up to a factor
\begin{equation}
    \mathcal{C}_{\text{CCLS}}(r) = 2^{d-1}\,\mathcal{C}_{\text{S}}(r)\,.
\end{equation}
Using the relation \eqref{eq:Sahakian_equals_covariant} to our covariant c-function \eqref{eq:covariant_c_function}, we thus obtain
\begin{equation}
    \mathcal{C}_{\text{CCLS}}(r) = \frac{2^{d-1}f(r)^{d-1}}{[(1-\mathcal{W}(r)^2)\,\mathcal{Z}(r)^{d-3}]^{1\slash 2}}\;\mathcal{C}(r)\,.
\end{equation}

\section{Tests of the covariant c-function}\label{sec:tests}
In this section we examine a sequence of holographic backgrounds and evaluate our covariant c-function \eqref{eq:covariant_c_function_V2} in each case. In all these cases, the holographic radial direction does not mix with the internal directions $\rho(r,y) = r$ and the c-function can be computed using its simplified form \eqref{eq:covariant_c_function_simplified}. Furthermore, these backgrounds fall into the restricted class \eqref{eq:restriction} in which $\widetilde{B} = \widetilde{B}(r)$ \eqref{eq:B_tilde} is independent of the internal directions. Thus the (non-covariant) flow c-function \eqref{eq:C_flow_first} is applicable to these backgrounds and it has been computed in previous works. We show that our covariant c-function reproduces all of these previous calculations providing a nontrivial explicit check of the validity of our proposal. In Section \ref{sec:KlebanovMurugan}, we test our covariant proposal in a holographic background which cannot be studied with the usual non-covariant methods~\cite{Bea:2015fja, Merrikin:2022yho}.

\subsection{Conformal models}\label{sec:conformalmodels}

In this subsection we consider generic CFTs in $d$ dimensions. While our primary focus will be on supersymmetric configurations, the analysis applies equally well to non-supersymmetric backgrounds. We begin with a simple example.

Let us consider Freund--Rubin geometries~\cite{Freund:1980xh} of the form AdS$_5\times \mathcal{X}_5$, where $\mathcal{X}_5$ is a five-dimensional Sasaki--Einstein manifold. These backgrounds are dual to four-dimensional CFTs, with the amount of supersymmetry determined by the properties of $\mathcal{X}_5$. Canonical examples include $\mathcal{X}_5= \{S^5, T^{1,1}, Y^{l,k}\}$, where $l$, $k$ are co-prime integer labels with $l < k$; see for example~\cite{Klebanov:1998hh,Gauntlett:2004hh, Franco:2005sm, Bertolini:2004xf}. The corresponding dual field theories are typically circular quiver gauge theories with nodes connected by bifundamental matter. They arise from stacks of D3-branes placed at singular points of Calabi--Yau manifolds.
The Einstein frame metric of the spacetime reads
\begin{eqnarray}
& & ds^2=\mathcal{H}(r)^{-\frac{1}{2}}(-dt^2+ dx_1^2+dx_2^2+dx_3^2)+ \mathcal{H}(r)^{\frac{1}{2}}(dr^2+ r^2ds^2_{\mathcal{X}_5})\,,
\end{eqnarray}
where the warp factor is
\begin{equation}
 \mathcal{H}(r) =\epsilon+\frac{L^4}{r^4}\,, \quad L^4= 4\pi g_s N\alpha'^2 \frac{\vol{(S^5)}}{\vol{(\mathcal{X}_5)}}\,.\label{quantitiesAdS5xX5}
\end{equation}
Since the dilaton is constant, Einstein and string frames are indistinguishable. There is a five-form that we do not display as it is not relevant for the discussion in this paper.

The energy scale is parametrized by $r$ so that the holographic radial coordinate is $\rho = r$ and we can use the formula \eqref{eq:covariant_c_function_simplified} to compute the covariant c-function \eqref{eq:covariant_c_function}. The necessary quantities are
\begin{equation}
    B(r) = \mathcal{H}(r)^{\frac{1}{2}}\,,\quad g(r) = \mathcal{H}(r)^{-\frac{3}{2}}\,,\quad h(r) = \mathcal{H}(r)^{\frac{5}{2}}\,r^{10}\det \mathcal{X}_5\,,\quad d = 4\,,\quad p = 5\,,
\end{equation}
where $\det \mathcal{X}_5$ is the determinant of the metric of $\mathcal{X}_5$. Substituting into \eqref{eq:covariant_c_function_simplified}, we obtain
\begin{equation}
 \mathcal{C}(r) = \frac{27\vol{(\mathcal{X}_5)}}{8 G_{\text{N}}^{(10)}} \frac{\left(L^4+\epsilon r^4 \right)^5}{\left( 3L^4 + 5 \epsilon r^4\right)^3}\ .\label{cnewAdS5xX5}  
\end{equation}
Let us test this expression. For the case $\epsilon=0$ and using  $G^{(10)}_{\text{N}}=8\pi^6 g_s^2\alpha'^4$ we find
\begin{equation}
\mathcal{C}(r)= \frac{N^2  }{4\pi }\frac{\vol (S^5)}{\vol (\mathcal{X}_5)}\,,
\end{equation}
which is a constant. When $\mathcal{X}_5=S^5$, we find $\mathcal{C}=\frac{N^2}{4\pi}$, which is the expected result for ${\cal N}=4$ SYM theory in the large-$N$ limit (and in the conventions used here); see for example~\cite{Gubser:1998vd}. If $\mathcal{X}_5=T^{1,1}$, we use $\vol (T^{1,1})=\frac{16\pi^3}{27}$ and obtain
\begin{equation}
\mathcal{C}_{ T^{1,1}}= \frac{27N^2}{64\pi} \ ,
\label{eq:C_T11}
\end{equation}
matching (in our conventions) with that in~\cite{Gubser:1998vd}. Other Einstein space examples follow from a similar analysis.

The parameter $\epsilon$ controls the asymptotics of the spacetime. For $\epsilon> 0$, the geometry is asymptotically flat rather than asymptotically AdS. From the field-theory point of view, this corresponds to deforming the low-energy theory, for instance ${\cal N}=4$ SYM when $\mathcal{X}_5=S^5$, by an irrelevant operator of dimension eight. Such a deformation requires a UV completion, which in this case is provided by coupling the field theory to 10-dimensional gravity. The monotonically divergent behavior of $\mathcal{C}$ in (\ref{cnewAdS5xX5}) then reflects the appearance of an infinite number of degrees of freedom at high energies, with $\mathcal{C}\sim r^8\sim E^8$ as $r\to\infty$.

Let us now move on to other canonical examples of backgrounds dual to conformal field theories.

\subsubsection{M5-branes, M2-branes, and the D1-D5 system}
We now briefly test our covariant c-function in other canonical backgrounds with an AdS factor.

\paragraph{M5-branes.}
Start with  the background generated by a stack of $N$ coincident M5-branes, see for example~\cite{Maldacena:1997re}. The 11-dimensional configuration consists of a metric and a four-form (which we do not display). The metric reads,
\begin{equation}
ds^2_{11} = \ell_{\text{P}}^2\biggl[\frac{r^2}{\mu_5}\,(-dt^2 + dx_1^2 + \ldots + dx_5^2) +4\mu^2_5\, \frac{dr^2}{r^2} +\mu^2_5\, ds^2_{S^4}\biggr]\,, ~~\mu^3_5=\pi N\ ,   \label{above}
\end{equation}
where $\ell_{\text{P}}$ is the Planck length in 11 dimensions. The four-form is proportional to the volume form of $S^4$. The holographic radial direction is simply $\rho = r$ so we can use the formula \eqref{eq:covariant_c_function_simplified} to compute the covariant c-function \eqref{eq:covariant_c_function}. We have 
\begin{equation}
    B(r) = \frac{4\mu_5^2\ell_{\text{P}}^2}{r^2}\,,\quad g(r) = \biggl(\ell_{\text{P}}^2\frac{r^2}{\mu_5}\biggr)^5\,,\quad h(r) = \ell_{\text{P}}^8\mu_5^{8}\,\det S^4\,,\quad d = 6\,,\quad p = 4\,,
\end{equation}
where $\det S^4 $ denotes the determinant of the round metric of $S^4$. Substituting into \eqref{eq:covariant_c_function_simplified}, we obtain
\begin{equation}
    \mathcal{C}(r) = \frac{N^3}{6\pi^2}\ ,
\end{equation}
where we have used $16\pi G_{\text{N}}^{(11)} = (2\pi)^8\ell_{\text{P}}^9$. This reproduces the well-known scaling $N^3$ indicating a non-Lagrangian CFT, in this case the $(2,0)$ SCFT.

\paragraph{M2-branes.} Let us now study the case of  M2-branes. In a suitable radial coordinate the metric representing a stack of $N$ coincident M2-branes reads~\cite{Maldacena:1997re},
\begin{equation}
ds_{11}^2= \frac{r^2}{\mu_2^4} \,(-dt^2 + dx_1^2 + dx_2^2) +\mu_2^2\,\frac{dr^2}{4 r^2}+\mu^2_2\, ds^2_{S^7},~~\mu_2^6= 32\pi^2 \ell_{\text{P}}^6 N\,.
\end{equation}
When accompanied by a four-form whose dual is proportional to the volume of $S^7$ we have a solution of 11-dimensional supergravity. The holographic radial direction is simply $\rho = r$ so we can use the formula \eqref{eq:covariant_c_function_simplified} to compute the covariant c-function \eqref{eq:covariant_c_function}. We have
\begin{equation}
    B(r) = \frac{\mu_2^2}{4r^2}\,,\quad g(r) = \frac{r^4}{\mu_2^8}\,,\quad h(r) = \mu_2^{14}\,\det S^7\,,\quad d = 3\,,\quad p = 7\,,
\end{equation}
where $\det S^7 $ denotes the determinant of the round metric of $S^7$. Substituting into \eqref{eq:covariant_c_function_simplified}, we obtain
\begin{equation}
    \mathcal{C}(r) = \frac{\sqrt{2}}{6}\,N^{\frac{3}{2}}\,.
\end{equation}
We find the characteristic scaling $N^{\frac{3}{2}}$ of M2-branes.

\paragraph{D1-D5 system.} Let us now study the D1-D5 system. In this case, the Type IIB metric of a configuration representing the  intersection of $N_1$ D1-branes and $N_5$ D5-branes  is of the form AdS$_3\times S^3\times T^4$ (the factor $T^4$ can be replaced with $\mathrm{K3}$ or other CY$_2$-fold). We reduce on $T^4$ to obtain a six-dimensional background metric~\cite{Maldacena:1997re} which reads,
\begin{equation}
ds_6^2=\ell_{\text{s}}^2\,\biggl[\frac{r^2}{g_6 \sqrt{N_1N_5}} \,(-dt^2+dx_1^2) + \frac{g_6\sqrt{N_1 N_5}\, dr^2}{r^2}+g_6\sqrt{N_1 N_5}\, ds^2_{S^3} \biggr]\,,
\label{eq:D1_D5_metric}
\end{equation}
where $\ell_{\text{s}}$ is the string length and the six-dimensional string coupling $g_6^2\equiv \frac{g_{\text{s}}^2}{\vol{(T^{4})}} $ and $\vol{(T^{4})} = (2\pi)^4$, where $g_{\text{s}}$ is the 10-dimensional string coupling, see~\cite{Maldacena:1997re}
for details. The quantities needed for the calculation of the c-function are
\begin{equation}
    B= \frac{g_6\sqrt{N_1 N_5}\,\ell_{\text{s}}^2}{r^2}\,,\quad g = \frac{\ell_{\text{s}}^2r^2}{g_6 \sqrt{N_1N_5}}\,,\quad h = (g_6\sqrt{N_1 N_5}\,\ell_{\text{s}}^2)^3\,\det S^3\,,\quad d = 2\,,\quad p = 3\,.
\end{equation}
Substituting into \eqref{eq:covariant_c_function_simplified} and using $G_{\text{N}}^{(6)} = 8\pi^6g_6^2\,\ell_{\text{s}}^4$, we obtain
\begin{equation}
    \mathcal{C}(r) =\frac{N_1 N_5}{8\pi^4}\,.\label{BHhere}
\end{equation}
Here we have computed the c-function using the six-dimensional geometry arising from dimensional reduction over the $T^4$, but we could have obtained the same result working directly in the 10-dimensional geometry as proven by \eqref{eq:c_bar_equals_c}.

\subsubsection{Generic conformal field theories}\label{subsec:genericCFT}

Let us now move to a more generic class of examples and discuss the covariant c-function for backgrounds dual to families of conformal field theories in different spacetime dimensions. We focus on Type II string theory, although the extension to 11-dimensional supergravity is immediate. The common structure of the relevant Einstein frame backgrounds is
\begin{equation}
  d s_{10}^2 =e^{-\frac{\Phi}{2}}f_1(y)\left[r^2(-d t^2+ \delta_{ab}\,dx^adx^b ) + \frac{d r^2}{r^2}\right] +h_{ij}(y)\,dy^idy^j\,,\quad \Phi=\Phi(y) \ ,\label{eq:metricaAdSgeneric}
\end{equation}
where the AdS$_{d+1}$ factor is warped over an internal manifold. Notice that the dilaton is also allowed to vary over the internal space, although it is independent of the holographic radial coordinate.

Backgrounds of this type arise in many dimensions. For $d=1$, infinite families of AdS${}_2$ solutions were constructed in~\cite{Lozano:2020txg, Lozano:2020sae, Lozano:2021rmk}, dual to supersymmetric conformal quantum mechanical systems. For $d=2$ the works~\cite{Lozano:2019zvg, Lozano:2020bxo, Lozano:2019emq, Couzens:2021veb} provide infinite classes of AdS${}_3$ backgrounds dual to ${\cal N}=(0,4)$ linear quiver CFTs. Warped AdS$_4$ solutions have been studied, for example, in~\cite{Assel:2011xz, Akhond:2021ffz, DHoker:2008lup}. Four-dimensional CFTs dual to warped AdS${}_5$ appear in~\cite{Gaiotto:2009gz, Aharony:2012tz, Reid-Edwards:2010vpm, Nunez:2019gbg, Lozano:2016kum, Nunez:2018qcj}. For AdS$_6$, together with the corresponding five-dimensional ${\cal N}=1$ SCFTs, see~\cite{DHoker:2016ysh, DHoker:2017zwj, DHoker:2016ujz, Legramandi:2021uds, Legramandi:2021aqv}. Finally, infinite families of AdS$_7$ solutions dual to six-dimensional ${\cal N}=(1,0)$ SCFTs were studied in 
\cite{Gaiotto:2014lca, Cremonesi:2015bld, Macpherson:2016xwk, Passias:2015gya, Apruzzi:2014qva, Nunez:2018ags, Filippas:2019puw} among many others. 

A common feature in these constructions is that they preserve eight Poincar\'e supercharges and describe duals to linear quiver theories. For these cases, one can use supersymmetric localization to reduce the field theory path integral to a matrix model. A central result of the works summarized in~\cite{Nunez:2023loo, Akhond:2022oaf} is the development of a formalism where this matrix model and the supergravity background are both determined by the same master partial differential equation. This correspondence allows for a direct map between the two sides, providing a sharp check on the validity of the proposed Type II dual pairs.

The holographic radial coordinate is again $r$ and we can use the formula \eqref{eq:covariant_c_function_simplified} to compute the covariant c-function \eqref{eq:covariant_c_function}. The relevant quantities are
\begin{equation}
    B(r,y)=\frac{e^{-\frac{\Phi}{2}}f_1(y)}{r^2}\,,\quad g(r,y) = (e^{-\frac{\Phi}{2}}f_1(y)\,r^2)^{d-1}\,,\quad p = 9-d\,.
\end{equation}
Substituting into \eqref{eq:covariant_c_function_simplified}, we obtain
\begin{equation}
 \mathcal{C}(r) = \frac{\widetilde{\mathcal{N}}}{2^{d-1} G_{\text{N}}^{(10)}} \,,\quad
 \widetilde{\mathcal{N}} \equiv  \int_{\mathcal{X}_p} d^{p}y \sqrt{h}\,e^{-\frac{(d-1)\Phi}{4}} f_1(y)^{\frac{d-1}{2}}\,. \label{CFT-ccov}
\end{equation}
This expression coincides with what is obtained in the paper~\cite{Jokela:2025cyz}. In fact, the authors of~\cite{Jokela:2025cyz} calculate the entanglement entropy for a spherical region and using the Liu--Mezei c-function~\cite{Liu:2012eea, Liu:2013una} found exactly the same result as in (\ref{CFT-ccov}), see Section 3 in~\cite{Jokela:2025cyz}. This gives us confidence that the covariant c-function introduced in Section \ref{section-definitions} certainly captures the known results for a plethora of CFT examples.

\begin{figure}[t]
\centering
{\tabulinesep=2mm
\renewcommand{\arraystretch}{1.2}
\begin{tabu} {|c|c|}
    \hline
    conformal model & covariant c-function\\
    \hline
    AdS$_5\times \mathcal{X}_5$ & 
    $\displaystyle
\mathcal{C}(r) = \frac{N^2}{4\pi }\frac{\vol (S^5)}{\vol (\mathcal{X}_5)}$ \\
    \hline
    M5-branes & 
    $\displaystyle
\mathcal{C}(r) = \frac{N^3}{6\pi^2}$ \\
    \hline
    M2-branes & 
    $\displaystyle
\mathcal{C}(r) = \frac{\sqrt{2}}{6}\,N^{\frac{3}{2}}$ \\
    \hline
    D1-D5 system & 
    $\displaystyle
\mathcal{C}(r) = \frac{N_1 N_5}{8\pi^4}$ \\
    \hline
    generic AdS$_{d+1}\times \mathcal{X}_p$  & 
    $\displaystyle
\mathcal{C}(r)  = \frac{1}{2^{d-1} G_{\text{N}}^{(10)}}\int_{\mathcal{X}_p} d^{p}y \sqrt{h}\,e^{-\frac{(d-1)\Phi}{4}} f_1(y)^{\frac{d-1}{2}}$ \\
    \hline
\end{tabu}
}
\caption{Summary of the covariant c-function in conformal models.}
\label{fig:conformal_models_summary}
\end{figure}

\subsection{Geometries induced by a stack of D-branes}\label{subsec:Dp_branes}
We now turn to non-conformal examples and consider the backgrounds induced by a stack of $N$ D$(d-1)$-branes. The dual field theories exhibit nontrivial RG flows. As discussed in~\cite{Itzhaki:1998dd}, the 10-dimensional supergravity description is not reliable over the full range $0\leq r<\infty$. Depending on the radial region, either the dilaton or the curvature becomes large, signaling a breakdown of the 10-dimensional description. In those regimes one must instead use an alternative description, which may be weakly coupled maximally supersymmetric Yang--Mills theory, an 11-dimensional uplift, or an S-dual frame. We refer the reader to~\cite{Itzhaki:1998dd} for a detailed discussion of the corresponding domains of validity.

Let us now collect the relevant ingredients. The dilaton and Einstein frame metric, supported by RR $F_{d+1}$ flux, solve the Type II equations of motion. For a D$(d-1)$-brane (with $2\leq d\leq 7$) in 10 dimensions, we have
\begin{equation}
ds_{10}^2= e^{-\frac{\Phi}{2}}\Big[\tilde{h}(r)^{-\frac{1}{2}}\,(-dt^2 + \delta_{ab}\,dx^adx^b) + \tilde{h}(r)^{\frac{1}{2}} dr^2+ \tilde{h}(r)^{\frac{1}{2}} r^2 ds^2_{S^{9-d}}\Big]\,,\quad e^{-4\Phi}=\tilde{h}(r)^{d-4}\,,\label{metric-E-Dp}
\end{equation}
with warp factor
\begin{equation}
    \tilde{h}(r)= \left( \frac{L}{r}\right)^{8-d}, \quad L^{8-d}= c_d\, g_{\text{YM}}^2 N\,,\quad c_{d} \equiv 2^{9-2d} \pi^{6-\frac{3d}{2}}\,\Gamma\left( \frac{8-d}{2}\right)\,.
\end{equation}
In this section, we set $\alpha' = g_{\text{s}} = 1$. The holographic radial coordinate is $r$ and we identify
\begin{equation}
    B=e^{-\frac{\Phi}{2}}\,\tilde{h}(r)^{\frac{1}{2}}\,,\quad g = (e^{-\frac{\Phi}{2}}\tilde{h}(r)^{-\frac{1}{2}})^{d-1}\,,\quad h = (e^{-\frac{\Phi}{2}}\tilde{h}(r)^{\frac{1}{2}}r^2)^{9-d} \,,\quad p = 9-d\,.
\end{equation}
Substituting into \eqref{eq:covariant_c_function_simplified}, we obtain
\begin{equation}
    \mathcal{C}(r) = \frac{c_d^{\frac{d}{2}}\vol{(S^{9-d})}}{G_{\text{N}}^{(10)}}\left( \frac{d-1}{10-d}\right)^{d-1}  g_{\text{YM}}^{d}\, N^{\frac{d}{2}}r^{\frac{(d-4)^2}{2}}\,.\label{central-Dp-r}
\end{equation}
It is instructive to compare this result with the proposal of~\cite{Sahakian:1999bd}. While agreement is found for some values of $d$, discrepancies arise in others. We attribute this difference to the absence of the $\sqrt{h}$ factor in the construction of~\cite{Sahakian:1999bd}.

Note the backgrounds in \eqref{metric-E-Dp} are singular, and the result \eqref{central-Dp-r} should therefore be interpreted with care. In particular, it is reliable only within the range of $r$ where the supergravity description remains valid~\cite{Itzhaki:1998dd}.

In the next subsection, we turn to backgrounds that are well-defined for all values of the radial coordinate. These provide a more controlled setting to study the behavior of the c-function in flows to a gapped IR.

\subsection{Generic confining models}\label{sec:confiningmodels}

We consider a family of confining backgrounds in various dimensions. This family is outlined in Appendix D of~\cite{Jokela:2025cyz} encoding lessons and developments in~\cite{Anabalon:2021tua, Anabalon:2024che, Anabalon:2024qhf, Chatzis:2024kdu, Kumar:2024pcz, Chatzis:2024top, Macpherson:2024qfi, Macpherson:2025pqi, Fatemiabhari:2024aua, Nunez:2023xgl, Nunez:2023nnl, Chatzis:2025dnu, Anabalon:2026yxk}. On the field theory side, we start from a CFT in $d$ dimensions compactified on a circle $S^1$ with periodic (bosonic) and antiperiodic (fermionic) boundary conditions, breaking SUSY. Supersymmetry can be restored by adding a Wilson line along $S^1$, which induces a ``spectral flow'' preserving SUSY when correlated with the circle size~\cite{Cassani:2021fyv, Kumar:2024pcz, Castellani:2024ial}. The theory thus flows from a CFT$_d$ with a given number of supercharges in the UV to a gapped QFT$_{d-1}$ with half as many supercharges in the IR. In this context we use ``confining'' in the holographic sense: the Wilson loop yields a quark-antiquark potential that grows linearly at large separation, while holographic entanglement entropy for slab regions saturates at large distances (see, for example,~\cite{Fatemiabhari:2024aua, Chatzis:2024top, Chatzis:2024kdu, Whittle:2025yog}), reflecting a finite correlation length~\cite{Jokela:2020wgs}.

As in previous sections, only the metric and dilaton are needed, while the NS and RR fields are omitted (see the papers ~\cite{Anabalon:2021tua, Anabalon:2024che, Anabalon:2024qhf, Chatzis:2024kdu, Kumar:2024pcz, Chatzis:2024top, Macpherson:2024qfi, Macpherson:2025pqi, Fatemiabhari:2024aua, Nunez:2023xgl, Nunez:2023nnl, Chatzis:2025dnu} for details). The Einstein frame metric and dilaton are
\begin{equation}
  ds_{10}^2 = e^{-\frac{\Phi}{2}}f_1(z)\left[r^2\,(-d t^2+ d\vec{x}_{{d-2}}^2 + f(r)\, d\phi^2) + \frac{d r^2}{r^2 f(r)}\right] +ds^2_{9-d}\,, \label{eq:metricaARgeneric}
\end{equation}
where the metric of the internal part and the dilaton are
\begin{equation}
    ds^2_{9-d} = v_{nm}(z)\, d z^n d z^m+ f_2(z)\left(d\xi + A_n(z)\,d z^n +A_\phi(r)\, d\phi\right)^2\,,\quad \Phi=\Phi(z)\,,\label{maxiro}
\end{equation}
and we have written the internal coordinates as $y = (z^1,\ldots,z^{p-1},\xi)$ and defined the indices $n,m$ to have the range $n,m=1,\ldots,p-1$. In the metric \eqref{eq:full_metric}, this corresponds to an $h_{ij}(r,y)$ with components 
\begin{equation}
h_{nm}=v_{nm}+f_2\,A_nA_m\,,\quad
h_{n\xi}=f_2\,A_n\,,\quad
h_{\xi\xi}=f_2\,,
\end{equation}
and therefore $\det h=f_2\det v$ using the Schur complement formula. The term $A_\phi(r)\,d\phi$ is encoded as the fibration component $A^\xi_\phi=A_\phi(r)$ in the general ansatz \eqref{eq:full_metric}. Furthermore, we have split the spatial coordinates as $x = (x^1,\ldots,x^{d-2},\phi)$ where the $\phi$-direction is compactified on a circle $S^1_\phi$ and fibered over a $U(1)$ R-symmetry generated by $\partial_\xi$. The function
\begin{equation}
    f(r)= 1- \frac{\mu}{r^{d}} -\frac{q^2}{r^{2(d-1)}} \ ,\label{eq:genericfR}
\end{equation}
captures the compactification on $S^1_\phi$. We focus on $\mu=0$, ensuring SUSY. The first zero of $f(r)$ marks the IR endpoint where $S^1_\phi$ shrinks. The parameter $q$ encodes the circle size in the dual QFT. The explicit one-forms $A_n$ and $A_\phi$ can be found in~\cite{Anabalon:2021tua, Anabalon:2024che, Anabalon:2024qhf, Chatzis:2024kdu, Kumar:2024pcz, Chatzis:2024top, Macpherson:2024qfi, Macpherson:2025pqi, Fatemiabhari:2024aua, Nunez:2023xgl, Nunez:2023nnl, Chatzis:2025dnu} for a variety of examples.

\begin{figure}[t]
\centering
{\tabulinesep=2mm
\renewcommand{\arraystretch}{1.2}
\begin{tabu} {|c|c|}
    \hline
    RG-flow & covariant c-function\\
    \hline
    D$(d-1)$-branes  &
    $\displaystyle
\mathcal{C}(r) = \frac{c_d^{\frac{d}{2}}\vol{(S^{9-d})}}{G_{\text{N}}^{(10)}}\left( \frac{d-1}{10-d}\right)^{d-1}  g_{\text{YM}}^{d}\, N^{\frac{d}{2}}r^{\frac{(d-4)^2}{2}}$ \\
    \hline
    \shortstack{generic confining model\\(flow from $d$ to $d-1$ dimensions)}
     &
    $\displaystyle
\mathcal{C}(r) = \frac{\widetilde{\mathcal{N}}}{2^{d-1} G_{\text{N}}^{(10)}} \Bigg[\frac{ f(r)^{-\frac{1}{2}}}{1+\frac{r f'(r)}{(2d-2)f(r)}} \Bigg]^{d-1}$ \\
    \hline
\end{tabu}
}
\caption{Summary of the covariant c-function in models with an RG flow.}
\label{fig:rg_flow_models_summary}
\end{figure}

The quantities relevant for the c-function are
\begin{equation}
    B(r,z)=\frac{e^{-\frac{\Phi}{2}}f_1(z)}{r^2 f(r)}\,,\quad g(r,z) = f(r)\,(e^{-\frac{\Phi}{2}}f_1(z)\,r^2)^{d-1}\,,\quad h(z) = f_2(z)\det\, (v_{nm}(z)) \,,\quad p = 9-d\,,
\end{equation}
where all the functions are independent of the internal coordinate $\xi$. Substituting into \eqref{eq:covariant_c_function_simplified}, we obtain
\begin{equation}
    \mathcal{C}(r) = \frac{\widetilde{\mathcal{N}}}{2^{d-1} G_{\text{N}}^{(10)}} \Bigg[\frac{ f(r)^{-\frac{1}{2}}}{1+\frac{r f'(r)}{(2d-2)f(r)}} \Bigg]^{d-1}\,,\quad
 \widetilde{\mathcal{N}} \equiv  \int_{\mathcal{X}_p} d^{p}y \sqrt{h}\,e^{-\frac{(d-1)\Phi}{4}} f_1(z)^{\frac{d-1}{2}}\,,
 \label{eq:c_function_confining}
\end{equation}
where $\widetilde{\mathcal{N}}$ is the same constant as in \eqref{CFT-ccov}. In particular, for $\mu = 0$, this simplifies to the simple expression
\begin{equation}
    \mathcal{C}(r) = \frac{\widetilde{\mathcal{N}}}{2^{d-1} G_{\text{N}}^{(10)}}\biggl(1 -\frac{q^2}{r^{2(d-1)}}\biggr)^{\frac{d-1}{2}} \,.
 \label{eq:c_function_confining_simple}
\end{equation}
The result \eqref{eq:c_function_confining} is a monotonic quantity as can be seen using $f(r)$ \eqref{eq:genericfR} and matches the flow c-function \eqref{eq:C_flow_first} computed in~\cite{Jokela:2025cyz} as expected. Notice that it vanishes in the IR where $f(r)=0$. We interpret this as an indication of the gapped character of the dual QFT. Also, note that at $r\to\infty$ (when $f(r)\to 1$ and $f'(r)\to 0$) we get the value of the UV CFT in \eqref{CFT-ccov}. We emphasize that when applied to backgrounds representing RG flows, the Liu--Mezei entanglement c-function~\cite{Liu:2012eea, Liu:2013una} fails to be monotonic as shown in~\cite{Jokela:2025cyz}. In this sense, the covariant c-function we have defined has all the desired properties: monotonicity, covariant definition, and coincides with the central charge at conformal points.

It turns out that the formula \eqref{eq:c_function_confining} is valid for more general backgrounds than only \eqref{eq:metricaARgeneric}. This includes backgrounds for which $f_1$, $f_2$, and $\det v$ are functions of both $r$ and $y$. Two such 10-dimensional backgrounds with $d = 4$ are presented in Appendix \ref{subapp:demanding_example} and \ref{subapp:very_demanding_example}. In both cases, the covariant c-function is given by \eqref{eq:c_function_confining} with all details of the background encoded in the overall coefficient $\widetilde{\mathcal{N}}$. Furthermore, for the two backgrounds in Appendix~\ref{appendix-details-confining}, our covariant c-function gives the same result as the flow c-function calculated in~\cite{Chatzis:2024kdu, Chatzis:2024top, Chatzis:2025dnu}. The reason is that in these examples the determinant of the codimension-two metric factorizes $H(r,y) = H_r(r)\,H_y(y)$ as discussed in Section \ref{subsec:relation_to_old}. This is consistent with the possibility of reducing the backgrounds to lower-dimensional gauged supergravity.

In Appendix \ref{appendix-details-KSBB}, we also discuss another similar background, the baryonic branch of the Klebanov--Strassler field theory. In this case, our covariant c-function matches the flow c-function computed in~\cite{Elander:2011mh}. The characteristic behaviour of that central function along the flow is that it vanishes in the far IR ($r\to 0$) and grows logarithmically with the energy for large energies, along the expectations from the dual QFT. See Appendix \ref{appendix-details-KSBB} for the detailed study.

\section{A top-down flow without dimensional reduction}\label{sec:KlebanovMurugan}

In this section we study a particularly instructive example of a holographic RG flow. From the field theory viewpoint, it describes a flow between two conformal fixed points in the same spacetime dimension. The novelty of the example lies in the 10-dimensional geometry: the warp factors depend non-factorizably on internal coordinates in such a way that the assumptions underlying the usual holographic c-functions are not satisfied. In particular, standard formulas based on an effective lower-dimensional description, such as those of~\cite{Freedman:1999gp}, are not applicable, and the background is not known to arise as the uplift of a solution of gauged supergravity. Our covariant definition in Section~\ref{section-definitions}, however, remains meaningful and makes it possible to extract nontrivial information about the c-function, in particular its asymptotic behavior. In this section we first comment on the dual field theory and then describe the background geometry, after which we study the corresponding covariant c-function and show that its asymptotics are physically sensible.

\subsection{The dual field theory and the supergravity background}

Before discussing the Type IIB background, let us briefly review the dual field theory. As anticipated, the solution describes a flow between two conformal fixed points. The UV theory is the Klebanov--Witten model~\cite{Klebanov:1998hh}, a two-node quiver
$SU(N)\times SU(N)$ with bifundamental matter $\mathbf{A}_i$ and $\mathbf{B}_j$ ($i,j=1,2$), transforming in the $(N,\bar{N})$ and $(\bar{N},N)$ representations, respectively. The theory is governed by a quartic superpotential $W\sim \text{Tr}\det[ \mathbf{A}_i\mathbf{B}_j]$. The IR fixed point corresponds to ${\cal N}=4 $ SYM theory~\cite{Klebanov:2007us}.

The flow is triggered by a vacuum expectation value (VEV) of the operator
\begin{equation}
 {\cal U}=\frac{1}{N}\text{Tr}\Big[|\mathbf{B}_1|^2 +|\mathbf{B}_2|^2-|\mathbf{A}_1|^2-|\mathbf{A}_2|^2 \Big]\,.
\end{equation}
We focus on a simple symmetric configuration in which 
\begin{eqnarray}
 \mathbf{B}_2 & = &  u \,\mathbf{1}_{N\times N} \label{eq:uintro}\\
 \mathbf{B}_1 & = & \mathbf{A}_1=\mathbf{A}_2=0 \ .     
\end{eqnarray}
It was proposed in~\cite{Klebanov:1999tb} that such deformations correspond to a resolution of the conifold. The Type IIB supergravity background describing this flow was constructed in~\cite{Klebanov:2007us}, where the expected field-theory features are reproduced. The Einstein-frame metric takes the form,
\begin{equation}
    ds^2_{10}=\mathcal{H}(r,y)^{-\frac{1}{2}}\,(-dt^2+ dx_1^2+dx_2^2+dx_3^2)+ \mathcal{H}(r,y)^{\frac{1}{2}}\,ds^2_6
    \label{eq:KM_metric} \ ,
\end{equation}
where $y=(\theta_1,\theta_2,\varphi_1,\varphi_2,\psi)$, the warp factor $\mathcal{H}(r,y)$ is defined momentarily and the metric on the resolved conifold is given by
\begin{align}
    ds_{6}^2 = \frac{dr^2}{\kappa(r)} &+ A_1^2(r)\,(d\theta_1^2 + \sin^2{\theta_1}\,d\varphi_1^2)+A_2^2(r)\, (d\theta_2^2 + \sin^2{\theta_2}\,d\varphi_2^2)\nonumber\\
    &+A_3^2(r)(d\psi +\cos{\theta_1}\,d\varphi_1+\cos{\theta_2}\,d\varphi_2)^2,
    \label{eq:resolved_conifold}
\end{align}
with
\begin{equation}
    \kappa(r)= \frac{r^2+ 9 a^2}{r^2+6 a^2}\,,\quad A_1^2(r)= \frac{r^2}{6}\,,\quad A_2^2(r)= \frac{r^2+6a^2}{6}\,,\quad A_3^2(r)=\frac{r^2}{9}\kappa(r) \ ,
\end{equation}
where $a$ sets the resolution scale. The angular coordinates have ranges $0\leq \theta_i\leq \pi $, $0\leq\varphi_i\leq 2\pi$, and $0\leq \psi\leq 4\pi$. Constant-$r$ slices are topologically $S^2\times S^3$, with the $S^3$ shrinking at $r=0$ while the $S^2$ remains of finite size set by $a$. The solution is completed by self-dual RR five-form flux (which we do not display), with a constant dilaton $\Phi=0$, while all other NSNS and RR field strengths vanish.

It is worth emphasizing that conformal invariance is broken along the flow despite the dilaton being constant. In holography, a constant dilaton usually implies that the gauge coupling does not run, but conformal symmetry can still be broken by the state. In the present case this scale is set by the resolution parameter $a$, which is dual to the VEV $u$ in (\ref{eq:uintro}) triggering the flow. From the field theory perspective, the VEV moves the theory away from the conformal point and selects a specific point in the space of vacua, inducing a Higgsing of the gauge group $SU(N)\times SU(N)$ down to the diagonal $SU(N)$, corresponding to the gauge group of the IR ${\cal N}=4$ theory. At the same time, the choice of the VEV selects a direction in the $\mathbf{B}_j$ doublet and breaks the global $SU(2)_\mathbf{B}$ symmetry to a $U(1)$ subgroup, while also modifying the R-symmetry along the flow. The intermediate regime is therefore not conformal; conformality is recovered only asymptotically in the UV and IR where the external geometry approaches AdS.

The warp factor $\mathcal{H}(r,y)$ is a solution to the Laplace equation on the resolved conifold
\begin{equation}
    -\nabla^2_{g_6}\mathcal{H}(r,y) = \frac{C}{\sqrt{g_6}}\,\delta(r-r_0)\,\delta^{(5)}(y-y_0) \ ,
    \label{eq:laplace_eq_resolved_conifold}
\end{equation}
where $\nabla_{g_6}^2$ is the Laplacian of the metric \eqref{eq:resolved_conifold} and the delta function has support at $r = r_0$, $y = y_0$, which is the position of the $N$ D3-brane sources, the number of which determines the coefficient to be $C = (2\pi)^4g_{\text{s}}\alpha'^2N $~\cite{Klebanov:2007us}. We assume that the branes are placed at $r = 0$ and $\theta_2 = 0$ (the north pole of the second $S^2$). In this case, $\mathcal{H} = \mathcal{H}(r,\theta_2)$ does not depend on angles other than $\theta_2$~\cite{Klebanov:2007us}.

The Laplace equation \eqref{eq:laplace_eq_resolved_conifold} can be solved by an infinite expansion in harmonics~\cite{Klebanov:2007us}
\begin{align}
    &\mathcal{H}(r,\theta_2)= L_{\text{UV}}^4 \sum_{l=0}^\infty
(2 l+1)\,H_l(r)\, P_l(\cos\theta_2) \ ,\  \mathcal{H}_l(r) =\frac{2 C_\beta}{9a^2 r^{2\beta+2}} ~{}_2 F_1\left(\beta,1+\beta,1+2\beta, -\frac{9a^2}{r^2}\right)
\nonumber\\
& C_{\beta}= (3a)^{2\beta} \frac{\Gamma(1+\beta)^2}{\Gamma(1+2\beta)},~~~\beta=\sqrt{1+\frac{3(l^2+l)}{2}}\ ,
\label{eq:hhat_series_solution}
\end{align}
with $\vol{(T^{1,1})} = \frac{16\pi^3}{27}$. The coefficient $L_{\text{UV}}^4$ can be fixed for example by matching with the known conifold solution of~\cite{Klebanov:1998hh} at $a = 0$ to give~\cite{Klebanov:2007us}
\begin{equation}
    L_{\text{UV}}^4 \equiv \frac{C}{4\vol{(T^{1,1})}} = \frac{27}{4}\,\pi g_{\text{s}}\alpha'^2 N\,.
    \label{eq:UV_radius}
\end{equation}
Keeping only the $l = 0$ zero-mode of the infinite series solution \eqref{eq:hhat_series_solution} produces the solution of~\cite{PandoZayas:2000ctr}
\begin{equation}
    \mathcal{H}(r,\theta_2)\vert_{l = 0} = L_{\text{UV}}^4\bigg[\frac{2}{9 a^2 r^2 } -\frac{2}{81a^4}\log\bigg(1+\frac{9a^2}{r^2}\bigg) \bigg]\ ,
    \label{eq:KM_zero_mode}
\end{equation}
which is independent of the internal coordinate $\theta_2$.

\paragraph{UV asymptotics.} To find the UV $r\rightarrow \infty$ asymptotics, we use the power series expansion of the hypergeometric function
\begin{equation}
    \mathcal{H}_l(r)=\frac{2C_\beta}{9a^2r^{2+2\beta}}\sum_{n=0}^{\infty}\frac{(\beta)_n(1+\beta)_n}{(1+2\beta)_n}\bigg(-\frac{9a^2}{r^2}\bigg)^n\frac{1}{n!}\ ,
\end{equation}
where $(\beta)_n=\frac{\Gamma(n+\beta)}{\Gamma(\beta)}$ denotes the Pochhammer symbol. It follows that the warp factor has the expansion
\begin{equation}
    \mathcal{H}(r,\theta_2) = \frac{L_{\text{UV}}^4}{r^4}\bigg[1+\frac{a^2}{r^2}(-6+9\cos\theta_2)+\frac{a^4}{r^4}\bigg(\frac{81}{2}-\frac{486}{5}\cos \theta_2\bigg)\bigg]+\ldots \,,\quad r\rightarrow \infty\,,
    \label{eq:KM_hhat_UV_expansion}
\end{equation}
where the ellipsis denote subleading terms. Thus at leading order $\mathcal{H} \sim \frac{L_{\text{UV}}^4}{r^4}$, which, together with $6A_1^2=6 A_2^2=9 A_3^2\sim r^2$, implies that the metric \eqref{eq:KM_metric} becomes
\begin{equation}
    ds^2_{10} \sim \frac{r^2}{L_{\text{UV}}^2}\,( -dt^2+ dx_1^2+dx_2^2+dx_3^2) + \frac{L_{\text{UV}}^2}{r^2}\,(dr^2 + r^2 ds_{T^{1,1}}^2 )\,,\quad r\rightarrow \infty\,,
\end{equation}
where the metric of $T^{1,1}$ is
\begin{equation}
    ds_{T^{1,1}}^2 = \frac{1}{6}\,(d\theta_1^2 + \sin^2{\theta_1}\,d\varphi_1^2) + \frac{1}{6}\,(d\theta_2^2 + \sin^2{\theta_2}\,d\varphi_2^2)+\frac{1}{9}\,(d\psi+\cos{\theta_1}\,d\varphi_1+\cos{\theta_2}\,d\varphi_2)^2\,.
\end{equation}
Thus the metric is asymptotically locally $\text{AdS}_5\times T^{1,1}$ with AdS radius \eqref{eq:UV_radius} dual to the Klebanov--Witten CFT~\cite{Klebanov:1998hh}.

\paragraph{IR asymptotics.} Since the warp factor depends on an internal coordinate $\theta_2$, in the Klebanov--Murugan solution, the radial direction characterizing the energy scale of the dual theory is given by the expression $\hat{r}\equiv \hat{r}(r,\theta_2)$, which in the UV reduces to $\hat{r} \sim r$. This is because the IR does not simply correspond to taking $r\rightarrow 0$, but to taking both $r\rightarrow 0$ and $\theta_2\rightarrow 0$ at the same time.

To find this radial coordinate $\hat{r}$, consider the resolved conifold metric \eqref{eq:resolved_conifold} in the IR regime. First taking $r\rightarrow 0$, we obtain
\begin{align}
    ds_{6}^2 \sim \frac{3}{2}\,dr^2 &+ \frac{r^2}{6}\,(d\theta_1^2 + \sin^2{\theta_1}\,d\varphi_1^2)+a^2\, (d\theta_2^2 + \sin^2{\theta_2}\,d\varphi_2^2)\nonumber\\
    &+\frac{r^2}{6}\,(d\psi +\cos{\theta_1}\,d\varphi_1+\cos{\theta_2}\,d\varphi_2)^2\,,\quad r\rightarrow 0\,.
    \label{eq:resolved_conifold_r_0}
\end{align}
Now, taking the limit $\theta_2\rightarrow 0$, we have $\sin{\theta_2}\sim \theta_2$ and $\cos{\theta_2}\sim 1$. Thus defining the coordinates $\chi = \psi + \varphi_2$ and $\varphi = \varphi_2$ (which defines an invertible coordinate transformation from $(\varphi_2,\psi)$ to $(\varphi,\chi)$), we obtain
\begin{equation}
    ds_{6}^2 \sim \frac{3}{2}\,dr^2 + \frac{r^2}{6}\,(d\theta_1^2 + \sin^2{\theta_1}\,d\varphi_1^2)+a^2\, (d\theta_2^2 + \theta_2^2\,d\varphi^2)+\frac{r^2}{6}\,(d\chi +\cos{\theta_1}\,d\varphi_1)^2\,,\quad r,\theta_2\rightarrow 0\ .
\end{equation}
Defining further $u\equiv \sqrt{\frac{2}{3}}\,r$, the metric becomes
\begin{equation}
    ds^2_6 \sim du^2+a^2\,(d\theta_2^2+\theta_2^2\,d\varphi^2)+\frac{u^2}{4}[(d\chi+\cos \theta_1d \varphi_1)^2+d\theta^2_1+\sin^2{\theta_1}\, d\varphi_1^2] \,,\quad r,\theta_2\rightarrow 0\ .
\end{equation}
The metric in the square brackets is the round unit $S^3$ metric and the range of the $\chi$-coordinate is $0\leq \chi < 4\pi$. Let us write $u=\hat{r} \sin{\alpha}$ and $a\theta_2=\hat{r}\cos{\alpha}$ so that
\begin{equation}
    \hat{r}^2 = \frac{2}{3}\,r^2 + a^2\theta_2^2\quad , \quad \tan{\alpha} = \sqrt{\frac{2}{3}}\frac{r}{a\theta_2}\ .
    \label{eq:tilde_r_alpha}
\end{equation}
We see that in the IR $(r,\theta_2) \rightarrow 0$, we have $\hat{r}\rightarrow 0$ while the ratio $r\slash \theta_2\geq 0$ is positive corresponding to $0\leq \alpha < \frac{\pi}{2}$. We obtain
\begin{equation}
    ds^2_6 \sim  d\hat{r}^2+\hat{r}^2\,(d\alpha^2+\sin^2{\alpha}\, ds^2_{S^3}+\cos^2{\alpha}\, d\varphi_2^2) = d\hat{r}^2 + \hat{r}^2 ds_{S^5}^2\,,\quad r,\theta_2\rightarrow 0\,,
\end{equation}
where the metric in the brackets is the round unit $S^5$ metric. Therefore, in the limit $r,\theta_2\rightarrow 0$ the resolved conifold is just a flat $\mathbb{R}^6$ with $\hat{r}$ being the radial distance from the origin and $(\alpha,\chi,\theta_1,\varphi_1,\varphi_2)$ being the angular coordinates.

In this regime, the Laplace equation \eqref{eq:laplace_eq_resolved_conifold} on the resolved conifold reduces to the Laplace equation on a flat $\mathbb{R}^6$ with a point-particle source located at the origin. We obtain, for the warp factor
\begin{equation}
    \mathcal{H}(r,\theta_2) \sim \frac{C}{4\pi^3}\frac{1}{\hat{r}^4} = \frac{C}{4\pi^3}\frac{1}{(\frac{2}{3}\,r^2 + a^2\theta_2^2)^2}\,,\quad r,\theta_2\rightarrow 0\,,
    \label{eq:warp_factor_IR}
\end{equation}
where $C = (2\pi)^4g_{\text{s}}\alpha'^2N $. This IR behavior of the warp factor can also be recovered directly from the series expansion \eqref{eq:hhat_series_solution} as argued in~\cite{Klebanov:2007us}. In this IR limit, the metric \eqref{eq:KM_metric} takes the form
\begin{equation}
    ds^2_{10} = \frac{\hat{r}^2}{L_{\text{IR}}^2}\,( -dt^2+ dx_1^2+dx_2^2+dx_3^2) + \frac{L_{\text{IR}}^2}{\hat{r}^2}\,(d\hat{r}^2 + \hat{r}^2 ds_{S^5}^2 )\,,
    \label{eq:KM_IR_metric}
\end{equation}
which is $\text{AdS}_5\times S^5$ with AdS radius
\begin{equation}
    L_{\text{IR}}^4 \equiv \frac{C}{4\vol{(S^5)}} = 4\pi g_{\text{s}}\alpha'^2 N
\end{equation}
with $\vol{(S^5)} = \pi^3$ describing a dual $\mathcal{N} = 4$ $SU(N)$ SYM as argued in~\cite{Klebanov:2007us}.

\subsection{The covariant c-function}

The analysis of the previous section shows that the IR of the Klebanov--Murugan solution is located at $(r,\theta_2)\rightarrow 0$ while the UV is at $r\rightarrow \infty$. Thus the holographic radial coordinate $\rho$ is not simply $r$, but it mixes with the internal coordinate $\theta_2$ as
\begin{equation}
    \rho\sim \begin{dcases}
        \frac{r}{L_{\text{UV}}}\,,\quad &r\rightarrow \infty\,,\quad \ \,\text{UV limit}\\
        \frac{\hat{r}}{L_{\text{IR}}}\,,\quad & r,\theta_2\rightarrow 0\,,\quad \text{IR limit}
    \end{dcases}\ ,
    \label{eq:limiting_rho_values}
\end{equation}
where $\hat{r}$ is defined in \eqref{eq:tilde_r_alpha}. In the UV, this gives the $\text{AdS}_5\times T^{1,1}$ Klebanov--Witten CFT and in the IR the $\text{AdS}_5\times S^5$ SYM theory. As a result, the normal vector $n^\mu$ \eqref{eq:n_vector_field} pointing along the holographic radial direction is not constant but varies along the flow.

We do not have a universal covariant definition for the holographic radial coordinate $\rho = \rho(r,\theta_2)$ which works for all top-down backgrounds. However, in this particular example, one can be found and it is given by
\begin{equation}
    \rho(r,\theta_2) = \sqrt{-\mathcal{G}_{tt}} = \mathcal{N}= \mathcal{H}(r,\theta_2)^{-\frac{1}{4}}\,,
    \label{eq:KM_rho}
\end{equation}
where $\mathcal{G}_{tt}$ is the time-time component of the 10-dimensional metric \eqref{eq:KM_metric}. This is covariantly defined as $\rho = \sqrt{-\mathcal{G}_{\mu\nu}\,\xi^\mu_t\xi^\nu_t}$ where $\xi^\mu_t$ is the timelike Killing vector present in a static spacetime. From the asymptotics \eqref{eq:KM_hhat_UV_expansion} and \eqref{eq:warp_factor_IR} of the warp factor, we can see that it correctly interpolates between the limits \eqref{eq:limiting_rho_values}. We will now use the foliation induced by \eqref{eq:KM_rho} to compute the covariant c-function \eqref{eq:covariant_c_function_V2}.

\paragraph{IR expansion of the c-function.} In the IR $\rho\rightarrow 0$, the Klebanov--Murugan metric \eqref{eq:KM_metric} approaches the $\text{AdS}_5\times S^5$ form \eqref{eq:KM_IR_metric} in coordinates $(\hat{r},\alpha)$ defined above \eqref{eq:tilde_r_alpha}. To leading order in $\rho\rightarrow 0$, the covariant c-function \eqref{eq:covariant_c_function_V2} along constant-$\rho$ slices in the metric \eqref{eq:KM_metric} can be computed along constant-$\hat{r}$ slices in the $\text{AdS}_5\times S^5$ metric \eqref{eq:KM_IR_metric}. Hence, the c-function approaches the $\mathcal{N}=4$ SYM theory value derived in Section \ref{sec:conformalmodels}:
\begin{equation}
    \lim_{\rho\rightarrow 0}\mathcal{C}(\rho) =\frac{N^2}{4\pi}\,.
\end{equation}
Hence the correct IR endpoint value is recovered once the change in the holographic radial direction is taken into account.

\paragraph{UV expansion of the c-function.} Using the UV asymptotics \eqref{eq:KM_hhat_UV_expansion} of the warp factor, we obtain
\begin{equation}
    \rho(r,\theta_2) =  \frac{r}{L_{\text{UV}}}\bigg[1+\frac{3}{4}\frac{a^2}{r^2}(2-3\cos\theta_2)+\frac{a^4}{r^4}\bigg(\frac{117}{64}+\frac{297 \cos \theta_2}{40}+\frac{405}{64} \cos 2 \theta_2\bigg)+\ldots\bigg]\ .
\end{equation}
The internal coordinate $\alpha$ transverse to $\rho$ satisfies the orthogonality condition
\begin{equation}
    n^\mu\,\partial_\mu \alpha(r,\theta_2) = 0
    \label{eq:alpha_diff_eg}
\end{equation}
as indicated in \eqref{eq:yhat}. This equation is solved by
\begin{equation}
    \alpha(r,\theta_2) = \theta_2+\frac{27}{4}\frac{a^2}{r^2}\sin{\theta_2} - \frac{a^4}{r^4}\sin \theta_2 \bigg(\frac{729}{20}+\frac{729}{32}\cos \theta_2\bigg) +\ldots\ .
\end{equation}
and $0 < \theta_2\leq \pi$ corresponds to $0 < \alpha\leq \pi$. In the limit $\rho\rightarrow \infty$, these equations invert as
\begin{equation}
\begin{aligned}
    r(\rho,\alpha) & = L_{\text{UV}}\rho\,\bigg[1+\frac{3}{4}\frac{a^2}{L_{\text{UV}}^2}\frac{3\cos{\alpha}-2}{\rho^2} +\frac{a^4}{L_{\text{UV}}^4\rho^4}\bigg(\frac{63}{64}-\frac{27 \cos \alpha }{40}-\frac{1053}{64} \cos 2 \alpha \bigg)+\ldots\bigg]\\  \theta_2(\rho,\alpha) & = \alpha - \frac{27}{4}\frac{a^2}{L_{\text{UV}}^2}\frac{\sin{\alpha}}{\rho^2}+\frac{a^4}{L_{\text{UV}}^4}\frac{\sin \alpha}{\rho^4} \bigg(\frac{81}{5}+\frac{3159}{32}\cos \alpha\bigg)+\ldots\ .
    \end{aligned}
\end{equation}
We have also kept track of and computed the leading terms hidden in the ellipsis, which contribute to the calculation below. Substituting into \eqref{eq:KM_metric}, we obtain the nine-dimensional spatial metric $G_{\alpha\beta}$ in $(\rho,\alpha)$ coordinates and it takes the form
\begin{equation}
     G_{\alpha\beta}\,dX^\alpha dX^\beta = \hat{g}_{ab}(\rho,\alpha)\,dx^adx^b+ \hat{B}(\rho,\alpha)\,d\rho^2 + \hat{h}_{ij}(\rho,\alpha)\,(d\hat{y}^i + \hat{A}^i_a\,dx^a)(d\hat{y}^j + \hat{A}^j_b\,dx^b)\ ,
     \label{eq:full_metric_KM}
\end{equation}
where by \eqref{eq:KM_rho} the field theory part is explicitly
\begin{equation}
    \hat{g}_{ab}(\rho,\alpha)\,dx^adx^b = \rho^2\,(dx_1^2+dx_2^2+dx_3^2)\,,
\end{equation}
the internal coordinates are $\hat{y}^i = (\theta_1,\alpha,\varphi_1,\varphi_2,\psi)$ and the radial component has the expansion
\begin{equation}
    \hat{B}(\rho,\alpha) = \frac{L_{\text{UV}}^2}{\rho^2}\biggl[1-\frac{9}{2}\frac{a^2}{L_{\text{UV}}^2}\frac{\cos{\alpha}}{\rho^2} + \frac{9}{80}\frac{a^4}{L_{\text{UV}}^4}\frac{205+216\cos{\alpha}+765\cos{2\alpha}}{\rho^4} +\ldots\biggr]\,.
    \label{eq:hat_B}
\end{equation}
The expansion of the determinant $\hat{h} = \det \hat{h}_{ij}$ of the internal part is explicitly
\begin{equation}
   \hat{h}(\rho,\alpha) = \frac{L_{\text{UV}}^{10}\sin^2{\alpha}\sin^2{\theta_1}}{11664}\biggl[1-\frac{9}{2}\frac{a^2}{L_{\text{UV}}^{2}}\frac{\cos{\alpha}}{\rho^2}+\frac{9}{40}\frac{a^4}{L_{\text{UV}}^4}\frac{205+216\cos{\alpha}+765\cos{(2\alpha)}}{\rho^4} +\ldots\biggr]\,,
   \label{eq:KM_dets}
\end{equation}
while the determinant of the field theory part is simply $\hat{g} = \det \hat{g}_{ab} = \rho^6$. There is no cross term $d\rho\,d\alpha$ in the metric since $\rho$ and $\alpha$ are orthogonal by \eqref{eq:alpha_diff_eg}. In these coordinates, $n^\alpha = \hat{B}^{-1\slash 2}\,\delta_\rho^\alpha$ and the extrinsic curvature is simply
\begin{equation}
    K = \frac{1}{2\sqrt{\hat{B}}}\partial_\rho\log{(\hat{g}\hat{h})}
\end{equation}
with which the covariant c-function \eqref{eq:covariant_c_function} becomes
\begin{equation}
    \mathcal{C}(\rho) = \frac{1}{G_{\text{N}}^{(10)}}\int_0^\pi d\alpha\int_0^\pi d\theta_1\int_0^{2\pi} d\varphi_1\int_0^{2\pi} d\varphi_2\int_0^{4\pi} d\psi\,\sqrt{\hat{h}}\,\biggl(\frac{3\sqrt{\hat{B}}}{\partial_\rho\log{(\hat{g}\hat{h})}}\biggr)^3\,.
\end{equation}
Substituting the expansions \eqref{eq:KM_dets} and \eqref{eq:hat_B} gives
\begin{equation}
    \mathcal{C}(\rho) = \frac{27N^2}{64\pi}\,\bigg(1 - \frac{2}{3\pi g_{\text{s}}N}\frac{a^4}{\ell_{\text{s}}^4}\frac{1}{\rho^4} +\ldots\bigg)\ ,
\end{equation}
where the leading term is the UV fixed-point value \eqref{eq:C_T11} of $\text{AdS}_5\times T^{1,1}$ and the subleading correction is negative in accordance with monotonicity. Physically, this $a$-dependent ${\cal O}(1/\rho^4)$ correction explicitly demonstrates how turning on the expectation value of the dimension-two operator $\cal U$ spontaneously breaks conformal symmetry: its quadratic backreaction on the geometry drives the theory away from the UV fixed point and initiates the downward RG flow; conversely, had the symmetry been explicitly broken by relevant deformations, we would have expected lower-order terms to dominate this asymptotic expansion. It would be desirable to prove monotonicity beyond the UV region. We have preliminary numerical evidence that this is the case, but a more detailed analysis is left for future work.

\paragraph{Difference to other backgrounds.} In the Klebanov--Murugan metric \eqref{eq:KM_metric}, the radial component $B(r,\theta_2) = \mathcal{H}(r,\theta_2)^{\frac{1}{2}}\slash \kappa(r)$ depends non-trivially on the internal coordinate $\theta_2$. This distinguishes it from all examples discussed in Section~\ref{sec:tests}, though not from the background considered in Appendix~\ref{subapp:very_demanding_example}. What truly differentiates the Klebanov--Murugan solution, or its generalizations~\cite{Cvetic:2007nv}, from all other backgrounds studied in this paper, including the example of Appendix~\ref{subapp:very_demanding_example}, is that it is not obtained as the uplift of a lower-dimensional metric via \eqref{eq:top_down_reduction}. This is precisely why previously proposed foliation-based c-functions, such as those of~\cite{Macpherson:2014eza,Bea:2015fja,Merrikin:2022yho}, cannot be applied directly to the Klebanov--Murugan solution, whereas our covariant definition remains well defined.

To see the obstruction to such an uplift, consider a $(4+1)$-dimensional metric of the form
\begin{equation}
    ds^2_{5} = \overbar{\Delta}(r)^2\,(-dt^2+ dx_1^2+dx_2^2+dx_3^2) + \overbar{B}(r)\,dr^2\,.
    \label{eq:KM_reduction}
\end{equation}
For the uplift \eqref{eq:top_down_reduction} to reproduce the Klebanov--Murugan metric \eqref{eq:KM_metric} in the field theory directions, the warp factor must be chosen as
\begin{equation}
    W(r,\theta_2) = \frac{\mathcal{H}(r,\theta_2)^{-\frac{1}{2}}}{\overbar{\Delta}(r)^2}\,.
\end{equation}
However, with this choice the radial component produced by the uplift is
\begin{equation}
   W(r,\theta_2)\,\overbar{B}(r) = \frac{\overbar{B}(r)}{\overbar{\Delta}(r)^2}\,\mathcal{H}(r,\theta_2)^{-\frac{1}{2}}\,.
\end{equation}
This cannot agree with the radial component of the Klebanov--Murugan metric $B(r,\theta_2) =\mathcal{H}(r,\theta_2)^{\frac{1}{2}}\slash \kappa(r)$ unless either $\overbar{\Delta}$ or $\overbar{B}$ is allowed to depend on the internal coordinate $\theta_2$ in contradiction with the assumption that \eqref{eq:KM_reduction} is a bottom-up metric.

\section{Discussion}\label{sec:discussion}

In this paper we proposed a covariant definition of a holographic c-function, \eqref{eq:covariant_c_function_V2}, formulated directly in terms of the extrinsic curvature of codimension-two slices in top-down geometries. The definition is coordinate independent, relying only on the staticity and $(d-1)$-dimensional translation invariance of the bulk metric in the field theory spatial directions, together with the existence of a spacelike holographic radial direction and a compact internal manifold. In particular, the construction does not rely on a special form of the metric in a specific coordinate system or on the existence of a consistent lower-dimensional truncation. We compared our definition to previous c-function proposals in the literature, including non-covariant top-down proposals and both covariant and non-covariant bottom-up proposals. We found that our definition unifies all these previous c-functions within a single covariant expression, reproducing them in various special cases. Furthermore, at conformal fixed points, our proposed function correctly evaluates to the standard central charges\footnote{In even dimensions, this would be the type A Weyl anomaly coefficient~\cite{Henningson:1998gx,Henningson:1998ey}, and in odd dimensions, the Euclidean partition function on a sphere.} and is stationary.

We tested our c-function in various explicit top-down examples, including generic backgrounds dual to fixed-point CFTs, such as D3-branes, M5-branes, M2-branes, and the D1-D5 system, and to holographic RG flows, including D-brane backgrounds and confining flows. In all cases, we find consistent monotonic behavior and agreement with previous results. Moreover, the construction remains applicable even when previous formulas cannot be used. The Klebanov--Murugan background provides a clear illustration: its warp factors depend on internal coordinates in a non-factorizable way, which obstructs c-functions based on lower-dimensional truncations, yet our covariant definition yields a well-defined quantity with the correct asymptotics.

\paragraph{Internal space integration.} Our covariant c-function involves an integration over the internal manifold which may at first seem unnatural. From the bulk perspective, however, it is unavoidable: it produces the effective volume factor that enters the holographic central charges and therefore accounts for the total number of degrees of freedom carried by the internal sector. This is familiar from dimensional reduction, where the internal volume sets the normalization of the lower-dimensional Newton's constant and hence the central charge (see for example~\cite{Kanitscheider:2009as,Couzens:2018wnk}). This fact underlies the ``invariance'' of the c-function under dimensional reduction allowing us to calculate the c-function using both the top-down or the bottom-up geometry (or any partially reduced geometry such as the D1-D5 metric \eqref{eq:D1_D5_metric}).

From the field-theory point of view, the internal manifold geometrically parametrizes a moduli space of degenerate vacua. Integrating over these internal directions can be interpreted as an effective averaging over the available vacuum configurations. In principle, one could perform this averaging using a judiciously chosen normalized weight distribution over the internal coordinates,
which would correspond to a specific weighting of these states. However, at conformal fixed points, the value of the c-function must exactly reproduce the universal central charge, imposing constraints on the allowed weight functions, but the constraints do admit non-uniform solutions in general. The central charge is a theory-dependent property, an inherent feature of the conformal fixed point itself. As such, it must be entirely blind to state-dependent features, such as a particular choice of vacuum. Away from the fixed points, the situation is different. The c-function is not itself an observable, but rather a scale-dependent diagnostic of the flow, similar in spirit to a beta function. Introducing a non-uniform weight would amount to a smooth redefinition of this diagnostic. As long as such deformations preserve monotonicity and endpoint values, they do not change the physical content.\footnote{Note that for \emph{any} function, a redefinition of the dependent variable leaves the values at the extrema unchanged, while altering the function away from them.} This ambiguity is expected and reflects the fact that only the fixed-point data are scheme-independent, while the interpolation depends on the chosen definition. Viewed in this light, our proposal of an unweighted, uniform geometric measure provides a covariant and natural representative within this class of valid schemes. Furthermore, this choice yields a c-function that is explicitly global-symmetry invariant at the fixed points.

\paragraph{Holographic radial direction.} The definition of a holographic measure for the number of degrees of freedom of the dual QFT consists of two steps. The first step identifies the holographic radial direction in the bulk, which is used to parametrize the RG flow and which is dual to the QFT energy scale, and the second step identifies a geometric quantity, which computes the number of degrees of freedom at a given scale. To define the energy scale of the dual theory covariantly, it is expected that there exists a scalar field $\rho$ in the bulk whose iso-surfaces $E = \rho(X)$, with $X^\mu=(t,x^a,r,y^i)$ being a bulk coordinate system, describe the region of the bulk which is encoded in the field theory at energy $E$.  Assuming such a covariant definition of the energy scale using a scalar field can be found, we have given a universal covariant proposal for the geometric quantity that computes the number of degrees of freedom, but we have not given a universal proposal for the scalar field itself.

From the bottom-up $(d+1)$-dimensional point of view, there are no ambiguities in determining the holographic radial direction: under the assumption of $d$-dimensional translation invariance, it is simply the direction orthogonal to the $d$ vector fields generating the symmetry. The issue arises in top-down geometries where the space orthogonal to the vector fields now includes the internal directions. Thus the radial direction can mix with the internal directions in different ways and no universal solution has been given in the literature. Most top-down geometries studied in the literature are obtained as an uplift of a solution of a lower-dimensional consistent truncation of the top-down theory. For such solutions, there is no ambiguity in the radial direction as it is determined by the uplift, however, this is simply the lamp post principle at play and there are also top-down geometries which do not have a well-defined dimensional reduction. A general definition of the radial direction for such geometries is preferred.

The Klebanov--Murugan geometry considered in Section \ref{sec:KlebanovMurugan} is an explicit example of a top-down geometry which does not have a well-defined dimensional reduction. To our knowledge, this is one of the only examples (see also~\cite{Cvetic:2007nv}) where the corresponding RG flow is well-controlled in the dual field theory as well: the UV is Klebanov--Witten theory while the IR is $\mathcal{N}=4$ SYM theory. Indeed, the IR $\text{AdS}_5\times S^5$ geometry is located at a different value for the internal angles compared to the UV, reflecting the evolution of the holographic radial coordinate. In this case, we also proposed that the energy scale is dual to the time-time component of the metric: in other words, the scalar field $\rho$ is simply the norm of the timelike Killing vector field present due to staticity. We showed that this is sufficient to match with the UV and IR central charges and also for our c-function to be monotonic near the UV, but a full analysis along the flow is out of reach.

One may ask whether the same definition of the radial direction holds in other situations, for example, in geometries where a dimensional reduction exists. The answer appears negative as it already fails in generic conformal models \eqref{eq:metricaAdSgeneric}, where the AdS part of the metric comes equipped with a $y$-dependent warp factor: in this case, the time-time component of the metric gives a radial direction that also has an internal component and not the $r$-direction that describes the energy scale of the dimensionally reduced theory. In the confining flow geometry of Appendix \ref{subapp:very_demanding_example}, which also has a dimensional reduction, but where the same warp factor depends on both $r$ and $y$ in a non-factorized manner, the problem is more imminent. We thus leave the issue of a universal definition of the radial direction in top-down geometries to future work.

\paragraph{Derivation from an action.}

A conceptual issue that our work leaves open is the \emph{bulk-action origin} of the covariant c-function. Our definition is written directly in terms of the Einstein frame metric and extrinsic curvature, while being proportional to the inverse Newton's constant and invariant under dimensional reduction. These are all notions attached to the Einstein--Hilbert action, which suggests that it might be possible to provide a derivation of the c-function functional directly from the bulk gravitational action. In this respect there is a suggestive parallel with black hole and holographic entanglement entropy functionals.

In Einstein gravity, the Bekenstein--Hawking area law for black hole entropy can be understood as arising from differentiating the bulk Lagrangian with respect to the Riemann tensor, as shown in~\cite{Wald:1993nt,Iyer:1994ys}. The idea that the resulting black hole entropy functional might serve, in a holographic setting, as a measure of the effective number of degrees of freedom has also been explored in the literature. In this direction, proposals for c-functions constructed from the higher-curvature gravity Lagrangian, under the assumption that the matter sector satisfies the NEC, were advanced in~\cite{Cremades:2006ke,Gullu:2010st}. A closely related line of development is the causal-horizon program, in which holographic RG flow is reinterpreted in terms of the thermodynamics of dynamical causal horizons~\cite{Banerjee:2015opa,Banerjee:2015uaa}. These bottom-up, gravity-side constructions are similar in spirit to our covariant c-function, although they are not explicitly related.

A second parallel with our proposal is holographic entanglement entropy (hEE). In AdS/CFT, the Ryu--Takayanagi (RT) prescription associates the entanglement entropy (EE) of a boundary region with the area of a codimension-two minimal surface in the bulk. Our c-function is also based on codimension-two geometry, albeit a different choice of surfaces. Moreover, the RT area functional can be related to the bulk gravitational action: in the replica approach of Lewkowycz and Maldacena and its extensions~\cite{Lewkowycz:2013nqa,Dong:2013qoa,Camps:2013zua}, the area term arises from differentiating the bulk Lagrangian with respect to the Riemann tensor. This provides a genuine bulk-action origin for the geometric object that computes hEE and it is therefore natural to ask whether an analogous construction exists for our covariant c-function. Such a derivation could help in determining the field theory dual of our c-function and a more detailed investigation of these questions is left for future work. We emphasize, however, that finding a fundamental origin for such geometric objects is difficult. It should be kept in mind that even the RT prescription itself does not yet have a derivation from string theory. A plausible reason is that the RT surface is not a fundamental object in string theory: it is an emergent semi-classical construct that appears only in the supergravity limit. This gap between the microscopic string description and the effective geometric picture is one reason why deriving RT, or a similar principle for our c-function, directly from first principles remains challenging. It is also worth noting that, even in theories where both sides of the duality are well understood, we currently lack fully controlled, interacting field-theory computations of EE that match its bulk counterpart, hEE. Nevertheless, the overall network of nontrivial checks, including universal terms and higher-curvature generalizations, makes the standard holographic entanglement prescription highly compelling, strongly inviting the search for similar derivations for other geometric probes.

\paragraph{Relation to entanglement c-functions.} Our foliation-based holographic c-function is not directly related to any previous entanglement-entropy-based c-function. One might expect such a connection, because entanglement entropy is designed to quantify how many degrees of freedom are entangled across a given scale. In two dimensions this intuition underlies entropic reformulations of Zamolodchikov's c-theorem~\cite{Casini:2004bw}, while in three dimensions it is reflected in F-type theorems~\cite{Casini:2015woa, Casini:2017vbe}, where universal terms of entanglement entropy play a central role. These facts have motivated many entanglement c-function proposals that have appeared in the literature~\cite{Liu:2012eea}. In addition, nonperturbative lattice studies of R\'enyi entropies and the associated entropic c-functions provide further support: for $(3+1)$-dimensional SU($N_c$) Yang--Mills theory and for $(2+1)$-dimensional $\mathbb{Z}_2$ gauge theory, slab-based entropic c-functions scale as $N_c^2-1$ at short distances and decrease towards zero at large separations, signaling the transition from deconfined to colorless states~\cite{Buividovich:2008kq,Nakagawa:2009jk,Rabenstein:2018bri,Rindlisbacher:2022bhe,Bulgarelli:2024onj}.  These results support the idea that entanglement based quantities can act as good measures of degrees of freedom in interacting gauge field theories.

However, there are holographic situations in which entanglement based c-functions fail to be valid measures of degrees of freedom. A known example concerns holographic confining models studied in~\cite{Jokela:2025cyz} where the Liu--Mezei entanglement c-function~\cite{Liu:2012eea} was found to be non-monotonic in flows such as Girardello--Petrini--Porrati--Zaffaroni~\cite{Girardello:1999bd} and Klebanov--Strassler~\cite{Klebanov:2000hb}. It turns out that our covariant c-function is monotonic in both cases through its equivalence with the flow c-function proven in Section~\ref{subsec:relation_to_old}: the flow c-function was found to be monotonic in~\cite{Jokela:2025cyz} (see also Appendix \ref{appendix-details-KSBB} for details regarding Klebanov--Strassler background). For confining models involving flows across dimensions, such as the ones considered in Section \ref{sec:confiningmodels} and Appendices \ref{subapp:demanding_example} and \ref{subapp:very_demanding_example}, the entanglement c-function of~\cite{Ishihara:2012jg} was also found to be non-monotonic in~\cite{Jokela:2025cyz}, while our covariant proposal is monotonic. Thus confining models appear to prefer the foliation-based definition considered in this work over entanglement based approaches.

\paragraph{Black hole and time-dependent backgrounds.} It would also be desirable to extend the analysis to other backgrounds. Backgrounds with inhomogeneities in spatial field theory directions and asymptotic anisotropic scaling violation were discussed in Section \ref{section-definitions}. Other generalizations include black hole backgrounds which come with an additional energy scale whose implications for irreversibility of the holographic RG flow remain to be clarified. Our covariant c-function is divergent at the horizon of a black hole where the extrinsic curvature vanishes. Indeed, this divergence occurs more generally at any codimension-two surface where $K=0$, such as minimal-area surfaces. This behavior differs from the bottom-up c-function of~\cite{Sahakian:1999bd}, based on the expansion of a null congruence, which is finite at the horizon of a $(d+1)$-dimensional black hole: as is visible from the formula \eqref{eq:Sahakian_equals_covariant}, the divergence coming from the extrinsic curvature in $\mathcal{C}(r)$ is canceled by the zero of the time-time component $\mathcal{N}$ of the metric. Whether it is possible to make sense of the divergence of our c-function or whether a modification is needed at finite temperature is left for future work.

Non-static spacetimes, \emph{i.e.}, stationary and general time-dependent spacetimes, constitute another interesting possible generalization. Our definition of the c-function relied on staticity, or more generally stationarity, and spatial translation invariance of the bulk metric in order to fix a preferred codimension-two foliation of the spacetime (together with the radial foliation). However, this restriction on the metric only concerns the foliation, and the c-function functional itself is also well-defined in general time-dependent geometries. In the time-dependent case, it might be possible to fix a preferred foliation along the lines of the covariant holographic entanglement entropy formula~\cite{Hubeny:2007xt}. A related direction is to relax the restriction to codimension-two slices altogether. Recent work in the large-$D$ limit suggests that hypersurfaces of arbitrary codimension can play a meaningful geometric role in holographic RG flows~\cite{Giataganas:2021jbj,Jain:2025xko}, and it would be interesting to understand whether our construction admits a sensible generalization in that context.

\paragraph{NEC violating flows.} It would be interesting to apply our c-function to the backgrounds explored in~\cite{Conde:2011aa,Hoyos:2021vhl}. In these cases, one encounters a set of seemingly fully consistent supergravity RG-flow solutions that solve the BPS equations, preserve supersymmetry, have no curvature singularities, and exhibit no signs of acausality, but can nevertheless violate the NEC in their dimensionally reduced descriptions. This NEC violation implies that all previous foliation-based holographic c-functions are non-monotonic. This has two possible implications: either the solutions are not UV complete and do not correspond to well-defined dual field theories obeying a posited c-theorem, or the solutions are UV complete and admit holographic duals, but none of the previous holographic c-functions correctly counts the true number of degrees of freedom. Since the monotonicity of our c-function is not directly determined by the NEC of the reduced theory, owing to the extra factor in \eqref{eq:Sahakian_equals_covariant}, it could remain monotonic in these setups.

\acknowledgments

We thank Elias Kiritsis, Vasilis Niarchos, and Achilleas Porfyriadis for useful discussions and Javier Subils for  collaboration at the initial stages of this work. N.~J. was supported in part by the Research Council of Finland through grant no. 354533 and the Centre of Excellence in Neutron-Star Physics (project 374062). J.~K. is supported by the Deutsche Forschungsgemeinschaft (DFG, German Research Foundation) through the German-Israeli Project Cooperation (DIP) grant `Holography and the Swampland', as well as under Germany's Excellence Strategy through the W\"urzburg-Dresden Cluster of Excellence on Complexity and Topology in Quantum Matter - ct.qmat (EXC 2147, project-id 390858490).  C.~N. is supported by STFC grants ST/Y509644-1, ST/X000648/1, and ST/T000813/1. J.~M.~P. was supported by the ERC starting grant 101078061 SINGinGR, under the European Union’s Horizon Europe program for research and innovation. He was also partially supported by  the H.F.R.I. call `Basic research Financing' (Horizontal support of all Sciences)
under the National Recovery and Resilience Plan `Greece 2.0' funded by the European Union --NextGenerationEU (H.F.R.I. Project Number: 15384),  by the H.F.R.I. Project Number: 23770  of the H.F.R.I call `3rd Call for H.F.R.I.'s Research Projects to Support Faculty Members \& Researchers' and the UoC grant number 12030. H.~R. was supported in part by the Magnus Ehrnrooth foundation.

\appendix

\section{Determinant of the codimension-two metric}\label{app:Schur_complement}

The metric \eqref{eq:H_metric} can be written as
\begin{equation}
    H_{AB}(Y)\,dY^A \,dY^B  = (g_{ab} + h_{ij}\,A^{i}_{a}A^{j}_{b})\,dx^adx^b+h_{ij}A^i_a\, dx^ady^j +h_{ij}A^j_b\, dx^bdy^i +h_{ij}\,dy^idy^j\,,
\end{equation}
which can be represented as a matrix
\begin{equation}
	H_{AB} = \begin{pmatrix}
		g+A^{T} h A & A^{T} h\\
		hA & h
	\end{pmatrix}
\end{equation}
where $A = (A^i_a)$ is regarded as a $p\times (d-1) $ matrix, with row index $i$ and column index $a$, $h = (h_{ij})$ is a $p\times p$ matrix, and we use the notation
\begin{equation}
	(A^{T} h A)_{ab}=h_{ij}\,A^{i}_{a}A^{j}_{b}\,,\quad (A^{T} h)_{aj} = h_{ij}A^i_a\,,\quad (h A)_{ia} = h_{ij}A^j_a\,.
\end{equation}
By the Schur complement formula, the determinant of a block matrix
\begin{equation}
	M = \begin{pmatrix}
		A & B\\
		C & D
	\end{pmatrix}
\end{equation}
is given by
\begin{equation}
	\det M = \det D\det (A - BD^{-1}C)\,.
\end{equation}
Thus we obtain
\begin{equation}
	\det H = \det h\det [g+A^{T} h A-(A^{T} h)\,h^{-1}\,(hA)] \,,
\end{equation}
where $(A^{T} h)\,h^{-1}\,(hA) = A^{T} hA$ so that
\begin{equation}
    \det H = \det h\det g\,.
\end{equation}

\section{Dimensional reduction of the Newton's constant}\label{app:newton_constant}

Consider a $(d+1+p)$-dimensional metric of the form
\begin{equation}
    ds_{d+1+p}^2 
    = W\,ds^2_{d+1}+ h_{ij}\,(dy^i + A^i_a\, dx^a)(dy^j + A^j_b\, dx^b)\,.
\end{equation}
where $ds^2_{d+1}= \overbar{\mathcal{G}}_{\overbar{\mu}\overbar{\nu}}\,d\overbar{X}^{\overbar{\mu}} d\overbar{X}^{\overbar{\nu}}$ is a $(d+1)$-dimensional metric with coordinates
\begin{equation}
    \overbar{X}^{\overbar{\mu}}=(r,t,x^a), \qquad a=1,\dots,d-1,
\end{equation}
and $y^i$, $i=1,\dots,p$, are internal coordinates. The top-down metric is $ ds_{d+1+p}^2 = \mathcal{G}_{\mu\nu}\,dX^\mu dX^\nu$ and we assume $X^{\overbar{\mu}} = \overbar{X}^{\overbar{\mu}}$. By translation invariance, we have $W=W(r,y)$, $h_{ij}=h_{ij}(r,y)$ and $A^i_a=A^i_a(r,y)$. The $(d+1+p)$-dimensional Einstein--Hilbert action is
\begin{equation}
    I_{d+1+p}[\mathcal{G}]
    = \frac{1}{16\pi G_{\text{N}}^{{(d+1+p)}}}
    \int d^{d+1}x\int_{\mathcal{X}_p} d^p y\, \sqrt{-\mathcal{G}}\, R[\mathcal{G}] \,.
\end{equation}
To identify the lower-dimensional Newton's constant, we isolate the term proportional to the Ricci scalar of $\overbar{\mathcal{G}}$. The determinant factorizes as
\begin{equation}
    \sqrt{-\mathcal{G}} = \sqrt{-\overbar{\mathcal{G}}}\,\sqrt{h}\;W^{\frac{d+1}{2}}
\end{equation}
by the Schur complement formula of Appendix \ref{app:Schur_complement}. Moreover, the part of the higher-dimensional Ricci scalar proportional to the $(d+1)$-dimensional Ricci scalar is
\begin{equation}
    R[\mathcal{G}] = W^{-1} R[\overbar{\mathcal{G}}] + \ldots,
\end{equation}
where the omitted terms involve derivatives of $W$, $h_{ij}$, and the field strengths of $A_a^i$. Therefore,
\begin{equation}
    \sqrt{-\mathcal{G}}\,R[\mathcal{G}]
   =\sqrt{-\overbar{\mathcal{G}}}\,\sqrt{h}\,W^{\frac{d-1}{2}} R[\overbar{\mathcal{G}}] + \ldots\,.
\end{equation}
Substituting into the action gives
\begin{equation}
    I_{d+1+p}[\mathcal{G}]
   =
    \frac{1}{16\pi G_{\text{N}}^{{(d+1+p)}}}
    \int d^p y\,\sqrt{h}\,W^{\frac{d-1}{2}}\int d^{d+1}x\,\sqrt{-\overbar{\mathcal{G}}}\, R[\overbar{\mathcal{G}}] + \ldots\,.
\end{equation}
Comparing with the $(d+1)$-dimensional Einstein--Hilbert term,
\begin{equation}
    I_{d+1}[\overbar{\mathcal{G}}]
   =
    \frac{1}{16\pi G_{\text{N}}^{{(d+1)}}}
    \int d^{d+1}x\,\sqrt{-\overbar{\mathcal{G}}}\,R[\overbar{\mathcal{G}}]
\end{equation}
yields the relation \eqref{eq:Newton_constant_relation}.

\section{Expansion of a null congruence}\label{app:null_expansion}

We consider the $(d+1)$-dimensional Lorentzian spacetime $(\overbar{\mathcal{M}},\overbar{\mathcal{G}})$ with a null vector $\overbar{k}^{\overbar{\mu}}$ that satisfies the affine condition \eqref{eq:affine_condition}. We define
\begin{equation}
	\overbar{\Theta}_{\overbar{\mu}\overbar{\nu}} = \overbar{H}_{\overbar{\mu}}^{
	\overbar{\rho}}\overbar{H}_{\overbar{\nu}}^{\overbar{\sigma}}\,\overbar{\nabla}_{\overbar{\rho}}k_{\overbar{\sigma}}\,,
\end{equation}
where the projector to the codimension-two slice is defined as 
\begin{equation}
	\overbar{H}_{\overbar{\nu}}^{\overbar{\mu}} \equiv \delta_{\overbar{\nu}}^{\overbar{\mu}}- \overbar{n}^{\overbar{\mu}} \overbar{n}_{\overbar{\nu}}+\overbar{t}^{\overbar{\mu}}\overbar{t}_{\overbar{\nu}} \equiv \overbar{G}_{\overbar{\nu}}^{\overbar{\mu}}- \overbar{n}^{\overbar{\mu}} \overbar{n}_{\overbar{\nu}}
    \label{eq:Hbar_app}
\end{equation}
and $\overbar{G}_{\overbar{\nu}}^{\overbar{\mu}}$ is the spatial projector. The expansion of the null congruence is then defined as
\begin{equation}
	\overbar{\Theta} \equiv \overbar{G}^{\overbar{\mu}\overbar{\nu}}\,\overbar{\Theta}_{\overbar{\mu}\overbar{\nu}} = \overbar{\nabla}_{\overbar{\mu}}k^{\overbar{\mu}}\,,
    \label{eq:expansion_app}
\end{equation}
where to obtain the second equality, we have used that $ \overbar{G}^{\overbar{\mu}\overbar{\nu}}\, H_{\overbar{\mu}}^{
	\overbar{\rho}}H_{\overbar{\nu}}^{\overbar{\sigma}} = H^{\overbar{\rho}\overbar{\sigma}}$ and that $H^{\overbar{\mu}\overbar{\nu}}\,\overbar{\nabla}_{\overbar{\mu}}k_{\overbar{\nu}} = \,\overbar{\nabla}_{\overbar{\mu}}k^{\overbar{\mu}}$ which follows from the affine condition \eqref{eq:affine_condition} after a straightforward calculation. Substituting the expression $\overbar{k}^{\overbar{\mu}} = \Omega\,( \overbar{n}^{\overbar{\mu}} + \overbar{t}^{\overbar{\mu}})$ for the null vector, we obtain
\begin{equation}
	\overbar{\Theta}_{\overbar{\mu}\overbar{\nu}} = \Omega\,\overbar{H}_{\overbar{\mu}}^{
	\overbar{\rho}}\overbar{H}_{\overbar{\nu}}^{\overbar{\sigma}}\,\overbar{\nabla}_{\overbar{\rho}}(\overbar{n}_{\overbar{\sigma}}+\overbar{t}_{\overbar{\sigma}})\,,
\end{equation}
where we used that $\overbar{H}_{\overbar{\mu}}^{
	\overbar{\rho}}\,\overbar{n}_{\overbar{\rho}} = \overbar{H}_{\overbar{\mu}}^{
	\overbar{\rho}}\,\overbar{t}_{\overbar{\rho}} = 0$. Substituting the second equation \eqref{eq:Hbar_app}, we obtain
\begin{equation}
 \overbar{\Theta}_{\overbar{\mu}\overbar{\nu}} = \Omega\,\left( \overbar{K}_{\overbar{\mu}\overbar{\nu}}+\overbar{L}_{\overbar{\mu}\overbar{\nu}}-\overbar{G}_{\overbar{\mu}}^{\overbar{\rho}}\,\overbar{n}^{\overbar{\sigma}} \overbar{n}_{\overbar{\nu}}\,\overbar{\nabla}_{\overbar{\rho}}\overbar{t}_{\overbar{\sigma}}-\overbar{n}_{\overbar{\mu}}\overbar{n}^{\overbar{\rho}} \,\overbar{G}_{\overbar{\nu}}^{\overbar{\sigma}}\,\overbar{\nabla}_{\overbar{\rho}}\overbar{t}_{\overbar{\sigma}}+ \overbar{n}_{\overbar{\mu}}\overbar{n}_{\overbar{\nu}}\overbar{n}^{\overbar{\rho}}\overbar{n}^{\overbar{\sigma}}\, \overbar{\nabla}_{\overbar{\rho}}\overbar{t}_{\overbar{\sigma}}\right)\,,
\end{equation}
where we have defined
\begin{equation}
    \overbar{K}_{\overbar{\mu}\overbar{\nu}} = \overbar{H}_{\overbar{\mu}}^{
	\overbar{\rho}}\overbar{H}_{\overbar{\nu}}^{\overbar{\sigma}}\,\overbar{\nabla}_{\overbar{\rho}}\overbar{n}_{\overbar{\sigma}} \,,\quad \overbar{L}_{\overbar{\mu}\overbar{\nu}} = \overbar{G}_{\overbar{\mu}}^{
	\overbar{\rho}}\overbar{G}_{\overbar{\nu}}^{\overbar{\sigma}}\,\overbar{\nabla}_{\overbar{\rho}}\overbar{t}_{\overbar{\sigma}}\,.
\end{equation}
Thus the expansion \eqref{eq:expansion_app} becomes
\begin{align}
	\overbar{\Theta} = \Omega\,(\overbar{K} + \overbar{L}-\overbar{n}^{\overbar{\sigma}} \overbar{n}^{\overbar{\rho}}\,\overbar{\nabla}_{\overbar{\rho}}t_{\overbar{\sigma}}) = \Omega\,(\overbar{K} + \overbar{L}-\overbar{n}^{\overbar{\mu}}\overbar{n}^{\overbar{\nu}}\,\overbar{L}_{\overbar{\mu}\overbar{\nu}})\,.
\end{align}
where the last equality follows from $\overbar{n}^{\overbar{\rho}}=\overbar{n}^{\overbar{\mu}}\,\overbar{G}^{\overbar{\rho}}_{\overbar{\mu}} $.

\section{More confining holographic backgrounds}\label{appendix-details-confining}
In this appendix we discuss several technical details that were omitted from 
Section~\ref{sec:confiningmodels} for conciseness. The best way to discuss these details is to give two examples that capture all the technical subtleties. Below we present two systems. The first one discusses the holographic dual to an infinite family of six-dimensional ${\cal N}=(1,0)$ SCFTs. These theories are first compactified on a compact hyperbolic Riemann surface $\mathbb{H}_2\slash \Gamma$ (obtained by compactifying the fundamental domain of the hyperbolic plane $\mathbb{H}_2$ under the quotient with a discrete subgroup $\Gamma$ of PSL$(2,\mathbb{R})$), leading to four dimensional ${\cal N}=1$ SCFTs. Subsequently, this family of four-dimensional SCFTs is compactified on a circle (as explained in Section~\ref{sec:confiningmodels}). The resulting field theory is a flow between a four dimensional strongly coupled SCFT and a three-dimensional gapped QFT. The details of the dual background are below.

\subsection{Demanding example}\label{subapp:demanding_example}

We consider the confining model that is obtained by starting from a family of 4d SCFTs and compactifying on a circle.
The background is discussed in~\cite{Giliberti:2024eii}. We use the conventions in
Section 4.1 of~\cite{Jokela:2025cyz}. The Einstein frame metric and dilaton read,
\begin{eqnarray}
& & ds_{10}^2= \mu \Bigg[r^2\left(-dt^2+dx_1^2+dx_2^2+ f(r) d\phi^2 \right) +\frac{dr^2}{r^2 f(r)} +\frac{4}{3}ds^2_{\mathbb{H}_2} -\frac{\alpha''}{6\alpha}dz^2-\frac{\alpha \alpha''}{3(2\alpha'^2-3\alpha\alpha'')}ds^2_{S^2}\Bigg],\nonumber\\
& & \mu= 3\pi\sqrt{6}
 e^{-\frac{\Phi}{2}}\sqrt{-\frac{\alpha}{\alpha''}}\,,~~~~~~e^{-4\Phi}= \frac{2^5 3^3}{(18\pi)^{10}}\left(-\frac{\alpha''}{\alpha}\right)^3 (2\alpha'^2-3\alpha\alpha'')^2\,,~~f(r)= 1-\frac{q}{r^6}\,,\label{6d4dCFTQFT}\end{eqnarray}
where $ds^2_{\mathbb{H}_2}$ is the two-dimensional hyperbolic metric, $\alpha = \alpha(z)$ is a function of the internal coordinate $z\in (0,P)$ and a prime denotes derivative with respect to $z$. The holographic radial direction is $r$ and the relevant quantities for the c-function are
\begin{equation}
    B(r,z)=\frac{\mu}{f(r)\, r^2}\,,\quad g(r,z) = f(r)\,(\mu\,r^2)^3\,,\quad h(r,z)=
\frac{\mu^5 2^3}{3^5}\frac{-\alpha \alpha''^{{3}}}{(2\alpha'^2-3\alpha\alpha'')^2} \det S^2 \det \mathbb{H}_2\,,\quad p = 5\,.
\label{eq:demanding_functions}
\end{equation}
Substituting into \eqref{eq:covariant_c_function_simplified}, the covariant c-function becomes
\begin{equation}
    \mathcal{C}(r) = \frac{\widetilde{\mathcal{N}}}{2^{3} G_{\text{N}}^{(10)}} \Bigg[\frac{ f(r)^{-\frac{1}{2}}}{1+\frac{r f'(r)}{6f(r)}} \Bigg]^{3}\,,\quad
 \widetilde{\mathcal{N}} \equiv  \frac{64\vol{(\mathbb{H}_2\slash \Gamma)}}{9}\int_0^P dz\, (-\alpha\alpha'')\,,
\end{equation}
which is of the form \eqref{eq:c_function_confining} with $d = 4$.

This covariant c-function matches the non-covariant flow c-function computed in~\cite{Jokela:2025cyz} since the determinants $g$ and $h$ factorize in $(r,z)$; matching in this case was shown in Section \ref{subsec:relation_to_old}. Moreover, the reason why the flow c-function can be computed for this background is because the $z$-dependence cancels in the ratio $\widetilde{B}(r) = B(r,z)\slash g(r,z)^{1\slash 3}$. Note that the c-function also factorizes into a part dependent only on the flow (the $r$-dependent part) and a constant $\widetilde{\cal N}$ that reflects the properties of the UV-SCFT. We believe this second factorization is a consequence of the conjecture by Gauntlett and Varela~\cite{Gauntlett:2007ma}, proved in~\cite{Cassani:2019vcl}.

\subsection{Very demanding example}\label{subapp:very_demanding_example}

In this section we study a more involved example which is a good test of the formula in equations \eqref{eq:covariant_c_function} and \eqref{eq:covariant_c_function-2} mostly because the warp functions in the metric depend both on the internal space coordinates and the radial direction in a non-factorizable way. Moreover, unlike the backgrounds discussed in Section~\ref{sec:confiningmodels},
the internal metric components $h_{ij}$ are not independent of the radial coordinate $r$. Nevertheless, the determinant $H$ of the codimension-two metric entering the covariant c-function factorizes. The solution was constructed in~\cite{Chatzis:2025dnu}, see Section 2.3 of that paper.

To present the background, we introduce notation. We let
\begin{equation}
 \begin{aligned}
  & 
 \lambda(r) = \Big( 1 +  \frac{\ell^2}{r^2}  \Big)^{\frac{1}{6}} 
 \\[5pt] 
  & F(r) = \frac{1}{L^2} - \frac{\ell^2 L^2 q^2}{ r^4} \Big( 1- \frac{1}{\lambda(r)^6} \Big) \, ,
 \end{aligned}\label{lionel1}
\end{equation}
where $  \ell , \, q $ are constants. We define $A_i \, (i = 1,2,3)$ three one-forms such that
\begin{eqnarray}\label{gauge_fields}
 & & A_1 = A_2 = A_{1\phi} d\phi= \, q \, \ell^2 \, L \, \frac{r^2 - r^2_\star}{r^2_\star \, r^2} \, d\phi \\ 
& & A_3 = A_{3\phi}d\phi= \, q_2 \, \ell^2 \, L \, \frac{r^2 - r^2_\star}{ (r^2_\star +  \, \ell^2) \, (r^2 + \, \ell^2) } \, d\phi \, .
\end{eqnarray}
The constant $r_\star$ is the value of the largest root of $F(r)$, and $q_2$ is another free parameter in this background. We also define,
\begin{eqnarray}
& & ds_5^2=\frac{r^2\lambda(r)^2}{L^2} \left(-dt^2+ dx_1^2+ dx_2^2+ L^2 F(r)\, d\phi^2 \right) +\frac{dr^2}{r^2\lambda(r)^{{4}} F(r)}\,,
\nonumber\\[5pt]
& & \mu_1= \sin\theta \cos\varphi\,,~~~~~~~\mu_2=\sin\theta \sin\varphi\,,~~~~~~~\mu_3= \cos\theta\,,
\nonumber\\[5pt]
& &D\mu_i= \left(d\mu_1+ 2\mu_2 A_{1\phi} d\phi\right)\delta_{i,1}+\left(d\mu_2 -2\mu_1 A_{1\phi}d\phi \right)\delta_{i,2}+ d\mu_3 \delta_{i,3}\,,
\label{betoalonso}
\end{eqnarray}
and the functions
\begin{equation}\label{IIA_functions}
    \begin{split}
        & \tilde{f}_1=\kappa^{2/3}\left(\frac{\dot{V}\tilde{\Delta}}{2V^{\prime \prime}}\right)^{1/3},\quad \tilde{f} = X^{-1} Z=\frac{Z}{\lambda^2(r)},\quad \tilde{f}_2=\frac{2\dot{V}V^{\prime\prime}}{Z^2\tilde{\Delta}}, \quad \tilde{f}_3 = \frac{4X^3\sigma^2V^{\prime\prime}}{2X^3\dot{V}-\ddot{V}}Z \, ,
        \\
        &\tilde{f}_4=\frac{2V^{\prime\prime}}{\dot{V}}Z\,,\quad \tilde{f}_5 = \frac{2(2X^3\dot{V}-\ddot{V})}{Z^2\dot{V}\tilde{\Delta}}\,, \quad  \tilde{\Delta}=(\dot{V}^{\prime})^2+V^{\prime\prime}(2\dot{V}-\ddot{V}) \,, 
        \\
        &
        Z=\left[ \frac{(\dot{V}^{\prime})^2+V^{\prime\prime}(2X^3\dot{V}-\ddot{V})}{(\dot{V}^{\prime})^2+V^{\prime\prime}(2\dot{V}-\ddot{V})}\right]^{1/3}\,, 
    \end{split}
\end{equation}
where $X(r)\equiv\lambda(r)^2$, $\kappa$ is a constant, $V = V(\sigma,\eta)$ is function for which a prime denotes the derivative with respect to $\eta$ (that is $\partial_\eta V=V'$) while a dot the derivative $\dot{V}=\sigma\partial_\sigma V$.

The family of Type IIA backgrounds (with RR and NSNS fields included) is written in~\cite{Chatzis:2025dnu}. In Einstein frame we find
\begin{equation}
    \begin{split}
        &\mathrm{d}s^2_{10}= \mu \left[ 4\tilde{f} \mathrm{d}s^2_5 + \tilde{f}_2 \mathrm{D}\mu_i\mathrm{D}\mu^i + \tilde{f}_3(\mathrm{d}\chi+A_3)^2 + \tilde{f}_4(\mathrm{d}\sigma^2 + \mathrm{d}\eta^2) \right]\,,\\
        &\mu = e^{-\frac{\Phi}{2}}\tilde{f}_1^{\frac{3}{2}} \tilde{f}_5^{\frac{1}{2}}\,,\quad  e^{\frac{4}{3}\Phi} = \tilde{f}_1\tilde{f}_5\,.\label{10dGM-2}
    \end{split}
\end{equation}
Compared to the other confining backgrounds, this solution is more complicated in its dependence on coordinates $y = (\sigma,\eta,\chi,\theta,\varphi)$ of the internal space: its components do not factorize in $r$ and $y$. The quantities relevant for the covariant c-function are
\begin{equation}
    B=\frac{4\mu\tilde{f}}{r^2\lambda(r)^{{4}} F(r)}\,,\quad g = \biggl(\frac{4\mu\tilde{f}r^2\lambda(r)^2}{L^2}\biggr)^3L^2F(r)\,,\quad h=
\mu^5  \tilde{f}_2^2 \tilde{f}_4^2 \tilde{f}_3\,\det S^2\det S^1\,,\quad p = 5\,.
\label{eq:very_demanding_functions}
\end{equation}
Substituting into \eqref{eq:covariant_c_function_simplified}, the covariant c-function becomes
\begin{equation}
    \mathcal{C}(r) =\frac{\widetilde{\mathcal{N}}}{2^3G_{\text{N}}^{(10)}}\left[
\frac{F(r)^{-1/2}}{\lambda(r)^2r\,\partial_r\log\!\left(r\lambda(r)F(r)^{1/6}\right)}
\right]^3\,,\quad
 \widetilde{\mathcal{N}} \equiv  32\kappa^3\vol{(S^2)}\vol{(S^1)} \int d\sigma d\eta\,\sigma \dot{V}V''\,.\label{eqD11}
\end{equation}
This result, obtained using the covariant definition, agrees with the one obtained by using the non-covariant approach in~\cite{Chatzis:2025dnu}. The reason for this match is that, although the individual metric components 
do not factorize in $r$ and $y$, the determinant $H=\det H_{AB}$ does (by the Schur complement formula of Appendix \ref{app:Schur_complement}):
\begin{equation}
    H(r,y) = g(r,y)\, h(r,y) = \kappa^{6}\,\frac{1024\,r^6\lambda(r)^6F(r)}{L^4}\,\sigma^2\,\dot{V}(\sigma,\eta)^2V''(\sigma,\eta)^2 \equiv H_r(r)\,H_y(y)\,.
\end{equation}
Note that the non-covariant flow c-function can be computed for this background because $\widetilde{B}$ \eqref{eq:C_flow_first} is independent of the internal coordinates
\begin{equation}
    \widetilde{B}(r) = \frac{B(r,y)}{g(r,y)^{1\slash 3}} = \frac{L^{4\slash 3}}{r^4F(r)^{4\slash 3}\lambda(r)^6}\,.
\end{equation}
This particularly demanding example gives us confidence in our procedure. As in previous examples, the result has the expected behaviors in the UV and in the IR and it is monotonic.

\subsection{The  baryonic branch of Klebanov--Strassler solution}\label{appendix-details-KSBB}

Here we study a family of solutions with a very interesting field theory interpretation. It contains the backgrounds for which the first reliable checks of holography for confining four-dimensional ${\cal N}=1$ SUSY field theories were made~\cite{Klebanov:2000hb, Maldacena:2000yy}. The system under study is a two-node quiver $SU(N)\times SU(N)$~\cite{Klebanov:1998hh}, that is conformal and strongly coupled. This is deformed by an imbalance in the quiver nodes $SU(N)\times SU(N+M)$~\cite{Klebanov:2000nc, Klebanov:2000hb, Gubser:2004qj, Butti:2004pk}, which might be thought of as a quasi-marginal operator deforming the dynamics. Some aspects of the strongly coupled dynamics flowing towards the IR can be studied with field theoretical methods~\cite{Strassler:2005qs, Dymarsky:2005xt}.
The relevant papers to understand the logic and connections with other dual geometries are~\cite{Gubser:2004qj, Butti:2004pk, Maldacena:2009mw, Gaillard:2010qg, Caceres:2011zn}. 

The Einstein frame metric and dilaton for these geometries are
\begin{eqnarray}
 ds^2_{10}& = & e^{\frac{\Phi}{2}}\mathcal{H}^{-\frac{1}{2}}\left[ -dt^2+ dx_1^2+dx_2^2+dx_3^2\right] + e^{\frac{\Phi}{2}}\mathcal{H}^{\frac{1}{2}}\big[e^{2k}dr^2+ e^{2\tilde{h}}(d\theta^2+\sin^2\theta d\varphi^2)\nonumber\\
& & +\frac{e^{2\tilde{g}}}{4}\left[(\omega_1+a d\theta)^2+(\omega_2-a\sin\theta d\varphi)^2 \right]+
\frac{e^{2k}}{4}(\omega_3-\cos\theta d\varphi)^2\big]\,,  ~~~\Phi=\Phi(r)\,.
\label{backgroundBB} 
\end{eqnarray}
We have used the left invariant forms of $SU(2)$,
\begin{equation}
\omega_1=\cos\psi_2 d\theta_2+ \sin\psi_2\sin\theta_2 d\varphi_2\,, ~~\omega_2= -\sin\psi_2 d\theta_2+ \cos\psi_2\sin\theta_2 d\varphi_2\,,~~\omega_3=d\psi_2 +\cos\theta_2d\varphi_2\,.
\end{equation}
The background is complemented by RR three- and five-form field strengths and a NS three-form, which as above are not needed for our calculation.
The functions $\{\Phi,k,\tilde{h},\tilde{g},a,\mathcal{H}\}$ depend only on the radial coordinate. Their expressions are written in terms of two functions $P(r)$ and $Q(r)$~\cite{Hoyos-Badajoz:2008znk, Nunez:2008wi}. For a summary of the notation, see Appendix A of~\cite{Conde:2011aa}. 
\begin{eqnarray}
& & Q=N_c(2r\,\coth(2r)-1)\,, ~~e^{2\tilde{h}}\,=\,\frac{1}{4}\,
\frac{P^2-Q^2}{P\coth(2r)-Q}\,,~~~ e^{2\tilde{g}}\,=\,P\,\coth(2r)\,-\,Q\,,\label{xxxxdd1.A}\\
& & e^{2k}=\frac{P'}{2}\,,~~~ a=\frac{P}{P\cosh(2r)-Q\sinh(2r)}\,,~~~ e^{4\Phi-4\Phi_0}= \frac{2\sinh(2r)^2}{(P^2-Q^2)P'}\,, ~~\mathcal{H}= 1-\kappa^2 e^{2\Phi}.\nonumber
\end{eqnarray}
Whilst the function
$P(r)$ is given as the solution of the following ``master equation''
\begin{equation}
P'' + P'\Big(\frac{P'+Q'}{P-Q} +\frac{P'-Q'}{P+Q} - 4 
\coth(2r)\Big)=0\,.
\label{master.A}
\end{equation}
Different solutions of the master equation (\ref{master.A}) were studied in~\cite{Nunez:2008wi,Gaillard:2010qg}. The range of the coordinates are $0\leq\theta\leq\pi$, $0\leq \theta_2\leq\pi$, $0\leq\varphi\leq 2\pi$ and $0\leq \varphi_2\leq 2\pi$. The circle coordinate $\psi_2$ ranges in $[0,4\pi]$.

The quantities relevant for the calculation of the covariant c-function are
\begin{equation}
    B(r)=e^{\frac{\Phi}{2}+2k}\mathcal{H}^{\frac{1}{2}}\,,\quad g(r) = e^{\frac{3\Phi}{2}}\mathcal{H}^{-\frac{3}{2}}\,,\quad h(r)=
\frac{1}{32}\,e^{\frac{5\Phi}{2} +4\tilde{h}+4\tilde{g}+2k}\mathcal{H}^{\frac{5}{2}} \det S^2 \det S^2\,,\quad p = 5\,.
\label{eq:klebanov_strassler}
\end{equation}
Substituting into \eqref{eq:covariant_c_function_simplified}, the covariant c-function becomes
\begin{equation}
\mathcal{C}(r)=  \frac{27}{G_{\text{N}}^{(10)}}\times 128\pi^3 \frac{e^{2\tilde{g}+2\tilde{h}+4k+2\Phi} \mathcal{H}^2}{\left[\frac{\mathcal{H}'}{\mathcal{H}} + 2(2\tilde{g}'+2\tilde{h}'+ k'+2\Phi') \right]^3} \ .\label{centralBB} 
\end{equation}
The expression in (\ref{centralBB}) coincides with the one derived (via non-covariant methods) in Section C of~\cite{Elander:2011mh}. Plots of this quantity  (for a given $P(r)$) can be found in~\cite{Elander:2011mh}. Qualitatively, this is a monotonic function of the radial coordinate (the energy) that vanishes at $r=0$ (the far IR) and grows logarithmically for large values of $r$. The vanishing in the far IR reflects the absence of interacting low-energy degrees of freedom (more precisely, there is a Goldstone multiplet~\cite{Gubser:2004qj,Strassler:2005qs, Dymarsky:2005xt}, but it decouples from the interacting low energy dynamics). The logarithmic growth in the UV is associated with the fact that the system does not reach a fixed point at high energies, but shows an unconventional UV behavior with growing number of degrees of freedom.

As we stated in the previous examples, the covariant c-function defined in Section~\ref{section-definitions} and the flow c-function (based on a non-covariant definition)~\cite{Bea:2015fja, Merrikin:2022yho} give the same results. 

\begin{figure}[t]
    \centering
    \includegraphics[width=0.5\linewidth]{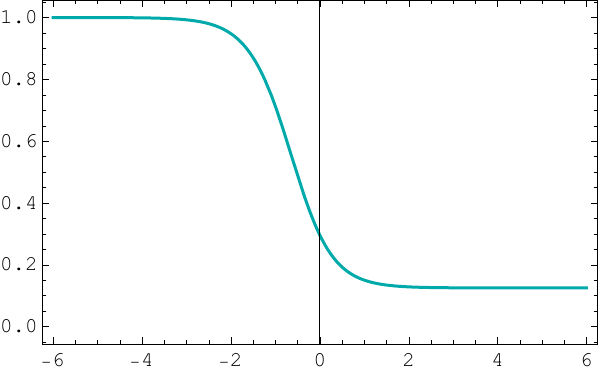}
    \put(-240,150){$\dfrac{G_{\text{N}}^{(10)}}{\widetilde{\mathcal{N}}}\,\mathcal{C}(r)$}
    \put(5,13){$r$}
    \caption{The c-function \eqref{eq:AdS3_H_2_c_function} for the flow from AdS$_5$ to AdS$_3 \times \mathbb{H}_2\slash \Gamma$, in $d=4$. Note that the IR is at $r\to -\infty$ and UV is at $r\to \infty$. The theory is reduced on the $\mathbb{H}_2\slash \Gamma$, a manifold that admits massless modes. These appear in the dynamics of the lower-dimensional theory, \emph{i.e.}, in the IR.}
    \label{fig:cflow-AdS5-AdS3H2}
\end{figure}

\section{\texorpdfstring{Flow from $\mathrm{AdS}_5$ to $\mathrm{AdS}_3\times \mathbb{H}_2$}{}}\label{app:C5}
Here we study a specific flow across dimensions, that begins and ends at a conformal fixed point.
The background written and studied below describes a stack of D3-branes on the tip of the conifold. This stack undergoes a compactification on a hyperbolic plane $\mathbb{H}_2\slash \Gamma$, more precisely, we compactify on the fundamental domain of the hyperbolic plane, quotient by a discrete subgroup of PSL$(2,\mathbb{R})$. The solution is written in 
\cite{Donos:2014eua} and further generalized in~\cite{Bea:2015fja}. Similar solutions were studied using the holographic entanglement entropy in~\cite{de-la-Cruz-Moreno:2023mew, GonzalezLezcano:2022mcd}. Note that although this solution is not translationally invariant in field theory directions due to the presence of the hyperbolic metric, the covariant c-function is well defined and independent of field theory directions.

In Type IIB, we have the metric
\begin{align}
    d s^2 &= e^{2A(r)}\left(-d t^2+d y^2\right) + e^{2B_w(r)}ds^2_{\mathbb{H}_2} + d r^2 + \frac{1}{6}\left(d\theta_1^2 + \sin^2\theta_1d\varphi_1^2+d\theta_2^2+\sin^2\theta_2 d \varphi_2^2\right) \nonumber \\
    &\qquad + \frac{1}{9}\left(d\psi+\cos\theta_1 d \varphi_1 +\cos\theta_2 d \varphi_2\right)^2 \ ,
\end{align}
and we take $F_3=H_3=\phi = \chi = 0$, for $F_5$ see~\cite{Bea:2015fja}. The functions $A(r)$ and $B_w(r)$ (not to confuse this warp factor with the $B(r)$ of the general metric in Section \ref{section-definitions}) read
\begin{equation}
    A(r) = \frac{3}{2}\,r - \frac{1}{4}\log\left( 1+e^{2r}\right) \,, \quad B_w(r) = \frac{1}{2}\log 3 + \frac{1}{2}\log\left(1+e^{2r}\right)\,.
\end{equation}
Notice that for $r\to \infty$, we recover the Klebanov--Witten metric and for $r\to -\infty$, we have the $\text{AdS}_3\times \mathbb{H}_2\slash \Gamma$ solution.

The quantities relevant for the calculation of the covariant c-function are
\begin{equation}
    B(r)=1\,,\quad g(r) = e^{2A(r)+4B_w(r)}\det \mathbb{H}_2\,,\quad h(r)=
\frac{1}{9\cdot 6^4}\, \det S^2 \det S^2\,,\quad p = 5\,.
\label{eq:AdS3_H2_flow}
\end{equation}
Substituting into \eqref{eq:covariant_c_function_simplified} with $d = 4$, the covariant c-function becomes
\begin{equation}
    \mathcal{C}(r)
	=
	\frac{16\pi^3}{27G_{\text{N}}^{(10)}}\biggl(\frac{3}{2}\biggr)^3
	\frac{1}{\left(A'(r)+2B_w'(r)\right)^3} = \frac{\widetilde{\mathcal{N}}}{G_{\text{N}}^{(10)}}\biggl(\frac{1+e^{2r}}{1+2e^{2r}}\biggr)^3\,,
    \label{eq:AdS3_H_2_c_function}
\end{equation}
where we have defined $\widetilde{\mathcal{N}} \equiv \frac{16\pi^3}{27}$. It has the IR and UV behaviors
\begin{equation}
    \lim_{r\to-\infty} \mathcal{C}(r) = \frac{\widetilde{\mathcal{N}}}{G_{\text{N}}^{(10)}}\,,\quad \lim_{r\to\infty} \mathcal{C}(r) = \frac{\widetilde{\mathcal{N}}}{8G_{\text{N}}^{(10)}}
\end{equation}
and it is plotted in Figure \ref{fig:cflow-AdS5-AdS3H2}. It is monotonically increasing towards the IR, but this is expected as additional massless modes appear.

\section{Sahakian c-function for some examples}\label{sahakian-examples}

In this appendix we discuss the expressions \eqref{eq:Sahakian_c_function_simplified} and \eqref{eq:Sahakian_equals_covariant}, when applied to some of the examples studied above. In particular, we focus our attention on backgrounds of the form
AdS$_5\times X^5$ in Section \ref{sec:conformalmodels}, on the generic CFTs in Section \ref{subsec:genericCFT} and on the very-demanding-example of Appendix \ref{subapp:very_demanding_example}. The c-function proposed by Sahakian~\cite{Sahakian:1999bd} works well when applied to a lower-dimensional gauge supergravity, characteristically a consistent truncation of Type II string or M-theory. The examples above can be understood within this logic.

To start with, consider an AdS$_5$ supergravity, with a solution of the form
\begin{equation}
 ds_5^2= \frac{r^2}{L^2}(-dt^2+d\vec{x}_3^2)+\frac{L^2}{r^2} dr^2 \ .   
\end{equation}
Comparing with the expressions in \eqref{eq:Sahakian_c_function_simplified} and \eqref{eq:Sahakian_equals_covariant}, we find
\begin{equation}
d=4,~~{\overbar{\cal N}}= \frac{r}{L},~~\overbar{B}=\frac{L^2}{r^2},~~\overbar{g}=\frac{r^6}{L^6} \ .   
\end{equation}
Using this, the Sahakian c-function reads
\begin{equation}
 {\cal C}_S=\frac{L^3}{8G_{\text{N}}^{(5)}}\ .   
\end{equation}
This should be equal to the result in \eqref{cnewAdS5xX5}, for the case $\epsilon=0$. This implies,
\begin{equation}
 \frac{L^3}{G_{\text{N}}^{(5)}}= \frac{\vol{(\mathcal{X}_5)}\, L^8}{G_{\text{N}}^{(10)}} \,,
\end{equation}
which is consistent with \eqref{eq:Newton_constant_relation_constant}.

Let us now discuss the case of generic CFTs in Section \ref{subsec:genericCFT}. In this case we assume the existence of an AdS$_{d+1}$ supergravity, whose lift to Type II and M-theory gives the backgrounds in \eqref{eq:metricaAdSgeneric}. The lower dimensional metric and other relevant quantities are,
\begin{eqnarray}
& &      ds_{d+1}^2= \frac{r^2}{L^2}(-dt^2+d\vec{x}_{d-1}^2)+\frac{L^2}{r^2} dr^2\nonumber\\
& & \overbar{\cal N}=\frac{r}{L}, ~~\sqrt{\overbar{B}}=\frac{L}{r}, ~~\overbar{g}= \left( \frac{r}{L}\right)^{2d-2} \ .
\end{eqnarray}
This gives a Sahakian c-function,
\begin{equation}
 {\cal C}_S= \frac{L^{d-1}}{2^{d-1} G_{\text{N}}^{(d+1)}}.   
\end{equation}
We compare this with the result in eq.(\ref{CFT-ccov}) and find,
\begin{equation}
 \frac{L^{d-1}}{G_{\text{N}}^{(d+1)}}= 
 \frac{\widetilde{\mathcal{N}}}{G_{\text{N}}^{(10)}} \,,\quad
 \widetilde{\mathcal{N}} \equiv  \int_{\mathcal{X}_p} d^{p}y \sqrt{h}\,e^{-\frac{(d-1)\Phi}{4}} f_1(y)^{\frac{d-1}{2}}\,.
\end{equation}
Notice that in both the examples above, the quantity
\begin{equation}
    \frac{\overbar{\cal N}^{d-1}}{\sqrt{\overbar{g}}}=1\,.
  \label{guardado}
  \end{equation}
Let us now study a situation for which the equality in \eqref{guardado} is not satisfied. In fact, this is what occurs in the very-demanding-example of Appendix \ref{subapp:very_demanding_example}. In this case, the lower (five) dimensional supergravity background is given in \eqref{betoalonso}. For ease of reference, we repeat it here
\begin{equation*}
  ds_5^2=\frac{r^2\lambda(r)^2}{L^2} \left(-dt^2+ dx_1^2+ dx_2^2+ L^2 F(r)\, d\phi^2 \right) +\frac{dr^2}{r^2\lambda(r)^{{4}} F(r)}\,.   
\end{equation*}
For this metric, the relevant quantities to calculate Sahakian's c-function are,
\begin{eqnarray}
& & d=4, ~~\overbar{\cal N}=\frac{r \lambda(r)}{L},~~\overbar{B}=\frac{1}{r^2\lambda^4(r) F(r)},~~\overbar{g}=\frac{r^6 \lambda^6(r) F(r)}{L^4}.
\end{eqnarray}
Note that in this case $\frac{\overbar{\cal N}^3}{\sqrt{\overbar{g}}}= \frac{1}{L\sqrt{F(r)}}$. We follow eqs.(\ref{eq:Sahakian_c_function_simplified})-(\ref{eq:Sahakian_equals_covariant}), and calculate
\begin{equation}
 {\cal C}_S(r)=\frac{1}{8G_{\text{N}}^{(5)}}\times \frac{1}{L\sqrt{F(r)}}\times \left[
\frac{F(r)^{-1/2}}{\lambda(r)^2r\,\partial_r\log\!\left(r\lambda(r)F(r)^{1/6}\right)}
\right]^3.  
\end{equation}
Imposing this to be equal to the result in \eqref{eqD11}, up to the relation in \eqref{eq:Sahakian_equals_covariant}, we find
\begin{equation}
    \frac{1}{G_{\text{N}}^{(5)}}=\frac{\widetilde{\cal N}}{G_{\text{N}}^{(10)}}.
\end{equation}
This is the expected lower-dimensional Newton's constant, consistent with the results above.

\bibliographystyle{JHEP}
\bibliography{refs}

@article{Cvetic:2007nv,
    author = "Cvetic, Mirjam and Vazquez-Poritz, Justin F.",
    title = "{Warped resolved L(a,b,c) cones}",
    eprint = "0705.3847",
    archivePrefix = "arXiv",
    primaryClass = "hep-th",
    reportNumber = "UPR-1181-T, MIFP-07-03",
    doi = "10.1103/PhysRevD.77.126003",
    journal = "Phys. Rev. D",
    volume = "77",
    pages = "126003",
    year = "2008"
}

@article{Kanitscheider:2009as,
    author = "Kanitscheider, Ingmar and Skenderis, Kostas",
    title = "{Universal hydrodynamics of non-conformal branes}",
    eprint = "0901.1487",
    archivePrefix = "arXiv",
    primaryClass = "hep-th",
    reportNumber = "ITFA-2009-01, NSF-KITP-2009-02",
    doi = "10.1088/1126-6708/2009/04/062",
    journal = "JHEP",
    volume = "04",
    pages = "062",
    year = "2009"
}

@article{Freund:1980xh,
    author = "Freund, Peter G. O. and Rubin, Mark A.",
    editor = "Salam, A. and Sezgin, E.",
    title = "{Dynamics of Dimensional Reduction}",
    reportNumber = "EFI 80/35-CHICAGO",
    doi = "10.1016/0370-2693(80)90590-0",
    journal = "Phys. Lett. B",
    volume = "97",
    pages = "233--235",
    year = "1980"
}

@article{Bousso:1999xy,
    author = "Bousso, Raphael",
    title = "{A Covariant entropy conjecture}",
    eprint = "hep-th/9905177",
    archivePrefix = "arXiv",
    reportNumber = "SU-ITP-99-23",
    doi = "10.1088/1126-6708/1999/07/004",
    journal = "JHEP",
    volume = "07",
    pages = "004",
    year = "1999"
}

@article{Alvarez:1998wr,
    author = "Alvarez, Enrique and Gomez, Cesar",
    title = "{Geometric holography, the renormalization group and the c theorem}",
    eprint = "hep-th/9807226",
    archivePrefix = "arXiv",
    reportNumber = "IFT-UAM-CSIC-98-20",
    doi = "10.1016/S0550-3213(98)00752-4",
    journal = "Nucl. Phys. B",
    volume = "541",
    pages = "441--460",
    year = "1999"
}

@article{Jain:2025xko,
    author = {Jain, Parul and J{\"a}rvinen, Matti},
    title = "{An analytic approach for holographic entanglement entropy at (quantum) criticality}",
    eprint = "2509.25360",
    archivePrefix = "arXiv",
    primaryClass = "hep-th",
    doi = "10.1007/JHEP03(2026)160",
    journal = "JHEP",
    volume = "03",
    pages = "160",
    year = "2026"
}

@article{Anabalon:2026yxk,
    author = "Anabal{\'o}n, Andr{\'e}s and Nastase, Horatiu and Nunez, Carlos and Oyarzo, Marcelo and Stuardo, Ricardo",
    title = "{Moduli space of ${\cal N}=4$ Super Yang-Mills from AdS/CFT}",
    eprint = "2603.18141",
    archivePrefix = "arXiv",
    primaryClass = "hep-th",
    month = "3",
    year = "2026"
}

@article{Giataganas:2021jbj,
    author = "Giataganas, Dimitrios and Pappas, Nikolaos and Toumbas, Nicolaos",
    title = "{Holographic observables at large d}",
    eprint = "2110.14606",
    archivePrefix = "arXiv",
    primaryClass = "hep-th",
    doi = "10.1103/PhysRevD.105.026016",
    journal = "Phys. Rev. D",
    volume = "105",
    number = "2",
    pages = "026016",
    year = "2022"
}

@article{Donos:2017ljs,
    author = "Donos, Aristomenis and Gauntlett, Jerome P. and Rosen, Christopher and Sosa-Rodriguez, Omar",
    title = "{Boomerang RG flows in M-theory with intermediate scaling}",
    eprint = "1705.03000",
    archivePrefix = "arXiv",
    primaryClass = "hep-th",
    reportNumber = "IMPERIAL-TP-2017-JG-03, DCPT-17-09",
    doi = "10.1007/JHEP07(2017)128",
    journal = "JHEP",
    volume = "07",
    pages = "128",
    year = "2017"
}

@book{Hawking:1973uf,
    author = "Hawking, Stephen W. and Ellis, George F. R.",
    title = "{The Large Scale Structure of Space-Time}",
    doi = "10.1017/9781009253161",
    isbn = "978-1-009-25316-1, 978-1-009-25315-4, 978-0-521-20016-5, 978-0-521-09906-6, 978-0-511-82630-6, 978-0-521-09906-6",
    publisher = "Cambridge University Press",
    series = "Cambridge Monographs on Mathematical Physics",
    month = "2",
    year = "2023"
}

@article{Kontou:2020bta,
    author = "Kontou, Eleni-Alexandra and Sanders, Ko",
    title = "{Energy conditions in general relativity and quantum field theory}",
    eprint = "2003.01815",
    archivePrefix = "arXiv",
    primaryClass = "gr-qc",
    doi = "10.1088/1361-6382/ab8fcf",
    journal = "Class. Quant. Grav.",
    volume = "37",
    number = "19",
    pages = "193001",
    year = "2020"
}

@article{Bulgarelli:2024onj,
    author = "Bulgarelli, Andrea and Panero, Marco",
    title = "{Duality transformations and the entanglement entropy of gauge theories}",
    eprint = "2404.01987",
    archivePrefix = "arXiv",
    primaryClass = "quant-ph",
    doi = "10.1007/JHEP06(2024)041",
    journal = "JHEP",
    volume = "06",
    pages = "041",
    year = "2024"
}

@article{Buividovich:2008kq,
    author = "Buividovich, P. V. and Polikarpov, M. I.",
    title = "{Numerical study of entanglement entropy in SU(2) lattice gauge theory}",
    eprint = "0802.4247",
    archivePrefix = "arXiv",
    primaryClass = "hep-lat",
    reportNumber = "ITEP-LAT-2008-07",
    doi = "10.1016/j.nuclphysb.2008.04.024",
    journal = "Nucl. Phys. B",
    volume = "802",
    pages = "458--474",
    year = "2008"
}

@article{Dong:2013qoa,
    author = "Dong, Xi",
    title = "{Holographic Entanglement Entropy for General Higher Derivative Gravity}",
    eprint = "1310.5713",
    archivePrefix = "arXiv",
    primaryClass = "hep-th",
    reportNumber = "SU-ITP-13-21",
    doi = "10.1007/JHEP01(2014)044",
    journal = "JHEP",
    volume = "01",
    pages = "044",
    year = "2014"
}

@article{Iyer:1994ys,
    author = "Iyer, Vivek and Wald, Robert M.",
    title = "{Some properties of Noether charge and a proposal for dynamical black hole entropy}",
    eprint = "gr-qc/9403028",
    archivePrefix = "arXiv",
    doi = "10.1103/PhysRevD.50.846",
    journal = "Phys. Rev. D",
    volume = "50",
    pages = "846--864",
    year = "1994"
}

@article{PandoZayas:2000ctr,
    author = "Pando Zayas, Leopoldo A. and Tseytlin, Arkady A.",
    title = "{3-branes on resolved conifold}",
    eprint = "hep-th/0010088",
    archivePrefix = "arXiv",
    reportNumber = "UM-TH-00-24, OHSTPY-HEP-T-00-019",
    doi = "10.1088/1126-6708/2000/11/028",
    journal = "JHEP",
    volume = "11",
    pages = "028",
    year = "2000"
}

@article{Klebanov:2007us,
    author = "Klebanov, Igor R. and Murugan, Arvind",
    title = "{Gauge/Gravity Duality and Warped Resolved Conifold}",
    eprint = "hep-th/0701064",
    archivePrefix = "arXiv",
    reportNumber = "PUPT-2221",
    doi = "10.1088/1126-6708/2007/03/042",
    journal = "JHEP",
    volume = "03",
    pages = "042",
    year = "2007"
}

@article{Sahakian:1999bd,
    author = "Sahakian, Vatche",
    title = "{Holography, a covariant c function, and the geometry of the renormalization group}",
    eprint = "hep-th/9910099",
    archivePrefix = "arXiv",
    reportNumber = "CLNS-99-1639",
    doi = "10.1103/PhysRevD.62.126011",
    journal = "Phys. Rev. D",
    volume = "62",
    pages = "126011",
    year = "2000"
}

@article{Donos:2017sba,
    author = "Donos, Aristomenis and Gauntlett, Jerome P. and Rosen, Christopher and Sosa-Rodriguez, Omar",
    title = "{Boomerang RG flows with intermediate conformal invariance}",
    eprint = "1712.08017",
    archivePrefix = "arXiv",
    primaryClass = "hep-th",
    reportNumber = "IMPERIAL-TP-2017-JG-05, DCPT-17-41, Imperial/TP/2017/JG/05; DCPT-17/41",
    doi = "10.1007/JHEP04(2018)017",
    journal = "JHEP",
    volume = "04",
    pages = "017",
    year = "2018"
}

@article{Nakagawa:2009jk,
    author = "Nakagawa, Y. and Nakamura, A. and Motoki, S. and Zakharov, V. I.",
    editor = "Liu, Chuan and Zhu, Yu",
    title = "{Entanglement entropy of SU(3) Yang-Mills theory}",
    eprint = "0911.2596",
    archivePrefix = "arXiv",
    primaryClass = "hep-lat",
    doi = "10.22323/1.091.0188",
    journal = "PoS",
    volume = "LAT2009",
    pages = "188",
    year = "2009"
}

@article{Zamolodchikov:1986gt,
    author = "Zamolodchikov, A. B.",
    title = "{Irreversibility of the Flux of the Renormalization Group in a 2D Field Theory}",
    journal = "JETP Lett.",
    volume = "43",
    pages = "730--732",
    year = "1986"
}

@article{Rabenstein:2018bri,
    author = {Rabenstein, Andreas and Bodendorfer, Norbert and Buividovich, Pavel and Sch\"afer, Andreas},
    title = "{Lattice study of R\'enyi entanglement entropy in $SU(N_c)$ lattice Yang-Mills theory with $N_c = 2, 3, 4$}",
    eprint = "1812.04279",
    archivePrefix = "arXiv",
    primaryClass = "hep-lat",
    doi = "10.1103/PhysRevD.100.034504",
    journal = "Phys. Rev. D",
    volume = "100",
    number = "3",
    pages = "034504",
    year = "2019"
}

@article{Rindlisbacher:2022bhe,
    author = {Rindlisbacher, Tobias and Jokela, Niko and P\"onni, Arttu and Rummukainen, Kari and Salami, Ahmed},
    title = "{Improved lattice method for determining entanglement measures in SU(N) gauge theories}",
    eprint = "2211.00425",
    archivePrefix = "arXiv",
    primaryClass = "hep-lat",
    reportNumber = "HIP-2022-26/TH",
    doi = "10.22323/1.430.0031",
    journal = "PoS",
    volume = "LATTICE2022",
    pages = "031",
    year = "2022"
}

@article{GonzalezLezcano:2022mcd,
    author = "Gonz\'alez Lezcano, Alfredo and Hong, Junho and Liu, James T. and Pando Zayas, Leopoldo A. and Uhlemann, Christoph F.",
    title = "{c-functions in flows across dimensions}",
    eprint = "2207.09360",
    archivePrefix = "arXiv",
    primaryClass = "hep-th",
    reportNumber = "LCTP-22-03",
    doi = "10.1007/JHEP10(2022)083",
    journal = "JHEP",
    volume = "10",
    pages = "083",
    year = "2022"
}

@article{Casini:2017vbe,
    author = "Casini, Horacio and Test\'e, Eduardo and Torroba, Gonzalo",
    title = "{Markov Property of the Conformal Field Theory Vacuum and the a Theorem}",
    eprint = "1704.01870",
    archivePrefix = "arXiv",
    primaryClass = "hep-th",
    doi = "10.1103/PhysRevLett.118.261602",
    journal = "Phys. Rev. Lett.",
    volume = "118",
    number = "26",
    pages = "261602",
    year = "2017"
}

@article{Liu:2012eea,
    author = "Liu, Hong and Mezei, Mark",
    title = "{A Refinement of entanglement entropy and the number of degrees of freedom}",
    eprint = "1202.2070",
    archivePrefix = "arXiv",
    primaryClass = "hep-th",
    reportNumber = "MIT-CTP-4336",
    doi = "10.1007/JHEP04(2013)162",
    journal = "JHEP",
    volume = "04",
    pages = "162",
    year = "2013"
}

@article{Freedman:1999gp,
    author = "Freedman, D. Z. and Gubser, S. S. and Pilch, K. and Warner, N. P.",
    title = "{Renormalization group flows from holography supersymmetry and a c theorem}",
    eprint = "hep-th/9904017",
    archivePrefix = "arXiv",
    reportNumber = "CERN-TH-99-86, HUTP-99-A015, USC-99-1, MIT-CTP-2846",
    doi = "10.4310/ATMP.1999.v3.n2.a7",
    journal = "Adv. Theor. Math. Phys.",
    volume = "3",
    pages = "363--417",
    year = "1999"
}

@article{Caceres:2022hei,
    author = "Caceres, Elena and Shashi, Sanjit",
    title = "{Anisotropic flows into black holes}",
    eprint = "2209.06818",
    archivePrefix = "arXiv",
    primaryClass = "hep-th",
    reportNumber = "UTWI-01-2022",
    doi = "10.1007/JHEP01(2023)007",
    journal = "JHEP",
    volume = "01",
    pages = "007",
    year = "2023"
}

@article{de-la-Cruz-Moreno:2023mew,
    author = "de-la-Cruz-Moreno, Jos{\'e} and Liu, James T. and Pando Zayas, Leopoldo A.",
    title = "{Discontinuity in RG flows across dimensions: entanglement, anomaly coefficients and geometry}",
    eprint = "2312.12382",
    archivePrefix = "arXiv",
    primaryClass = "hep-th",
    reportNumber = "LCTP-23-18",
    doi = "10.1007/JHEP08(2024)158",
    journal = "JHEP",
    volume = "08",
    pages = "158",
    year = "2024"
}

@article{Hoyos:2020zeg,
    author = {Hoyos, Carlos and Jokela, Niko and Pen\'\i{}n, Jos\'e Manuel and Ramallo, Alfonso V.},
    title = "{Holographic spontaneous anisotropy}",
    eprint = "2001.08218",
    archivePrefix = "arXiv",
    primaryClass = "hep-th",
    reportNumber = "HIP-2019-39/TH",
    doi = "10.1007/JHEP04(2020)062",
    journal = "JHEP",
    volume = "04",
    pages = "062",
    year = "2020"
}

@article{Hoyos:2021vhl,
    author = {Hoyos, Carlos and Jokela, Niko and Pen\'\i{}n, Jos\'e Manuel and Ramallo, Alfonso V. and Tarr\'\i{}o, Javier},
    title = "{Risking your NEC}",
    eprint = "2104.11749",
    archivePrefix = "arXiv",
    primaryClass = "hep-th",
    reportNumber = "HIP-2021-16/TH",
    doi = "10.1007/JHEP10(2021)112",
    journal = "JHEP",
    volume = "10",
    pages = "112",
    year = "2021"
}

@article{Casini:2015woa,
    author = "Casini, Horacio and Huerta, Marina and Myers, Robert C. and Yale, Alexandre",
    title = "{Mutual information and the F-theorem}",
    eprint = "1506.06195",
    archivePrefix = "arXiv",
    primaryClass = "hep-th",
    doi = "10.1007/JHEP10(2015)003",
    journal = "JHEP",
    volume = "10",
    pages = "003",
    year = "2015"
}

@article{Lewkowycz:2013nqa,
    author = "Lewkowycz, Aitor and Maldacena, Juan",
    title = "{Generalized gravitational entropy}",
    eprint = "1304.4926",
    archivePrefix = "arXiv",
    primaryClass = "hep-th",
    doi = "10.1007/JHEP08(2013)090",
    journal = "JHEP",
    volume = "08",
    pages = "090",
    year = "2013"
}

@article{Myers:2010xs,
    author = "Myers, Robert C. and Sinha, Aninda",
    title = "{Seeing a c-theorem with holography}",
    eprint = "1006.1263",
    archivePrefix = "arXiv",
    primaryClass = "hep-th",
    doi = "10.1103/PhysRevD.82.046006",
    journal = "Phys. Rev. D",
    volume = "82",
    pages = "046006",
    year = "2010"
}

@article{Myers:2010tj,
    author = "Myers, Robert C. and Sinha, Aninda",
    title = "{Holographic c-theorems in arbitrary dimensions}",
    eprint = "1011.5819",
    archivePrefix = "arXiv",
    primaryClass = "hep-th",
    doi = "10.1007/JHEP01(2011)125",
    journal = "JHEP",
    volume = "01",
    pages = "125",
    year = "2011"
}

@article{Macpherson:2016xwk,
    author = "Macpherson, Niall T. and Tomasiello, Alessandro",
    title = "{Minimal flux Minkowski classification}",
    eprint = "1612.06885",
    archivePrefix = "arXiv",
    primaryClass = "hep-th",
    doi = "10.1007/JHEP09(2017)126",
    journal = "JHEP",
    volume = "09",
    pages = "126",
    year = "2017"
}

@article{Gaiotto:2014lca,
    author = "Gaiotto, Davide and Tomasiello, Alessandro",
    title = "{Holography for (1,0) theories in six dimensions}",
    eprint = "1404.0711",
    archivePrefix = "arXiv",
    primaryClass = "hep-th",
    doi = "10.1007/JHEP12(2014)003",
    journal = "JHEP",
    volume = "12",
    pages = "003",
    year = "2014"
}

@article{Cremonesi:2015bld,
    author = "Cremonesi, Stefano and Tomasiello, Alessandro",
    title = "{6d holographic anomaly match as a continuum limit}",
    eprint = "1512.02225",
    archivePrefix = "arXiv",
    primaryClass = "hep-th",
    doi = "10.1007/JHEP05(2016)031",
    journal = "JHEP",
    volume = "05",
    pages = "031",
    year = "2016"
}

@article{Apruzzi:2014qva,
    author = "Apruzzi, Fabio and Fazzi, Marco and Passias, Achilleas and Rosa, Dario and Tomasiello, Alessandro",
    title = "{AdS$_{6}$ solutions of type II supergravity}",
    eprint = "1406.0852",
    archivePrefix = "arXiv",
    primaryClass = "hep-th",
    doi = "10.1007/JHEP11(2014)099",
    journal = "JHEP",
    volume = "11",
    pages = "099",
    year = "2014",
    note = "[Erratum: JHEP 05, 012 (2015)]"
}

@article{Liu:2013una,
    author = "Liu, Hong and Mezei, M\'ark",
    title = "{Probing renormalization group flows using entanglement entropy}",
    eprint = "1309.6935",
    archivePrefix = "arXiv",
    primaryClass = "hep-th",
    reportNumber = "MIT-CTP-4500",
    doi = "10.1007/JHEP01(2014)098",
    journal = "JHEP",
    volume = "01",
    pages = "098",
    year = "2014"
}

@article{Nunez:2018ags,
    author = "N\'u\~nez, Carlos and Pen\'\i{}n, Jos\'e Manuel and Roychowdhury, Dibakar and Van Gorsel, Jeroen",
    title = "{The non-Integrability of Strings in Massive Type IIA and their Holographic duals}",
    eprint = "1802.04269",
    archivePrefix = "arXiv",
    primaryClass = "hep-th",
    doi = "10.1007/JHEP06(2018)078",
    journal = "JHEP",
    volume = "06",
    pages = "078",
    year = "2018"
}

@article{Macpherson:2014eza,
    author = "Macpherson, Niall T. and N\'u\~nez, Carlos and Pando Zayas, Leopoldo A. and Rodgers, Vincent G. J. and Whiting, Catherine A.",
    title = "{Type IIB supergravity solutions with AdS$_{5}$ from Abelian and non-Abelian T dualities}",
    eprint = "1410.2650",
    archivePrefix = "arXiv",
    primaryClass = "hep-th",
    reportNumber = "MCTP-14-28",
    doi = "10.1007/JHEP02(2015)040",
    journal = "JHEP",
    volume = "02",
    pages = "040",
    year = "2015"
}

@article{Merrikin:2022yho,
    author = "Merrikin, Paul and N\'u\~nez, Carlos and Stuardo, Ricardo",
    title = "{Compactification of 6d N=(1,0) quivers, 4d SCFTs and their holographic dual Massive IIA backgrounds}",
    eprint = "2210.02458",
    archivePrefix = "arXiv",
    primaryClass = "hep-th",
    doi = "10.1016/j.nuclphysb.2023.116356",
    journal = "Nucl. Phys. B",
    volume = "996",
    pages = "116356",
    year = "2023"
}

@article{Fatemiabhari:2024aua,
    author = "Fatemiabhari, Ali and Nunez, Carlos",
    title = "{From conformal to confining field theories using holography}",
    eprint = "2401.04158",
    archivePrefix = "arXiv",
    primaryClass = "hep-th",
    doi = "10.1007/JHEP03(2024)160",
    journal = "JHEP",
    volume = "03",
    pages = "160",
    year = "2024"
}

@article{Passias:2015gya,
    author = "Passias, Achilleas and Rota, Andrea and Tomasiello, Alessandro",
    title = "{Universal consistent truncation for 6d/7d gauge/gravity duals}",
    eprint = "1506.05462",
    archivePrefix = "arXiv",
    primaryClass = "hep-th",
    doi = "10.1007/JHEP10(2015)187",
    journal = "JHEP",
    volume = "10",
    pages = "187",
    year = "2015"
}

@article{Komargodski:2011vj,
    author = "Komargodski, Zohar and Schwimmer, Adam",
    title = "{On Renormalization Group Flows in Four Dimensions}",
    eprint = "1107.3987",
    archivePrefix = "arXiv",
    primaryClass = "hep-th",
    doi = "10.1007/JHEP12(2011)099",
    journal = "JHEP",
    volume = "12",
    pages = "099",
    year = "2011"
}

@article{Nunez:2018qcj,
    author = "N\'u\~nez, Carlos and Roychowdhury, Dibakar and Thompson, Daniel C.",
    title = "{Integrability and non-integrability in $ \mathcal{N}=2 $ SCFTs and their holographic backgrounds}",
    eprint = "1804.08621",
    archivePrefix = "arXiv",
    primaryClass = "hep-th",
    doi = "10.1007/JHEP07(2018)044",
    journal = "JHEP",
    volume = "07",
    pages = "044",
    year = "2018"
}

@article{Lozano:2016kum,
    author = "Lozano, Yolanda and N\'u\~nez, Carlos",
    title = "{Field theory aspects of non-Abelian T-duality and $ \mathcal{N}  =$ 2 linear quivers}",
    eprint = "1603.04440",
    archivePrefix = "arXiv",
    primaryClass = "hep-th",
    doi = "10.1007/JHEP05(2016)107",
    journal = "JHEP",
    volume = "05",
    pages = "107",
    year = "2016"
}

@article{Nunez:2023loo,
    author = "Nunez, Carlos and Santilli, Leonardo and Zarembo, Konstantin",
    title = "{Linear Quivers at Large-N}",
    eprint = "2311.00024",
    archivePrefix = "arXiv",
    primaryClass = "hep-th",
    doi = "10.1007/s00220-024-05186-1",
    journal = "Commun. Math. Phys.",
    volume = "406",
    number = "1",
    pages = "6",
    year = "2025"
}

@article{Bea:2015fja,
    author = "Bea, Yago and Edelstein, Jose D. and Itsios, Georgios and Kooner, Karta S. and N\'u\~nez, Carlos and Schofield, Daniel and Sierra-Garcia, J. Anibal",
    title = "{Compactifications of the Klebanov-Witten CFT and new AdS$_{3}$ backgrounds}",
    eprint = "1503.07527",
    archivePrefix = "arXiv",
    primaryClass = "hep-th",
    doi = "10.1007/JHEP05(2015)062",
    journal = "JHEP",
    volume = "05",
    pages = "062",
    year = "2015"
}

@article{Giliberti:2024eii,
    author = "Giliberti, Mauro and Fatemiabhari, Ali and N\'u\~nez, Carlos",
    title = "{Confinement and screening via holographic Wilson loops}",
    eprint = "2409.04539",
    archivePrefix = "arXiv",
    primaryClass = "hep-th",
    doi = "10.1007/JHEP11(2024)068",
    journal = "JHEP",
    volume = "11",
    pages = "068",
    year = "2024"
}

@article{Lozano:2020txg,
    author = "Lozano, Yolanda and N\'u\~nez, Carlos and Ramirez, Anayeli and Speziali, Stefano",
    title = "{New AdS$_{2}$ backgrounds and $ \mathcal{N} $ = 4 conformal quantum mechanics}",
    eprint = "2011.00005",
    archivePrefix = "arXiv",
    primaryClass = "hep-th",
    doi = "10.1007/JHEP03(2021)277",
    journal = "JHEP",
    volume = "03",
    pages = "277",
    year = "2021"
}

@article{Lozano:2019zvg,
    author = "Lozano, Yolanda and Macpherson, Niall T. and N\'u\~nez, Carlos and Ramirez, Anayeli",
    title = "{Two dimensional ${\cal N}=(0,4)$ quivers dual to AdS$_3$ solutions in massive IIA}",
    eprint = "1909.10510",
    archivePrefix = "arXiv",
    primaryClass = "hep-th",
    doi = "10.1007/JHEP01(2020)140",
    journal = "JHEP",
    volume = "01",
    pages = "140",
    year = "2020"
}

@article{Akhond:2021ffz,
    author = "Akhond, Mohammad and Legramandi, Andrea and N\'u\~nez, Carlos",
    title = "{Electrostatic description of 3d $ \mathcal{N} $ = 4 linear quivers}",
    eprint = "2109.06193",
    archivePrefix = "arXiv",
    primaryClass = "hep-th",
    doi = "10.1007/JHEP11(2021)205",
    journal = "JHEP",
    volume = "11",
    pages = "205",
    year = "2021"
}

@article{Nunez:2019gbg,
    author = "N\'u\~nez, Carlos and Roychowdhury, Dibakar and Speziali, Stefano and Zacar\'\i{}as, Salom\'on",
    title = "{Holographic aspects of four dimensional ${\cal N }=2$ SCFTs and their marginal deformations}",
    eprint = "1901.02888",
    archivePrefix = "arXiv",
    primaryClass = "hep-th",
    doi = "10.1016/j.nuclphysb.2019.114617",
    journal = "Nucl. Phys. B",
    volume = "943",
    pages = "114617",
    year = "2019"
}

@article{Filippas:2019puw,
    author = "Filippas, Kostas and N\'u\~nez, Carlos and Van Gorsel, Jeroen",
    title = "{Integrability and holographic aspects of six-dimensional $ \mathcal{N}=\left(1,\ 0\right) $ superconformal field theories}",
    eprint = "1901.08598",
    archivePrefix = "arXiv",
    primaryClass = "hep-th",
    doi = "10.1007/JHEP06(2019)069",
    journal = "JHEP",
    volume = "06",
    pages = "069",
    year = "2019"
}

@article{Lozano:2020bxo,
    author = "Lozano, Yolanda and N\'u\~nez, Carlos and Ramirez, Anayeli and Speziali, Stefano",
    title = "{$M$-strings and AdS$_3$ solutions to M-theory with small $\mathcal{N}=(0,4)$ supersymmetry}",
    eprint = "2005.06561",
    archivePrefix = "arXiv",
    primaryClass = "hep-th",
    doi = "10.1007/JHEP08(2020)118",
    journal = "JHEP",
    volume = "08",
    pages = "118",
    year = "2020"
}

@article{Legramandi:2021uds,
    author = "Legramandi, Andrea and N\'u\~nez, Carlos",
    title = "{Electrostatic description of five-dimensional SCFTs}",
    eprint = "2104.11240",
    archivePrefix = "arXiv",
    primaryClass = "hep-th",
    doi = "10.1016/j.nuclphysb.2021.115630",
    journal = "Nucl. Phys. B",
    volume = "974",
    pages = "115630",
    year = "2022"
}

@article{DHoker:2016ysh,
    author = "D'Hoker, Eric and Gutperle, Michael and Uhlemann, Christoph F.",
    title = "{Holographic duals for five-dimensional superconformal quantum field theories}",
    eprint = "1611.09411",
    archivePrefix = "arXiv",
    primaryClass = "hep-th",
    doi = "10.1103/PhysRevLett.118.101601",
    journal = "Phys. Rev. Lett.",
    volume = "118",
    number = "10",
    pages = "101601",
    year = "2017"
}

@article{Assel:2011xz,
    author = "Assel, Benjamin and Bachas, Costas and Estes, John and Gomis, Jaume",
    title = "{Holographic Duals of D=3 N=4 Superconformal Field Theories}",
    eprint = "1106.4253",
    archivePrefix = "arXiv",
    primaryClass = "hep-th",
    doi = "10.1007/JHEP08(2011)087",
    journal = "JHEP",
    volume = "08",
    pages = "087",
    year = "2011"
}

@article{Castellani:2024ial,
    author = "Castellani, Federico and N\'u\~nez, Carlos",
    title = "{Holography for confined and deformed theories: TsT-generated solutions in type IIB supergravity}",
    eprint = "2410.00094",
    archivePrefix = "arXiv",
    primaryClass = "hep-th",
    doi = "10.1007/JHEP12(2024)155",
    journal = "JHEP",
    volume = "12",
    pages = "155",
    year = "2024"
}

@article{Kumar:2024pcz,
    author = "Kumar, S. Prem and Stuardo, Ricardo",
    title = "{Twisted circle compactification of $ \mathcal{N} $ = 4 SYM and its holographic dual}",
    eprint = "2405.03739",
    archivePrefix = "arXiv",
    primaryClass = "hep-th",
    doi = "10.1007/JHEP08(2024)089",
    journal = "JHEP",
    volume = "08",
    pages = "089",
    year = "2024"
}

@article{Anabalon:2021tua,
    author = "Anabalon, Andres and Ross, Simon F.",
    title = "{Supersymmetric solitons and a degeneracy of solutions in AdS/CFT}",
    eprint = "2104.14572",
    archivePrefix = "arXiv",
    primaryClass = "hep-th",
    doi = "10.1007/JHEP07(2021)015",
    journal = "JHEP",
    volume = "07",
    pages = "015",
    year = "2021"
}

@article{Nunez:2023xgl,
    author = "N\'u\~nez, Carlos and Oyarzo, Marcelo and Stuardo, Ricardo",
    title = "{Confinement and D5-branes}",
    eprint = "2311.17998",
    archivePrefix = "arXiv",
    primaryClass = "hep-th",
    doi = "10.1007/JHEP03(2024)080",
    journal = "JHEP",
    volume = "03",
    pages = "080",
    year = "2024"
}

@article{Anabalon:2024qhf,
    author = "Anabal\'on, A. and Astefanesei, D. and Gallerati, A. and Oliva, J.",
    title = "{Supersymmetric smooth distributions of M2-branes as AdS solitons}",
    eprint = "2402.00880",
    archivePrefix = "arXiv",
    primaryClass = "hep-th",
    doi = "10.1007/JHEP05(2024)077",
    journal = "JHEP",
    volume = "05",
    pages = "077",
    year = "2024"
}

@article{Anabalon:2024che,
    author = "Anabal\'on, Andr\'es and Nastase, Horatiu and Oyarzo, Marcelo",
    title = "{Supersymmetric AdS solitons and the interconnection of different vacua of $ \mathcal{N} $ = 4 Super Yang-Mills}",
    eprint = "2402.18482",
    archivePrefix = "arXiv",
    primaryClass = "hep-th",
    doi = "10.1007/JHEP05(2024)217",
    journal = "JHEP",
    volume = "05",
    pages = "217",
    year = "2024"
}

@article{Chatzis:2024top,
    author = "Chatzis, Dimitrios and Fatemiabhari, Ali and N\'u\~nez, Carlos and Weck, Peter",
    title = "{Conformal to confining SQFTs from holography}",
    eprint = "2405.05563",
    archivePrefix = "arXiv",
    primaryClass = "hep-th",
    doi = "10.1007/JHEP08(2024)041",
    journal = "JHEP",
    volume = "08",
    pages = "041",
    year = "2024"
}

@article{Chatzis:2024kdu,
    author = "Chatzis, Dimitrios and Fatemiabhari, Ali and N\'u\~nez, Carlos and Weck, Peter",
    title = "{SCFT deformations via uplifted solitons}",
    eprint = "2406.01685",
    archivePrefix = "arXiv",
    primaryClass = "hep-th",
    doi = "10.1016/j.nuclphysb.2024.116659",
    journal = "Nucl. Phys. B",
    volume = "1006",
    pages = "116659",
    year = "2024"
}

@article{Macpherson:2024qfi,
    author = "Macpherson, Niall T. and Merrikin, Paul and Stuardo, Ricardo",
    title = "{Circle compactifications of Minkowski$_{D}$ solutions, flux vacua and solitonic branes}",
    eprint = "2412.15102",
    archivePrefix = "arXiv",
    primaryClass = "hep-th",
    doi = "10.1007/JHEP08(2025)143",
    journal = "JHEP",
    volume = "08",
    pages = "143",
    year = "2025"
}

@article{Nunez:2023nnl,
    author = "N\'u\~nez, Carlos and Oyarzo, Marcelo and Stuardo, Ricardo",
    title = "{Confinement in (1 + 1) dimensions: a holographic perspective from I-branes}",
    eprint = "2307.04783",
    archivePrefix = "arXiv",
    primaryClass = "hep-th",
    doi = "10.1007/JHEP09(2023)201",
    journal = "JHEP",
    volume = "09",
    pages = "201",
    year = "2023"
}

@article{Jokela:2020wgs,
    author = "Jokela, Niko and Subils, Javier G.",
    title = "{Is entanglement a probe of confinement?}",
    eprint = "2010.09392",
    archivePrefix = "arXiv",
    primaryClass = "hep-th",
    reportNumber = "HIP-2020-29, ICCUB-20-023",
    doi = "10.1007/JHEP02(2021)147",
    journal = "JHEP",
    volume = "02",
    pages = "147",
    year = "2021"
}

@article{Klebanov:2000hb,
    author = "Klebanov, Igor R. and Strassler, Matthew J.",
    title = "{Supergravity and a confining gauge theory: Duality cascades and chi SB resolution of naked singularities}",
    eprint = "hep-th/0007191",
    archivePrefix = "arXiv",
    reportNumber = "IASSNS-HEP-00-56, PUPT-1944",
    doi = "10.1088/1126-6708/2000/08/052",
    journal = "JHEP",
    volume = "08",
    pages = "052",
    year = "2000"
}

@article{Girardello:1999bd,
    author = "Girardello, L. and Petrini, M. and Porrati, M. and Zaffaroni, A.",
    title = "{The Supergravity dual of N=1 superYang-Mills theory}",
    eprint = "hep-th/9909047",
    archivePrefix = "arXiv",
    reportNumber = "BICOCCA-FT-99-27, NYU-TH-99-09-02",
    doi = "10.1016/S0550-3213(99)00764-6",
    journal = "Nucl. Phys. B",
    volume = "569",
    pages = "451--469",
    year = "2000"
}

@article{Ishihara:2012jg,
    author = "Ishihara, Masafumi and Lin, Feng-Li and Ning, Bo",
    title = "{Refined Holographic Entanglement Entropy for the AdS Solitons and AdS black Holes}",
    eprint = "1203.6153",
    archivePrefix = "arXiv",
    primaryClass = "hep-th",
    doi = "10.1016/j.nuclphysb.2013.04.003",
    journal = "Nucl. Phys. B",
    volume = "872",
    pages = "392--426",
    year = "2013"
}

@article{Gaiotto:2009gz,
    author = "Gaiotto, Davide and Maldacena, Juan",
    title = "{The Gravity duals of N=2 superconformal field theories}",
    eprint = "0904.4466",
    archivePrefix = "arXiv",
    primaryClass = "hep-th",
    doi = "10.1007/JHEP10(2012)189",
    journal = "JHEP",
    volume = "10",
    pages = "189",
    year = "2012"
}

@article{Aharony:2012tz,
    author = "Aharony, Ofer and Berdichevsky, Leon and Berkooz, Micha",
    title = "{4d N=2 superconformal linear quivers with type IIA duals}",
    eprint = "1206.5916",
    archivePrefix = "arXiv",
    primaryClass = "hep-th",
    reportNumber = "WIS-11-12-JUNE-DPPA",
    doi = "10.1007/JHEP08(2012)131",
    journal = "JHEP",
    volume = "08",
    pages = "131",
    year = "2012"
}

@article{Reid-Edwards:2010vpm,
    author = "Reid-Edwards, R. A. and Stefanski, jr., B.",
    title = "{On Type IIA geometries dual to N = 2 SCFTs}",
    eprint = "1011.0216",
    archivePrefix = "arXiv",
    primaryClass = "hep-th",
    doi = "10.1016/j.nuclphysb.2011.04.002",
    journal = "Nucl. Phys. B",
    volume = "849",
    pages = "549--572",
    year = "2011"
}

@article{Casini:2004bw,
    author = "Casini, H. and Huerta, M.",
    title = "{A Finite entanglement entropy and the c-theorem}",
    eprint = "hep-th/0405111",
    archivePrefix = "arXiv",
    doi = "10.1016/j.physletb.2004.08.072",
    journal = "Phys. Lett. B",
    volume = "600",
    pages = "142--150",
    year = "2004"
}

@article{Henningson:1998gx,
    author = "Henningson, M. and Skenderis, K.",
    title = "{The Holographic Weyl anomaly}",
    eprint = "hep-th/9806087",
    archivePrefix = "arXiv",
    reportNumber = "CERN-TH-98-188, KUL-TF-98-21",
    doi = "10.1088/1126-6708/1998/07/023",
    journal = "JHEP",
    volume = "07",
    pages = "023",
    year = "1998"
}

@article{Henningson:1998ey,
    author = "Henningson, Mans and Skenderis, Kostas",
    editor = "Lust, D. and Otto, H. J.",
    title = "{Holography and the Weyl anomaly}",
    eprint = "hep-th/9812032",
    archivePrefix = "arXiv",
    reportNumber = "GOTEBORG-ITP-98-14, SPIN-1998-08, SPIN-1998-8",
    doi = "10.1002/(SICI)1521-3978(20001)48:1/3<125::AID-PROP125>3.0.CO;2-B",
    journal = "Fortsch. Phys.",
    volume = "48",
    pages = "125--128",
    year = "2000"
}

@article{Gubser:2004qj,
    author = "Gubser, Steven S. and Herzog, Christopher P. and Klebanov, Igor R.",
    title = "{Symmetry breaking and axionic strings in the warped deformed conifold}",
    eprint = "hep-th/0405282",
    archivePrefix = "arXiv",
    reportNumber = "PUPT-2120, NSF-KITP-04-71",
    doi = "10.1088/1126-6708/2004/09/036",
    journal = "JHEP",
    volume = "09",
    pages = "036",
    year = "2004"
}

@article{Butti:2004pk,
    author = "Butti, Agostino and Grana, Mariana and Minasian, Ruben and Petrini, Michela and Zaffaroni, Alberto",
    title = "{The Baryonic branch of Klebanov-Strassler solution: A supersymmetric family of SU(3) structure backgrounds}",
    eprint = "hep-th/0412187",
    archivePrefix = "arXiv",
    reportNumber = "BICOCCA-FT-04-18, CPHT-RR-070-1204, LPTENS-04-52",
    doi = "10.1088/1126-6708/2005/03/069",
    journal = "JHEP",
    volume = "03",
    pages = "069",
    year = "2005"
}

@article{Maldacena:2009mw,
    author = "Maldacena, Juan and Martelli, Dario",
    title = "{The Unwarped, resolved, deformed conifold: Fivebranes and the baryonic branch of the Klebanov-Strassler theory}",
    eprint = "0906.0591",
    archivePrefix = "arXiv",
    primaryClass = "hep-th",
    doi = "10.1007/JHEP01(2010)104",
    journal = "JHEP",
    volume = "01",
    pages = "104",
    year = "2010"
}

@article{Gaillard:2010qg,
    author = "Gaillard, Jerome and Martelli, Dario and Nunez, Carlos and Papadimitriou, Ioannis",
    title = "{The warped, resolved, deformed conifold gets flavoured}",
    eprint = "1004.4638",
    archivePrefix = "arXiv",
    primaryClass = "hep-th",
    doi = "10.1016/j.nuclphysb.2010.09.011",
    journal = "Nucl. Phys. B",
    volume = "843",
    pages = "1--45",
    year = "2011"
}

@article{Caceres:2011zn,
    author = "Caceres, Elena and Nunez, Carlos and Pando-Zayas, Leopoldo A.",
    title = "{Heating up the Baryonic Branch with U-duality: A Unified picture of conifold black holes}",
    eprint = "1101.4123",
    archivePrefix = "arXiv",
    primaryClass = "hep-th",
    reportNumber = "MCTP-11-01, UTTG-01-11",
    doi = "10.1007/JHEP03(2011)054",
    journal = "JHEP",
    volume = "03",
    pages = "054",
    year = "2011"
}

@article{Conde:2011aa,
    author = "Conde, Eduardo and Gaillard, Jerome and Nunez, Carlos and Piai, Maurizio and Ramallo, Alfonso V.",
    title = "{A Tale of Two Cascades: Higgsing and Seiberg-Duality Cascades from type IIB String Theory}",
    eprint = "1112.3350",
    archivePrefix = "arXiv",
    primaryClass = "hep-th",
    doi = "10.1007/JHEP02(2012)145",
    journal = "JHEP",
    volume = "02",
    pages = "145",
    year = "2012"
}

@article{Elander:2011mh,
    author = "Elander, Daniel and Gaillard, Jerome and Nunez, Carlos and Piai, Maurizio",
    title = "{Towards multi-scale dynamics on the baryonic branch of Klebanov-Strassler}",
    eprint = "1104.3963",
    archivePrefix = "arXiv",
    primaryClass = "hep-th",
    doi = "10.1007/JHEP07(2011)056",
    journal = "JHEP",
    volume = "07",
    pages = "056",
    year = "2011"
}

@article{Jokela:2025cyz,
    author = "Jokela, Niko and Kastikainen, Jani and Nunez, Carlos and Pen{\'\i}n, Jos{\'e} Manuel and Ruotsalainen, Helime and Subils, Javier G.",
    title = "{On entanglement c-functions in confining gauge field theories}",
    eprint = "2505.14397",
    archivePrefix = "arXiv",
    primaryClass = "hep-th",
    reportNumber = "HIP-2025-16/TH",
    doi = "10.1007/JHEP11(2025)101",
    journal = "JHEP",
    volume = "11",
    pages = "101",
    year = "2025"
}

@article{Chatzis:2025dnu,
    author = "Chatzis, Dimitrios and Hammond, Madison and Itsios, Georgios and Nunez, Carlos and Zoakos, Dimitrios",
    title = "{Universal observables, SUSY RG-flows and holography}",
    eprint = "2506.10062",
    archivePrefix = "arXiv",
    primaryClass = "hep-th",
    reportNumber = "HU-EP-25/19",
    doi = "10.1007/JHEP08(2025)134",
    journal = "JHEP",
    volume = "08",
    pages = "134",
    year = "2025"
}

@article{Gubser:1998vd,
    author = "Gubser, Steven S.",
    title = "{Einstein manifolds and conformal field theories}",
    eprint = "hep-th/9807164",
    archivePrefix = "arXiv",
    reportNumber = "PUPT-1805",
    doi = "10.1103/PhysRevD.59.025006",
    journal = "Phys. Rev. D",
    volume = "59",
    pages = "025006",
    year = "1999"
}

@article{Lozano:2020sae,
    author = "Lozano, Yolanda and Nunez, Carlos and Ramirez, Anayeli and Speziali, Stefano",
    title = "{AdS$_{2}$ duals to ADHM quivers with Wilson lines}",
    eprint = "2011.13932",
    archivePrefix = "arXiv",
    primaryClass = "hep-th",
    doi = "10.1007/JHEP03(2021)145",
    journal = "JHEP",
    volume = "03",
    pages = "145",
    year = "2021"
}

@article{Lozano:2021rmk,
    author = "Lozano, Yolanda and Nunez, Carlos and Ramirez, Anayeli",
    title = "{$\text{AdS}_2\times \text{S}^2\times \text{CY}_2$ solutions in Type IIB with 8 supersymmetries}",
    eprint = "2101.04682",
    archivePrefix = "arXiv",
    primaryClass = "hep-th",
    doi = "10.1007/JHEP04(2021)110",
    journal = "JHEP",
    volume = "04",
    pages = "110",
    year = "2021"
}

@article{Lozano:2019emq,
    author = "Lozano, Yolanda and Macpherson, Niall T. and Nunez, Carlos and Ramirez, Anayeli",
    title = "{AdS$_3$ solutions in Massive IIA with small $\mathcal{N}=(4,0)$ supersymmetry}",
    eprint = "1908.09851",
    archivePrefix = "arXiv",
    primaryClass = "hep-th",
    doi = "10.1007/JHEP01(2020)129",
    journal = "JHEP",
    volume = "01",
    pages = "129",
    year = "2020"
}

@article{Couzens:2021veb,
    author = "Couzens, Christopher and Lozano, Yolanda and Petri, Nicol{\`o} and Vandoren, Stefan",
    title = "{N=(0,4) black string chains}",
    eprint = "2109.10413",
    archivePrefix = "arXiv",
    primaryClass = "hep-th",
    doi = "10.1103/PhysRevD.105.086015",
    journal = "Phys. Rev. D",
    volume = "105",
    number = "8",
    pages = "086015",
    year = "2022"
}

@article{DHoker:2008lup,
    author = "D'Hoker, Eric and Estes, John and Gutperle, Michael and Krym, Darya",
    title = "{Exact Half-BPS Flux Solutions in M-theory. I: Local Solutions}",
    eprint = "0806.0605",
    archivePrefix = "arXiv",
    primaryClass = "hep-th",
    reportNumber = "UCLA-08-TEP-16",
    doi = "10.1088/1126-6708/2008/08/028",
    journal = "JHEP",
    volume = "08",
    pages = "028",
    year = "2008"
}

@article{DHoker:2016ujz,
    author = "D'Hoker, Eric and Gutperle, Michael and Karch, Andreas and Uhlemann, Christoph F.",
    title = "{Warped $AdS_6\times S^2$ in Type IIB supergravity I: Local solutions}",
    eprint = "1606.01254",
    archivePrefix = "arXiv",
    primaryClass = "hep-th",
    doi = "10.1007/JHEP08(2016)046",
    journal = "JHEP",
    volume = "08",
    pages = "046",
    year = "2016"
}

@article{DHoker:2017zwj,
    author = "D'Hoker, Eric and Gutperle, Michael and Uhlemann, Christoph F.",
    title = "{Warped $AdS_6\times S^2$ in Type IIB supergravity III: Global solutions with seven-branes}",
    eprint = "1706.00433",
    archivePrefix = "arXiv",
    primaryClass = "hep-th",
    doi = "10.1007/JHEP11(2017)200",
    journal = "JHEP",
    volume = "11",
    pages = "200",
    year = "2017"
}

@article{Legramandi:2021aqv,
    author = "Legramandi, Andrea and Nunez, Carlos",
    title = "{Holographic description of SCFT$_{5}$ compactifications}",
    eprint = "2109.11554",
    archivePrefix = "arXiv",
    primaryClass = "hep-th",
    doi = "10.1007/JHEP02(2022)010",
    journal = "JHEP",
    volume = "02",
    pages = "010",
    year = "2022"
}

@article{Akhond:2022oaf,
    author = "Akhond, Mohammad and Legramandi, Andrea and Nunez, Carlos and Santilli, Leonardo and Schepers, Lucas",
    title = "{Matrix models and holography: Mass deformations of long quiver theories in 5d and 3d}",
    eprint = "2211.13240",
    archivePrefix = "arXiv",
    primaryClass = "hep-th",
    doi = "10.21468/SciPostPhys.15.3.086",
    journal = "SciPost Phys.",
    volume = "15",
    number = "3",
    pages = "086",
    year = "2023"
}

@article{Cassani:2021fyv,
    author = "Cassani, Davide and Komargodski, Zohar",
    title = "{EFT and the SUSY Index on the 2nd Sheet}",
    eprint = "2104.01464",
    archivePrefix = "arXiv",
    primaryClass = "hep-th",
    doi = "10.21468/SciPostPhys.11.1.004",
    journal = "SciPost Phys.",
    volume = "11",
    pages = "004",
    year = "2021"
}

@article{Macpherson:2025pqi,
    author = "Macpherson, Niall T. and Merrikin, Paul and Nunez, Carlos and Stuardo, Ricardo",
    title = "{Twisted-circle compactifications of SQCD-like theories and holography}",
    eprint = "2506.15778",
    archivePrefix = "arXiv",
    primaryClass = "hep-th",
    doi = "10.1007/JHEP08(2025)146",
    journal = "JHEP",
    volume = "08",
    pages = "146",
    year = "2025"
}

@article{Maldacena:2000yy,
    author = "Maldacena, Juan Martin and Nunez, Carlos",
    title = "{Towards the large N limit of pure N=1 superYang-Mills}",
    eprint = "hep-th/0008001",
    archivePrefix = "arXiv",
    doi = "10.1103/PhysRevLett.86.588",
    journal = "Phys. Rev. Lett.",
    volume = "86",
    pages = "588--591",
    year = "2001"
}

@inproceedings{Strassler:2005qs,
    author = "Strassler, Matthew J.",
    title = "{The Duality cascade}",
    booktitle = "{Theoretical Advanced Study Institute in Elementary Particle Physics (TASI 2003): Recent Trends in String Theory}",
    eprint = "hep-th/0505153",
    archivePrefix = "arXiv",
    reportNumber = "UW-PT-05-13",
    doi = "10.1142/9789812775108_0005",
    pages = "419--510",
    month = "5",
    year = "2005"
}

@article{Hoyos-Badajoz:2008znk,
    author = "Hoyos-Badajoz, Carlos and Nunez, Carlos and Papadimitriou, Ioannis",
    title = "{Comments on the String dual to N=1 SQCD}",
    eprint = "0807.3039",
    archivePrefix = "arXiv",
    primaryClass = "hep-th",
    doi = "10.1103/PhysRevD.78.086005",
    journal = "Phys. Rev. D",
    volume = "78",
    pages = "086005",
    year = "2008"
}

@article{Nunez:2008wi,
    author = "Nunez, Carlos and Papadimitriou, Ioannis and Piai, Maurizio",
    title = "{Walking Dynamics from String Duals}",
    eprint = "0812.3655",
    archivePrefix = "arXiv",
    primaryClass = "hep-th",
    doi = "10.1142/S0217751X10049189",
    journal = "Int. J. Mod. Phys. A",
    volume = "25",
    pages = "2837--2865",
    year = "2010"
}

@article{Dymarsky:2005xt,
    author = "Dymarsky, Anatoly and Klebanov, Igor R. and Seiberg, Nathan",
    title = "{On the moduli space of the cascading SU(M+p) x SU(p) gauge theory}",
    eprint = "hep-th/0511254",
    archivePrefix = "arXiv",
    reportNumber = "PUPT-2183, ITEP-TH-62-05",
    doi = "10.1088/1126-6708/2006/01/155",
    journal = "JHEP",
    volume = "01",
    pages = "155",
    year = "2006"
}

@article{Klebanov:2000nc,
    author = "Klebanov, Igor R. and Tseytlin, Arkady A.",
    title = "{Gravity duals of supersymmetric SU(N) x SU(N+M) gauge theories}",
    eprint = "hep-th/0002159",
    archivePrefix = "arXiv",
    reportNumber = "PUPT-1919, OHSTPY-HEP-T-00-002",
    doi = "10.1016/S0550-3213(00)00206-6",
    journal = "Nucl. Phys. B",
    volume = "578",
    pages = "123--138",
    year = "2000"
}

@article{Gauntlett:2007ma,
    author = "Gauntlett, Jerome P. and Varela, Oscar",
    title = "{Consistent Kaluza-Klein reductions for general supersymmetric AdS solutions}",
    eprint = "0707.2315",
    archivePrefix = "arXiv",
    primaryClass = "hep-th",
    doi = "10.1103/PhysRevD.76.126007",
    journal = "Phys. Rev. D",
    volume = "76",
    pages = "126007",
    year = "2007"
}

@article{Cassani:2019vcl,
    author = "Cassani, Davide and Josse, Gr{\'e}goire and Petrini, Michela and Waldram, Daniel",
    title = "{Systematics of consistent truncations from generalised geometry}",
    eprint = "1907.06730",
    archivePrefix = "arXiv",
    primaryClass = "hep-th",
    doi = "10.1007/JHEP11(2019)017",
    journal = "JHEP",
    volume = "11",
    pages = "017",
    year = "2019"
}

@article{Gauntlett:2004hh,
    author = "Gauntlett, Jerome P. and Martelli, Dario and Sparks, James F. and Waldram, Daniel",
    title = "{A New infinite class of Sasaki-Einstein manifolds}",
    eprint = "hep-th/0403038",
    archivePrefix = "arXiv",
    reportNumber = "IMPERIAL-TP-03-04-9",
    doi = "10.4310/ATMP.2004.v8.n6.a3",
    journal = "Adv. Theor. Math. Phys.",
    volume = "8",
    number = "6",
    pages = "987--1000",
    year = "2004"
}

@article{Franco:2005sm,
    author = "Franco, Sebastian and Hanany, Amihay and Martelli, Dario and Sparks, James and Vegh, David and Wecht, Brian",
    title = "{Gauge theories from toric geometry and brane tilings}",
    eprint = "hep-th/0505211",
    archivePrefix = "arXiv",
    reportNumber = "MIT-CTP-3646, CERN-PH-TH-2005-084, HUTP-05-A0027",
    doi = "10.1088/1126-6708/2006/01/128",
    journal = "JHEP",
    volume = "01",
    pages = "128",
    year = "2006"
}

@article{Bertolini:2004xf,
    author = "Bertolini, M. and Bigazzi, F. and Cotrone, A. L.",
    title = "{New checks and subtleties for AdS/CFT and a-maximization}",
    eprint = "hep-th/0411249",
    archivePrefix = "arXiv",
    reportNumber = "SISSA-89-2004-EP, LPTHE-04-31, CPHT-RR063-1104",
    doi = "10.1088/1126-6708/2004/12/024",
    journal = "JHEP",
    volume = "12",
    pages = "024",
    year = "2004"
}

@article{Klebanov:1999tb,
    author = "Klebanov, Igor R. and Witten, Edward",
    title = "{AdS / CFT correspondence and symmetry breaking}",
    eprint = "hep-th/9905104",
    archivePrefix = "arXiv",
    reportNumber = "PUPT-1863, IASSNS-HEP-99-49",
    doi = "10.1016/S0550-3213(99)00387-9",
    journal = "Nucl. Phys. B",
    volume = "556",
    pages = "89--114",
    year = "1999"
}

@article{Maldacena:1997re,
    author = "Maldacena, Juan Martin",
    title = "{The Large $N$ limit of superconformal field theories and supergravity}",
    eprint = "hep-th/9711200",
    archivePrefix = "arXiv",
    reportNumber = "HUTP-97-A097, HUTP-98-A097",
    doi = "10.4310/ATMP.1998.v2.n2.a1",
    journal = "Adv. Theor. Math. Phys.",
    volume = "2",
    pages = "231--252",
    year = "1998"
}

@article{Itzhaki:1998dd,
    author = "Itzhaki, Nissan and Maldacena, Juan Martin and Sonnenschein, Jacob and Yankielowicz, Shimon",
    title = "{Supergravity and the large N limit of theories with sixteen supercharges}",
    eprint = "hep-th/9802042",
    archivePrefix = "arXiv",
    reportNumber = "TAUP-2474-98, HUTP-98-A003",
    doi = "10.1103/PhysRevD.58.046004",
    journal = "Phys. Rev. D",
    volume = "58",
    pages = "046004",
    year = "1998"
}

@article{Caceres:2023mqz,
    author = "C{\'a}ceres, Elena and Castillo V{\'a}squez, Rodrigo and Landsteiner, Karl and Salazar Landea, Ignacio",
    title = "{Holographic a-functions and Boomerang RG flows}",
    eprint = "2310.15983",
    archivePrefix = "arXiv",
    primaryClass = "hep-th",
    doi = "10.1007/JHEP02(2024)019",
    journal = "JHEP",
    volume = "02",
    pages = "019",
    year = "2024"
}

@article{deBoer:1999tgo,
    author = "de Boer, Jan and Verlinde, Erik P. and Verlinde, Herman L.",
    title = "{On the holographic renormalization group}",
    eprint = "hep-th/9912012",
    archivePrefix = "arXiv",
    reportNumber = "PUPT-1898, ITFA-99-39, SPIN-1999-29",
    doi = "10.1088/1126-6708/2000/08/003",
    journal = "JHEP",
    volume = "08",
    pages = "003",
    year = "2000"
}

@article{Camps:2013zua,
    author = "Camps, Joan",
    title = "{Generalized entropy and higher derivative Gravity}",
    eprint = "1310.6659",
    archivePrefix = "arXiv",
    primaryClass = "hep-th",
    doi = "10.1007/JHEP03(2014)070",
    journal = "JHEP",
    volume = "03",
    pages = "070",
    year = "2014"
}

@article{Giataganas:2017koz,
    author = {Giataganas, Dimitrios and G{\"u}rsoy, Umut and Pedraza, Juan F.},
    title = "{Strongly-coupled anisotropic gauge theories and holography}",
    eprint = "1708.05691",
    archivePrefix = "arXiv",
    primaryClass = "hep-th",
    reportNumber = "NCTS-TH/1712, NCTS-TH-1712",
    doi = "10.1103/PhysRevLett.121.121601",
    journal = "Phys. Rev. Lett.",
    volume = "121",
    number = "12",
    pages = "121601",
    year = "2018"
}

@article{Cremonini:2020rdx,
    author = "Cremonini, Sera and Li, Li and Ritchie, Kyle and Tang, Yuezhang",
    title = "{Constraining nonrelativistic RG flows with holography}",
    eprint = "2006.10780",
    archivePrefix = "arXiv",
    primaryClass = "hep-th",
    doi = "10.1103/PhysRevD.103.046006",
    journal = "Phys. Rev. D",
    volume = "103",
    number = "4",
    pages = "046006",
    year = "2021"
}

@article{Whittle:2025yog,
    author = "Whittle, Jonathan",
    title = "{Holographic entanglement entropy and c-functions in conformal and confining backgrounds}",
    eprint = "2511.20891",
    archivePrefix = "arXiv",
    primaryClass = "hep-th",
    month = "11",
    year = "2025"
}

@article{Donos:2014eua,
    author = "Donos, Aristomenis and Gauntlett, Jerome P.",
    title = "{Flowing from AdS$_{5}$ to AdS$_{3}$ with T$^{1,1}$}",
    eprint = "1404.7133",
    archivePrefix = "arXiv",
    primaryClass = "hep-th",
    reportNumber = "IMPERIAL-TP-2014-JG-02",
    doi = "10.1007/JHEP08(2014)006",
    journal = "JHEP",
    volume = "08",
    pages = "006",
    year = "2014"
}

@article{Klebanov:1998hh,
    author = "Klebanov, Igor R. and Witten, Edward",
    title = "{Superconformal field theory on three-branes at a Calabi-Yau singularity}",
    eprint = "hep-th/9807080",
    archivePrefix = "arXiv",
    reportNumber = "IASSNS-HEP-98-64, PUPT-1804",
    doi = "10.1016/S0550-3213(98)00654-3",
    journal = "Nucl. Phys. B",
    volume = "536",
    pages = "199--218",
    year = "1998"
}

@article{Gullu:2010st,
    author = "Gullu, Ibrahim and Sisman, Tahsin Cagri and Tekin, Bayram",
    title = "{c-functions in the Born-Infeld extended New Massive Gravity}",
    eprint = "1005.3214",
    archivePrefix = "arXiv",
    primaryClass = "hep-th",
    doi = "10.1103/PhysRevD.82.024032",
    journal = "Phys. Rev. D",
    volume = "82",
    pages = "024032",
    year = "2010"
}

@article{Cremades:2006ke,
    author = "Cremades, Daniel and Lozano-Tellechea, Ernesto",
    title = "{Holography, the second law and a {\textasciitilde}C-function in higher curvature gravity}",
    eprint = "hep-th/0608174",
    archivePrefix = "arXiv",
    reportNumber = "DAMTP-2006-66",
    doi = "10.1088/1126-6708/2007/01/045",
    journal = "JHEP",
    volume = "01",
    pages = "045",
    year = "2007"
}

@article{Banerjee:2015opa,
    author = "Banerjee, Shamik",
    title = "{RG Flow and Thermodynamics of Causal Horizons in AdS}",
    eprint = "1508.01343",
    archivePrefix = "arXiv",
    primaryClass = "hep-th",
    doi = "10.1007/JHEP10(2015)098",
    journal = "JHEP",
    volume = "10",
    pages = "098",
    year = "2015"
}

@article{Banerjee:2015uaa,
    author = "Banerjee, Shamik and Bhattacharyya, Arpan",
    title = "{RG Flow and Thermodynamics of Causal Horizons in Higher-Derivative AdS Gravity}",
    eprint = "1509.08475",
    archivePrefix = "arXiv",
    primaryClass = "hep-th",
    doi = "10.1007/JHEP05(2016)126",
    journal = "JHEP",
    volume = "05",
    pages = "126",
    year = "2016"
}

@article{Wald:1993nt,
    author = "Wald, Robert M.",
    title = "{Black hole entropy is the Noether charge}",
    eprint = "gr-qc/9307038",
    archivePrefix = "arXiv",
    reportNumber = "EFI-93-42",
    doi = "10.1103/PhysRevD.48.R3427",
    journal = "Phys. Rev. D",
    volume = "48",
    number = "8",
    pages = "R3427--R3431",
    year = "1993"
}

@article{Jafferis:2010un,
    author = "Jafferis, Daniel L.",
    title = "{The Exact Superconformal R-Symmetry Extremizes Z}",
    eprint = "1012.3210",
    archivePrefix = "arXiv",
    primaryClass = "hep-th",
    doi = "10.1007/JHEP05(2012)159",
    journal = "JHEP",
    volume = "05",
    pages = "159",
    year = "2012"
}

@article{Jafferis:2011zi,
    author = "Jafferis, Daniel L. and Klebanov, Igor R. and Pufu, Silviu S. and Safdi, Benjamin R.",
    title = "{Towards the F-Theorem: N=2 Field Theories on the Three-Sphere}",
    eprint = "1103.1181",
    archivePrefix = "arXiv",
    primaryClass = "hep-th",
    reportNumber = "PUPT-2366",
    doi = "10.1007/JHEP06(2011)102",
    journal = "JHEP",
    volume = "06",
    pages = "102",
    year = "2011"
}

@article{Klebanov:2011gs,
    author = "Klebanov, Igor R. and Pufu, Silviu S. and Safdi, Benjamin R.",
    title = "{F-Theorem without Supersymmetry}",
    eprint = "1105.4598",
    archivePrefix = "arXiv",
    primaryClass = "hep-th",
    reportNumber = "PUPT-2377",
    doi = "10.1007/JHEP10(2011)038",
    journal = "JHEP",
    volume = "10",
    pages = "038",
    year = "2011"
}

@article{Couzens:2018wnk,
    author = "Couzens, Christopher and Gauntlett, Jerome P. and Martelli, Dario and Sparks, James",
    title = "{A geometric dual of $c$-extremization}",
    eprint = "1810.11026",
    archivePrefix = "arXiv",
    primaryClass = "hep-th",
    reportNumber = "Imperial/TP/2018/JG/03",
    doi = "10.1007/JHEP01(2019)212",
    journal = "JHEP",
    volume = "01",
    pages = "212",
    year = "2019"
}

@article{Hubeny:2007xt,
    author = "Hubeny, Veronika E. and Rangamani, Mukund and Takayanagi, Tadashi",
    title = "{A Covariant holographic entanglement entropy proposal}",
    eprint = "0705.0016",
    archivePrefix = "arXiv",
    primaryClass = "hep-th",
    reportNumber = "DCPT-07-13, KUNS-2069",
    doi = "10.1088/1126-6708/2007/07/062",
    journal = "JHEP",
    volume = "07",
    pages = "062",
    year = "2007"
}

\end{document}